%% file: min_max_cbf_arXiv1.tex
\documentclass[twocolumn,final]{IEEEtran}

\usepackage{ifthen}
\usepackage[cmex10]{amsmath}
\usepackage{amsthm,amssymb}
\usepackage{epsfig}
\usepackage{xcolor}
\usepackage{subfigure}

\newtheorem{theorem}{Theorem}

\newtheorem{proposition}{Proposition}

\newtheorem{lemma}{Lemma}

\newtheorem{corollary}{Corollary}

\newcommand \bol {\mbox{\boldmath $\lambda$}}
\newcommand \bomu {\mbox{\boldmath $\mu$}}
\newcommand \bovx {\mbox{\boldmath $x$}}

\newcommand \boz {\mbox{\boldmath $z$}}

\newcommand \bone {\mbox{\boldmath $1$}}

\newcommand \bal {\bar{\lambda}}

\newcommand \bobl {\mbox{\boldmath $\bar{\lambda}$}}
\newcommand \bobmu {\mbox{\boldmath $\bar{\mu}$}}

\newcommand \bobld {\mbox{\boldmath $\bar{{\lambda}}$}(\delta)}

\newcommand \bobmud {\mbox{\boldmath $\bar{{\mu}}$}(\delta)}
\newcommand \bobmumd {\mbox{\boldmath $\bar{\mu}$}(-\delta)}

\newcommand \bamu {\bar{\mu}}
\newcommand \bog {\mbox{\boldmath $\gamma$}}

\newcommand \tl {\mbox{$\tilde{\lambda}$}}

\newcommand \botld {\mbox{\boldmath $\tilde{\lambda}$}(\delta)}
\newcommand \botmud {\mbox{\boldmath $\tilde{\mu}$}(\delta)}
\newcommand \botgd {\mbox{\boldmath $\tilde{\gamma}$}(\delta)}

\begin{document}

\title{Min-max fair coordinated beamforming in cellular systems via large systems analysis}

\author
{Randa Zakhour and Stephen V. Hanly
\thanks {
    R. Zakhour is with the Department of Electrical and Computer Engineering,
    University of Texas at Austin: randa.zakhour@gmail.com, S. V. Hanly is with the Department of Electronic Engineering,
    Macquarie University, Australia: svhanly@gmail.com.  This work was supported by the Australian Research Council (ARC) under grant DP0984862, and NUS grant R-263-000-572-133.}
}

\maketitle

\begin{abstract}
This paper considers base station (BS) cooperation in the form of coordinated beamforming, focusing on min-max fairness in the power usage subject to target SINR constraints. We show that the optimal beamforming strategies have an interesting nested zero-forcing structure. In the asymptotic regime where the number of antennas at each BS and the number of users in each cell both grow large with their ratio tending to a finite constant, the dimensionality of the optimization is greatly reduced, and only knowledge of statistics is required to solve it. The optimal solution is characterized in general, and an algorithm is proposed that converges to the optimal transmit parameters, for feasible SINR targets. For the two cell case, a simple single parameter characterization is obtained. These asymptotic results provide insights into the average performance, as well as simple but efficient beamforming strategies for the finite system case. In particular, the optimal beamforming strategy from the large systems analysis only requires the base stations to have local instantaneous channel state information; the remaining parameters of the beamformer can be calculated using channel statistics which can easily be shared amongst the base stations.
\end{abstract}

\section{Introduction}
Base station cooperation in cellular networks has received much recent attention, both in academia and the industry, as a means to raise overall data rate capacity, and as a mechanism to provide improved fairness, in particular by helping the cell boundary mobiles. Different levels of cooperation have been proposed, each requiring differing levels of communication and coordination of the base stations (BSs) transmissions.

One such cooperative scheme on the downlink (DL) is so-called coordinated beamforming (CBf). This was first proposed in \cite{dahrouj_iss08}, and different power related optimizations have since been considered and algorithms developed in \cite{dahrouj_iss08, dahrouj_tw10}. By transforming the precoding design at the BSs into a centralized optimization problem, CBf allows BSs to each serve a disjoint set of users in a much more efficient way than conventional schemes. In conventional schemes, a BS is oblivious, when optimizing its precoder, to the interference it generates in other cells,  although it does take into account the interference experienced at its own mobiles (which does ultimately couple the cells). In CBf, the coordination is explicit in that the precoders are designed jointly by the base stations, at the cost of the increased overhead required to provide each BS with the network-wide channel state information (CSI).

In this paper, we will be focusing on the CBf problem of min-max fairness in transmit power consumption subject to SINR targets at the users. A related formulation appeared in \cite{yu_sp07} in the context of the MIMO broadcast channel with per antenna power constraints, but the analysis and the structure of the optimal precoding strategies is different. As the analysis below will show, the optimal beamforming strategies turn out to have an interesting \emph{nested zero-forcing structure}\footnote{Zero-forcing strategies are usually only optimal in the limit of high SNR, but in our formulation they can be optimal even at finite SNR.}. At the top level, the BSs can be divided into two disjoint groups:
\begin{itemize}
\item a ``selfish'' group (which could consist of all the BSs) whose beamforming follows from solving the Lagrangian dual problem, ignoring the existence of the altruistic (non-selfish) cells altogether; and
\item an ``altruistic'' group (which could consist of all but one BS) whose precoders are orthogonal to the channels of users in the ``selfish'' group: given this constraint, the transmission of the ``altruistic'' group is designed by solving further min-max fairness problems (of reduced dimension) in a recursive manner\footnote{Such recursively defined optimizations also arise in max-min formulations in computer networking (see \cite{bertsekas}).}. This process is repeated until the set of ``altruistic'' BSs is exhausted.
\end{itemize}

These problems are convex and can thus be solved efficiently. However, they require centralized channel knowledge and the optimal beamforming vectors need to be recomputed every time the instantaneous channel changes. Moreover, it is difficult to get insights into average performance without resorting to Monte Carlo (MC) simulations. We thus turn to random matrix theory (RMT) for a way to circumvent these difficulties.

RMT has received considerable attention in the wireless communications literature over the past 15 years, ever since Telatar \cite{telatar_99} applied it to characterizing the capacity region of a point to point MIMO link. RMT large system analysis (LSA) has also allowed analysis of the uplink (UL) of cellular systems \cite{tse_it1999, dai_sp03, tulino_it05, aktas_it06, dumont_it2010}.
Recently, LSA has also been applied to the DL: in \cite{Couillet_2010}, duality between the multiple access channel (MAC) and the broadcast channel (BC) is used to characterize and optimize asymptotic ergodic capacity for correlated channels; Focusing on beamforming strategies, a LSA of regularized ZF (RZF) beam\-forming was undertaken to find its limiting performance for a single cell, allowing the optimization of the regularization parameter \cite{nguyen_globecom08}; \cite{couillet_WDS08} considers ZF and RZF for correlated channel models.

Here, we generalize the results in \cite{zakhour_allerton10,ZH10_archive} to provide a characterization of the asymptotically optimal solution of the min-max power optimization problem and the strategies that achieve it: the latter provide simple to compute precoding vectors for the finite system case which rely on \emph{local} instantaneous CSI only, and involve parameters which depend on the channel statistics alone. More specifically, we focus on the large system regime where the number of antennas at each BS and the number of users in each cell grow large at the same finite rate. Our results assume users in a given cell have independent and identically distributed (iid) channels, but these results can easily be extended to the case where several groups of iid users exist per cell.

Note that a very recent result \cite{lakshminaryana_pimrc10} has applied RMT to a different CBf setup: more particularly, they consider the problem of weighted sum of the transmit powers minimization CBf problem initially formulated in \cite{dahrouj_iss08}, and propose a strategy which also requires instantaneous local CSI and sharing channel statistics.  MC simulations are resorted to in order to claim asymptotical optimality of the results; these are however derived for a more general channel model than we use in the present paper.

The paper is structured as follows. Section \ref{sec:model_pb_formulation} presents the system model and formulates the particular version of the coordinated optimization problem that we consider in the present paper. We then analyze the Karush-Kuhn-Tucker (KKT) conditions \cite{boyd} of the optimization problem and obtain the optimal beamforming structure, from which the result that zero-forcing is sometimes optimal emerges. We then proceed to derive the large system equivalent; an algorithm to solve the latter is proposed, and numerical simulations illustrate the usefulness of applying large system optimal precoding to the finite system, leading to reduced CSI exchange and simplified precoding design.

An important contribution of the paper is the formulation of a convex optimization problem \eqref{eq:SINR_LAS_constraint}-\eqref{eq:mkDef} which fully characterizes the optimal precoding structure in the large system limit. This result is a nontrivial extension of two cell results obtained in \cite{zakhour_allerton10,ZH10_archive}.

\section{System Model}\label{sec:model_pb_formulation}
\begin{figure}[htp]
\begin{center}
\scalebox{0.55}{
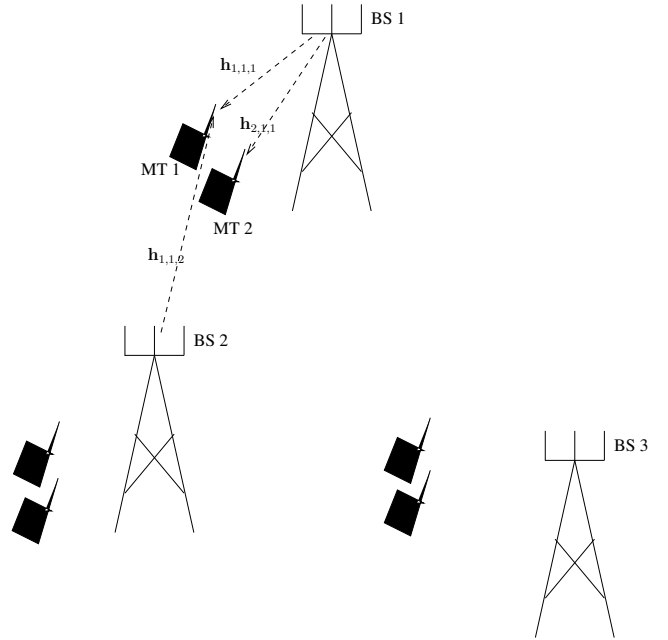}
    \end{center}
\caption{System model for $L = 3$, $N_t = 3$, $U_k = 2$ for $k =1, \ldots, 3$.}
\label{fig:system_model}
\end{figure}

Figure \ref{fig:system_model} depicts the system considered, where $L$ cells each with a base station endowed with $N_t$ antennas and different numbers of single-antenna mobile users, so that cell $k$ has $U_k$ users: we refer to the ratio $\frac{U_k}{N_t}$ as the cell loading of cell $k$. We assume flat fading and denote the channel vector from BS $k$ to user $u$ in cell $j$ by $\mathbf{h}_{u,j,k} \in \mathbb{C}^{1\times N_t}$.

The baseband representation of the received signal at user $u$ in cell $k$ is given by
\begin{align}
y_{u,k} = \sum_{j=1}^L \mathbf{h}_{u,k,j} \mathbf{x}_j + n_{u,k}, \label{eq:rx_signal}
\end{align}
where $\mathbf{x}_{j} \in \mathbb{C}^{N_t}$ denotes BS $j$'s transmit signal, consisting of the sum of the linearly precoded $\mathcal{CN}(0,1)$\footnote{$\mathcal{CN}(0, \sigma^2)$ denotes a zero-mean, variance $\sigma^2$, complex, circularly symmetric Gaussian scalar random variable.} symbols of the users it serves; $n_{u,k} \sim \mathcal{CN}(0, \sigma^2)$ is the receiver noise. Each BS's transmission is subject to a power constraint $P$, so that $\mathbb{E} \left[\mathbf{x}_{j}^H\mathbf{x}_{j}\right] \le P$.
Denoting by $s_{u,j}$ the data symbols intended for user $u$ in cell $j$ and $\mathbf{w}_{u,j}$ the corresponding beamforming vector,
\begin{align}
\mathbf{x}_j = \sum_{u=1}^{U_j} \mathbf{w}_{u,j} s_{u,j}.
\end{align}
Equation \eqref{eq:rx_signal} can thus be rewritten as
\begin{align}
y_{u,k} = \mathbf{h}_{u,k,k} \mathbf{w}_{u,k} s_{u,k} + \sum_{ (u', j) \neq (u, k)} \mathbf{h}_{u,k,j} \mathbf{w}_{u',j} s_{u',j} + n_{u,k},\nonumber
\end{align}
from which we can get the SINR attained at user $u$ in cell $k$, under the assumption that each user treats interference as noise,
\begin{align}
\textrm{SINR}_{u,k} = \frac{\left|\mathbf{h}_{u,k,k} \mathbf{w}_{u,k}\right|^2}{\sigma^2 + \sum_{ (u', j) \neq (u, k)} \left|\mathbf{h}_{u,k,j} \mathbf{w}_{u',j}\right|^2}.\label{eq:SINRexpression}
\end{align}

\subsection{Coordinated Beamforming}

CBf allows the BSs to jointly design their transmissions, a strategy which allows performance gains over the conventional approach in which each BS locally designs its transmission, based only on local CSI, and interference feedback from its own users. A potential downside to CBf is the overhead of passing instantaneous CSI between the base stations. In this paper, we show that large systems analysis allows the simplification in which only the {\it statistics} of the CSI need be passed between base stations (see also \cite{zakhour_allerton10} for a two cell result, and \cite{lakshminaryana_pimrc10} for weighted power minimization).

Let $\gamma_k$ denote the target SINR of users in cell $k$. We formulate the beamforming design problem, $\mathcal{P}_{\textrm{primal}}$, as follows (see \cite{yu_sp07} for a similar objective function in the context of the broadcast channel with per-antenna power constraints):
\begin{align}
\mathcal{P}_{\textrm{primal}}: ~ & \begin{array}{cc} \textrm{min.} & \phi  \sum_k P \nonumber \\ \phi, \{\mathbf{w}_{u,k}\} & \end{array} \nonumber \\
& \textrm{such that (s.t.) } \sum_{u=1}^{U_k} \|\mathbf{w}_{u,k}\|^2 \le \phi P , \forall k= 1, \ldots, L, \nonumber \\
& \quad \quad \quad \quad \quad \quad \textrm{SINR}_{u,k} \ge \gamma_k, \forall k= 1, \ldots, L, \nonumber \\
& \quad \quad \quad \quad \quad \quad \quad \quad \quad \quad \quad \quad \quad u=1, \ldots, U_k. \label{eq:sinr_constraint}
\end{align}
Thus $\phi P$ effectively corresponds to the maximum power consumed at any of the BSs, and the optimal $\phi$ must be  $\le 1$ for the set of SINRs to be attainable in the actual system.

\subsection{Channel model}\label{sec:channel_model}
Users in each cell are assumed to have iid channels, yet be sufficiently distant for their channels to be uncorrelated. Antennas at each BS are also sufficiently apart for uncorrelated Rayleigh fading to model the channel. Thus, entries of $\mathbf{h}_{u,k,j}$ are iid $\mathcal{CN}\left(0, \epsilon_{k,j}\right)$.

\section{Lagrangian Duality and Asymptotic Dual problem}
The SINR equation \eqref{eq:sinr_constraint} can be converted into the following constraint \cite{wiesel_sp06, yu_sp07, dahrouj_tw10}, which, by imposing that $\mathbf{h}_{u,k,k} \mathbf{w}_{u,k}$ be real and strictly positive\footnote{As noted in the references, this does not affect the optimum.}, is equivalent to a second order cone constraint
\begin{align}
& \frac{1}{\gamma_k} |\mathbf{h}_{u,k,k} \mathbf{w}_{u,k}|^2 \ge \sigma^2 + \sum_{(\bar{u}, j) \neq (u,k)} |\mathbf{h}_{u,k,j} \mathbf{w}_{\bar{u},j}|^2. \label{eq:SINR2}
\end{align}

Following \cite{yu_sp07}, the resulting problem is convex and since Slater's condition holds (unless the set of target SINRs is not achievable, even with infinite power), the KKT conditions are necessary and sufficient for optimality, and the duality gap \footnote{the difference between the primal problem's optimum and that of its dual} is zero.

Let $\frac{\lambda_{u,k}}{N_t}$ be the Lagrange coefficient associated with SINR constraint at user $u$ in cell $k$, $\mu_k$ the Lagrange coefficient associated with the power constraint at BS $k$. The Lagrangian $L(\phi, \mathbf{w}_{u,k}, \{\frac{\lambda_{u,k}}{N_t}\}, \{\mu_k\})$ is given by
\begin{align}
&\phi \sum_k P+ \sum_k \mu_k \left[\sum_{u=1}^{U_k} \|\mathbf{w}_{u,k}\|^2 - \phi P\right] + \nonumber \\
&\sum_{u,k} \frac{\lambda_{u,k}}{N_t} \left[\sigma^2 + \sum_{(\bar{u}, j) \neq (u,k)} |\mathbf{h}_{u,k,j} \mathbf{w}_{\bar{u},j}|^2-\frac{1}{\gamma_k} |\mathbf{h}_{u,k,k} \mathbf{w}_{u,k}|^2\right] \nonumber \\
&= \sum_{u,k} \frac{\lambda_{u,k}}{N_t} \sigma^2 + \phi P \sum_k (1-\mu_k) \nonumber \\
&+
\sum_{u,k} \mathbf{w}_{u,k}^H \left[\boldsymbol{\Sigma}_{u,k}-\frac{\lambda_{u,k}}{\gamma_{k}N_t}\mathbf{h}_{u,k,k}^H\mathbf{h}_{u,k,k}\right]\mathbf{w}_{u,k}, \label{eq:Lagrangian}
\end{align}
where
\begin{align}
\boldsymbol{\Sigma}_{u,k} = \mu_k \mathbf{I} + \sum_{(\bar{u}, j) \neq (u,k)} \frac{\lambda_{\bar{u}, j}}{N_t}
\mathbf{h}_{\bar{u},j,k}^H\mathbf{h}_{\bar{u},j,k}.
\end{align}

\subsection{KKT conditions and Optimal precoding structures}
Analyzing the KKT conditions provides insights into the structure of the optimal solution, and reveals cases where zero-forcing a subset of the users is optimal.

In particular, investigation of the possible solutions to the stationarity constraint
\begin{align}
&\left[\boldsymbol{\Sigma}_{u,k}-\frac{\lambda_{u,k}}{\gamma_{k}N_t}\mathbf{h}_{u,k,k}^H\mathbf{h}_{u,k,k}\right]\mathbf{w}_{u,k}  = \boldsymbol{0}, \label{eq:KKT_delL}
\end{align}
leads to the following two lemmas, which distinguish between two types of optimal precoding structures.
For the sake of completeness, the full set of KKT conditions is listed in Appendix \ref{app:KKT}.

\begin{lemma}\label{lemma:nonsingular}
If the optimal $\boldsymbol{\Sigma}_{u,k}$ is nonsingular, the optimal $\lambda_{u,k}$ will be strictly positive and satisfy
\begin{align}
\lambda_{u,k} = \frac{\gamma_k}{\frac{1}{N_t} \mathbf{h}_{u,k,k} \boldsymbol{\Sigma}_{u,k}^{-1} \mathbf{h}_{u,k,k}^H}, \label{eq:lambda_with_equality}
\end{align}
and the optimal $\mathbf{w}_{u,k}$, are of the form
\begin{align}
\mathbf{w}_{u,k} = \delta_{u,k} \boldsymbol{\Sigma}_{u,k}^{-1} \mathbf{h}_{u,k,k}^H, \label{eq:w_opt_nonsing}
\end{align}
where $\delta_{u,k}$ is a scalar to be determined.
\end{lemma}
\begin{IEEEproof}
Refer to Appendix \ref{proof:opt_nonsingular}.
\end{IEEEproof}

\begin{lemma}\label{lemma:singular}
If the optimal $\boldsymbol{\Sigma}_{u,k}$ is rank-deficient, then the following properties hold:
\begin{enumerate}
\item the optimal $\mu_k$ is equal to $0$,
\item $\forall u'$ in cell $k$, $\lambda_{u',k} = 0$,
\item $\forall u'$ in cell $k$, $\boldsymbol{\Sigma}_{u',k} = \boldsymbol{\Sigma}_{k}$
\item $\forall u'$ in cell $k$, $\mathbf{w}_{u',k}$ lies in the null space of $\boldsymbol{\Sigma}_{k}$.
\end{enumerate}
\end{lemma}
\begin{IEEEproof}
The first property is trivial to show, since having a strictly positive $\mu_k$ guarantees nonsingularity of  $\boldsymbol{\Sigma}_{u,k}$.
The remaining properties are proven in Appendix \ref{proof:opt_singular}.
\end{IEEEproof}

This leads to the following corollary, in which we recognize that $\mathbf{w}_{u,k}$ lying in the null space of $\boldsymbol{\Sigma}_{u,k}$, when the latter is singular, is equivalent to the base station zero-forcing to the subset of users with strictly positive $\lambda$'s. Since this does not necessarily include all other mobiles, we shall call this \emph{partial} zero-forcing.
\begin{corollary}\label{lemma:zf_opt}
Partial zero-forcing beamforming at BS $k$ may be optimal only if $\mu_k = 0$. If this is the case for any user in cell $k$, it will be the case for all users in that cell.
\end{corollary}
\begin{IEEEproof}
By Lemma \ref{lemma:nonsingular},  the non-singularity of $\boldsymbol{\Sigma}_{u,k}$ implies beamforming vectors of the form \eqref{eq:w_opt_nonsing} will interfere with all other users. On the other hand, by Lemma \ref{lemma:singular}, if $\boldsymbol{\Sigma}_{u,k}$ is singular, then the beamforming vectors used in cell $k$ lie in the null space of $\boldsymbol{\Sigma}_{k}$. Since in this case $\mu_k = 0$ and $\lambda_{u,k} = 0$ $\forall u \in \textrm{cell } k$, this implies that base station $k$ zero-forces all other-cell mobiles who have strictly positive $\lambda$'s. Indeed, grouping the channels from base station $k$ to the mobiles with positive $\lambda$'s into a matrix $\mathbf{H}_{k, sel}$ ($sel$ stands for selfish, explained below) the beamforming constraint \eqref{eq:KKT_delL} becomes
\begin{align}
\mathbf{H}_{k, sel} \mathbf{w}_{u,k} = \boldsymbol{0},
\end{align}
which is a zero-forcing constraint.
\end{IEEEproof}

Thus, the BSs split into two groups: a ``selfish'' group, whose users have strictly positive optimal $\lambda_{u,k}$'s, and an ``altruistic'' group, whose users have optimal $\lambda_{u,k}$ equal to zero. The altruistic group may be empty, but if not, this group zero-forces the interference to the selfish group, \emph{while still requiring less power than the ``selfish" BSs, at each altruistic BS.}

If we can identify the altruistic group, the problem decomposes into a high level optimization over the selfish base stations (who are oblivious to the zero-forcing altruistic base stations) followed by optimizations over the base stations in the altruistic group (who are impacted by the interference from the optimized selfish base stations). As with the concept of max-min fairness in networking \cite{bertsekas}, these further optimizations are required in order to provide a network-wide min-max fair power allocation: They provide the optimal power and beamforming allocations in the altruistic cells, but have no bearing on the overall min-max power allocation value, as determined in the highest level optimization problem. In summary, we will find
\begin{itemize}
\item ``selfish'' BSs that design their precoding while only considering other selfish base stations; denote this set by $\mathcal{K}_{1, sel}$; their optimal precoding vectors are obtained by solving the highest level of the optimization problem.
\item ``altruistic'' BSs for which zero-forcing with respect to the ``selfish'' set is optimal; their optimal precoding strategies, in a min-max fair power sense, out of all possible ones satisfying this zero-forcing constraint, are obtained by solving, in a reduced dimensional space, a problem of identical structure to $\mathcal{P}_{\textrm{primal}}$.
 Denote the set of ``altruistic'' BSs by $\mathcal{K}_{1, alt}$, so that $\mathcal{K}_{1, sel} \bigcap \mathcal{K}_{1, alt} = \emptyset$ and $\mathcal{K}_{1, sel} \bigcup \mathcal{K}_{1, alt} = \{1, \ldots, L\}$.
 \end{itemize}

To obtain the reduced dimensional problem for the altruistic base stations, let $\boldsymbol{\Pi}^{\perp}_{k, sel}$ denote the projection matrix onto the null space of  $\mathbf{H}_{k, sel}$,
\begin{align}
\boldsymbol{\Pi}^{\perp}_{k, sel} = \mathbf{I}_{N_t} - \mathbf{H}_{k, sel}^H \left(\mathbf{H}_{k, sel}\mathbf{H}_{k, sel}^H \right)^{-1} \mathbf{H}_{k, sel}.
\end{align}
Let $r_{k,sel} = \textrm{rank}\left(\mathbf{H}_{k,sel}\right)$ and $\mathbf{U}_{k, sel}^{\perp} \mathbf{D}_{k, sel}^{\perp} \left(\mathbf{U}_{k, sel}^{\perp}\right)^H$ denote the eigenvalue decomposition of $\boldsymbol{\Pi}^{\perp}_{k, sel}$; the first $N_t - r_{k,sel}$ diagonal elements of $\mathbf{D}_{k, sel}^{\perp}$ are strictly positive, and the remaining elements are zero.\footnote{Statistically, the event that the different user's channels are linearly dependent has measure zero, and the rank of $\mathbf{H}_{k,sel}$ will be $\sum_{j \in \mathcal{K}_{1, sel}} U_j$.}

The beamforming vector $\mathbf{w}_{u,k}$, for a user in cell $k \in \mathcal{K}_{1, alt}$, is such that the last $\textrm{rank}\left(\mathbf{H}_{k,sel}\right)$ elements of $\tilde{\mathbf{w}}_{u,k} = \left(\mathbf{U}_{k, sel}^{\perp}\right)^H\mathbf{w}_{u,k}$ will be zero. We can thus restrict the optimization problem to determining what the nonzero elements of $\tilde{\mathbf{w}}_{u,k}$ should be: let $\bar{\mathbf{w}}_{u,k}$ consist of the first $N_t - r_{k,sel}$ elements. Let $\bar{\mathbf{U}}_{k, sel}^{\perp}$ consist of the first $N_t -r_{k,sel}$ columns of
$\mathbf{U}_{k, sel}^{\perp}$. The corresponding reduced dimension min-max problem is given by
\begin{align}
\quad & \begin{array}{cc} \textrm{min.} & \phi \sum_{k \in \mathcal{K}_{1, alt}} P \\ \phi, \{\mathbf{\bar{w}}_{u,k}\} & \end{array}  ~~ \nonumber \\
& \textrm{s.t. } \quad \quad \quad \quad \sum_{u=1}^{U_k} \|\mathbf{\bar{w}}_{u,k}\|^2 \le \phi P , \forall k \in \mathcal{K}_{1, alt}, \nonumber \\
& \quad \quad\quad \quad \quad \quad \textrm{SINR}_{u,k} \ge \gamma_k, \forall k \in \mathcal{K}_{1, alt}, \nonumber \\
& \quad \quad \quad \quad \quad \quad \quad \quad \quad \quad \quad \quad \quad
 u=1, \ldots, U_k. \label{eq:L1Opt}
\end{align}
where
\begin{align}
\textrm{SINR}_{u,k} = \frac{\left|\bar{\mathbf{h}}_{u,k,k} \bar{\mathbf{w}}_{u,k}\right|^2}{\sigma_{u,k}^2 + \sum_{ (u', j) \neq (u, k), k,j \in \mathcal{K}_{1, alt}} \left|\bar{\mathbf{h}}_{u,k,j} \bar{\mathbf{w}}_{u',j}\right|^2},\nonumber
\end{align}
where $\sigma^2_{u,k}$ is the noise plus interference from the BSs in $\mathcal{K}_{1, sel}$, $\bar{\mathbf{h}}_{u,k,k} = \mathbf{h}_{u,k,k} \bar{\mathbf{U}}_{k, sel}^{\perp} \in \mathbb{C}^{N_t - r_{k,sel}}$.
Due to the independence of the user channels, the fact that entries in ${\mathbf{h}}_{u,k,k}$ are circularly symmetric iid random variables, and since $\mathbf{U}_{k, sel}^{\perp}$ is a unitary matrix, $r_{k,sel} = \sum_{j \in \mathcal{K}_{1,sel}} U_j$, entries of $\bar{\mathbf{h}}_{u,j,k}$ will also be $\mathcal{CN}(0, \epsilon_{j,k})$, and we will be able to apply similar large system results to this subproblem as we will do for the original problem (see Section~\ref{sec:LAS_dual}).

Note that this defines a recursive way to  solve the problem, where at each stage the optimal precoding vectors of the selfish BSs\footnote{These are only selfish with respect to the current level problem; at the highest level, they are altruistic BSs.} are determined and the channels of the corresponding users are used to reduce the dimension of the problem solved at the next level down. The recursion stops when all base stations are selfish. This can occur at the first stage, depending on the parameter settings of the original problem. We now present a dual problem formulation that enables the altruistic base stations to be identified, and which provides the optimal precoding solution for the selfish base stations.

\subsection{Dual Problem Formulation}
The Lagrangian in \eqref{eq:Lagrangian} provides the convex dual program to $\mathcal{P}_{\textrm{primal}}$:
\begin{align}
\mathcal{P}_{\textrm{dual}}: & ~~\begin{array}{cc} \textrm{max. } & \sigma^2 \sum_{u,k} \frac{\lambda_{u,k}}{N_t} \\ \bol, \bomu & \end{array}  \nonumber \\
& \textrm{s.t. } \quad \mu_k \ge 0, \lambda_{u,k} \ge 0  \nonumber \\
& \quad  \quad  \sum_{k} \left(1-\mu_k\right) \ge 0 \nonumber \\
& \quad  \quad \mathbf{v}_{u,k}^H\boldsymbol{\Sigma}_{u,k}\mathbf{v}_{u,k} \ge \frac{\lambda_{u,k}}{\gamma_{k}N_t}|\mathbf{v}_{u,k}^H\mathbf{h}_{u,k,k}^H|^2, \forall \mathbf{v}_{u,k}, \label{eq:SINR_CBf_UL}
\end{align}
where \eqref{eq:SINR_CBf_UL} corresponds to the positive-semidefiniteness constraint  on $\boldsymbol{\Sigma}_{u,k}-\frac{\lambda_{u,k}}{\gamma_{k}N_t}\mathbf{h}_{u,k,k}^H\mathbf{h}_{u,k,k}$. The optimization is over $\{\mu_k\}, \left\{\lambda_{u,k}\right\}, \{\mathbf{v}_{u,k}\}$.

If $\boldsymbol{\Sigma}_{u,k}$ is rank-deficient, the optimal $\lambda_{u,k}$ is zero: since \eqref{eq:SINR_CBf_UL} must hold for any $\mathbf{v}_{u,k}$, rank-deficiency of $\boldsymbol{\Sigma}_{u,k}$ implies we can choose $\mathbf{v}_{u,k}$ such that its left-hand side is zero, but the right-hand side is zero only if $\lambda_{u,k}$ is also zero\footnote{This would not be necessary iff $\mathbf{h}_{u,k,k}$ lies entirely in the range of $\boldsymbol{\Sigma}_{u,k}$, an event that occurs with zero probability.}.
Otherwise, the $\mathbf{v}_{u,k}$ corresponding to the strictest constraint, up to a scalar multiplication, is given by
\begin{align}
\mathbf{v}_{u,k} = \boldsymbol{\Sigma}_{u,k}^{-1}\mathbf{h}_{u,k,k}^H.\label{eq:Vuk}
\end{align}
Plugging this in \eqref{eq:SINR_CBf_UL}, we get
\begin{align}
\frac{1}{\frac{1}{N_t}\mathbf{h}_{u,k,k}\boldsymbol{\Sigma}_{u,k}^{-1}\mathbf{h}_{u,k,k}^H} \ge \frac{\lambda_{u,k}}{\gamma_{k}}.
\end{align}
This constraint will hold with equality at the optimum.

\subsubsection{Downlink power allocation}\label{sec:power_allocation}
Once the dual is solved, the SINR constraints can be used to fully determine the beamforming vectors of users with strictly positive $\lambda_{u,k}$'s. From Lemma \ref{lemma:nonsingular}, we can write
\begin{align}
\mathbf{w}_{u,k} = \sqrt{\frac{p_{u,k}}{N_t}} \frac{\mathbf{v}_{u,k}}{\|\mathbf{v}_{u,k}\|},\label{eq:Wuk_finite}
\end{align}
where $\frac{p_{u,k}}{N_t}$ is the power allocation to user $u$ in cell $k$ and $\mathbf{v}_{u,k}$ is as given by \eqref{eq:Vuk}.
Thus, for user $u$ in cell $k$ belonging to $\mathcal{K}_{1,sel}$, $\textrm{SINR}_{u,k} = \gamma_k$ can be rewritten as
\begin{align}
&\frac{p_{u,k}}{N_t} \frac{\left|\mathbf{h}_{u,k,k} \mathbf{v}_{u,k}\right|^2}{\| \mathbf{v}_{u,k}\|^2} \nonumber \\
&= \gamma_k \left[\sigma^2 + \sum_{ (u', j) \neq (u, k), \in \mathcal{K}_{1,sel}} \frac{p_{u',j}}{N_t}\frac{\left|\mathbf{h}_{u,k,j} \mathbf{v}_{u',j}\right|^2}{\|\mathbf{v}_{u',j}\|^2}\right].\label{eq:SINRConstraintPA}
\end{align}

It is important to note that this solution for power levels at the selfish base stations is the same as that which would be obtained if we were to ignore the altruistic cells altogether. The solution to {\it this} problem is unique: Uniqueness holds because if we fix the $\mu_k$ values assigned to the selfish cells (the only cells in this formulation), the solution for the $\lambda_{u,k}$ variables is unique \cite{yates95}. The dual objective function is then a strictly concave function of the $\mu_k$ variables and so has a unique maximizing solution. Uniqueness in the primal problem follows from uniqueness in the dual. So we can first solve the highest level optimization problem, find the unique optimal power allocation for the selfish base stations, and then use that power allocation to determine the noise levels in the lower level optimizations, where we solve for the power allocation for the altruistic base stations.

Once the power levels in the highest level optimization are determined, the interference generated at users in $\mathcal{K}_{1,alt}$ can be computed. For user $u$ in cell $k$  belonging to $\mathcal{K}_{1,alt}$,
\begin{align}
\sigma^2_{u,k} = \sigma^2 + \sum_{(u',j) \in \mathcal{K}_{1,sel}} \frac{p_{u,k}}{N_t} \frac{\left|\mathbf{h}_{u,k,j}\mathbf{v}_{u',j}\right|^2}{\|\mathbf{v}_{u',j}\|^2},\label{eq:Sigma_ukPA}
\end{align}
is needed to solve the reduced dimension min-max problem in \eqref{eq:L1Opt}. Once the dual of the latter is solved, a similar approach can be used to determine power levels for users there with strictly positive dual Lagrange coefficients. And so on, until all beamforming vectors have been determined.

\subsection{Large System Dual Problem}
\label{sec:LAS_dual}
We now consider the large system regime in which the number of antennas at each base station, $N_t$, grows large ($N_t \rightarrow \infty$), with the ratio $\frac{U_k}{N_t}$, i.e. the cell loading, tending to a finite constant $\beta_k > 0$, with user channels satisfying the model given in Section \ref{sec:channel_model}. In this case, as shown in the following theorem, the number of dual variables to optimize over reduces from $L+\sum_{k=1}^L U_k$ to $2L$.  Moreover, as will later be shown, once these are found, the asymptotically optimal beamformers can be determined, and can be computed using local instantaneous CSI alone.
\begin{theorem}\label{theo:LAS_dual}
If feasible, the optimal $\{\mu_k\}$'s and the empirical distribution (e.d.) of the (normalized) dual variables (i.e. the $\lambda_{u,k}$'s) converge weakly, as $N_t \rightarrow \infty$ with $\frac{U_k}{N_t}\rightarrow \beta_k > 0$, for $k = 1, \ldots, L$,  to the constants obtained by solving the following problem
\begin{align}
\mathcal{P}_{\textrm{dual}}^{\infty}: \quad \quad \quad \quad  & \textrm{max. } \sigma^2 \sum_{k=1}^L \beta_k \lambda_{k} \nonumber  \\
& \textrm{s.t. } \quad \mu_k \ge 0, \lambda_{k} \ge 0  \nonumber \\
& \quad \quad \sum_{k=1}^L \left(1-\mu_k\right) \ge 0  \nonumber \\
& \quad \quad \lambda_k \le F_k(\boldsymbol{\lambda}, \mu_k, \gamma_k), \label{eq:SINR_LAS_constraint}
\end{align}
where
\begin{align}
F_k(\boldsymbol{\lambda}, \mu_k, \gamma_k) = \frac{\gamma_k}{\epsilon_{k,k} \bar{m}_k\left(-\mu_k, \boldsymbol{\lambda}\right)} \label{eq:defFk}
\end{align}
and\footnote{
$I_{x}$ is the indicator corresponding to $x > 0$.}
\begin{align}
\bar{m}_k(-\mu_k, \boldsymbol{\lambda}) &= \left\{\begin{array}{ll}
m_k(-\mu_k, \boldsymbol{\lambda}) & \mu_k > 0,  \textrm{ or } \\
& \left(\mu_k = 0 \textrm{ and } \sum_{j}  \beta_j I_{\lambda_j} > 1 \right) \\
\infty & \textrm{otherwise}
\end{array}\right. \label{eq:mkbarDef} \\
m_k(-\mu_k, \boldsymbol{\lambda}) &= \frac{1}{\mu_k + \sum_{j} \frac{ \beta_j\lambda_j \epsilon_{j,k}}{1+\lambda_j \epsilon_{j,k} m_k(-\mu_k, \boldsymbol{\lambda})}}.\label{eq:mkDef}
\end{align}

\end{theorem}
\begin{IEEEproof}
Refer to Appendix \ref{proof:LAS_dual}.
\end{IEEEproof}

\begin{lemma}
\label{lem:dual_uniqueness}
$\mathcal{P}_{\textrm{dual}}^{\infty}$ is a convex optimization problem with a unique maximizer $(\bomu^*, \bol^*)$.
\end{lemma}
\begin{IEEEproof}
The objective function and all but constraints \eqref{eq:SINR_LAS_constraint} are linear, so trivially convex. That \eqref{eq:SINR_LAS_constraint} is a convex constraint is shown in
 Appendix~\ref{proof:SINR_LAS_constraint}. Uniqueness is shown in Appendix~\ref{proof:dual_uniqueness}.
\end{IEEEproof}

In the following, we will denote the unique maximizer by $(\bomu^*, \bol^*)$, and let $\mathcal{K}^{\infty}_{sel}$ denote the set of cells with $\lambda_k^* > 0$.

To find $(\bomu^*, \bol^*)$ it is useful to write down the Lagrangian for the problem $\mathcal{P}_{\textrm{dual}}^{\infty}$:
\begin{align}
L(\bomu,\bol,\bovx,\boz,z) &= \sigma^2 \sum_{k} \lambda_k \beta_k  + \sum_{k} \mu_k x_k +z\sum_k(1-\mu_k)  \nonumber \\
&~~ + \sum_k z_k \left(F_k(\boldsymbol{\lambda}, \mu_k, \gamma_k)-\lambda_k\right),\label{eq:Lag}
\end{align}
where $\{x_k\}$ are the Lagrange coefficients corresponding to the positivity constraints on $\{\mu_k\}$, $z$ is the Lagrange coefficient corresponding to
$\sum_k(1-\mu_k) \ge 0$ and the $\{z_k\}$ correspond to the $\lambda_k \le F_k(\boldsymbol{\lambda}, \mu_k, \gamma_k)$ constraints. Since $\mathcal{P}_{\textrm{dual}}^{\infty}$ is convex, the KKT conditions are necessary and sufficient for optimality.
\begin{lemma}
\label{lem:KKT_for_dual}
At the optimal solution, $(\bomu^*, \bol^*)$, to $\mathcal{P}_{\textrm{dual}}^{\infty}$, Lagrange variables $(\bovx,\boz,z)$ satisfying the KKT conditions must satisfy the following equations:
\begin{enumerate}
\item $\forall k, x_k \geq 0$
\item $\forall k, \mu_k^* x_k = 0$
\item $\forall k, \lambda_k^* = F_k\left(\bol^*, \mu^*_k, \gamma_k\right)$
\item $\sum_k \mu_k^* = L$
\item $z = \frac{\sigma^2}{L} \sum_{k=1}^L \beta_k \lambda_k^*$
\item $\forall k \in \mathcal{K}^{\infty}_{sel}, x_k = z - P_k$, where  $\{P_k, k \in \mathcal{K}^{\infty}_{sel}\}$ are the unique solution to the linear equations:
\begin{align}
\frac{P_k \epsilon_{k,k}}{\beta_k \left(\sigma^2 + \sum_{j \in \mathcal{K}^{\infty}_{sel}} \frac{P_j \epsilon_{k,j}}{\left(1 + \lambda_k^* \epsilon_{k,j} \frac{\gamma_j}{\epsilon_{j,j} \lambda_j^*}\right)^2}\right)} & = \gamma_k'
\label{eq:downlink_BSpower}
\end{align}
where
\begin{align}
\gamma_k' & = \gamma_k \left[\frac{\mu_k^*+\sum_{j\in \mathcal{K}^{\infty}_{sel}} \frac{\beta_j\lambda_j^* \epsilon_{j,k}}{1+\lambda_j^* \epsilon_{j,k}\frac{\gamma_k}{\epsilon_{k,k} \lambda_k^*}}}{\mu_k^*+\sum_{j\in \mathcal{K}^{\infty}_{sel}} \frac{\beta_j\lambda_j^* \epsilon_{j,k}}{(1+\lambda_j^* \epsilon_{j,k}\frac{\gamma_k}{\epsilon_{k,k} \lambda_k^*})^2}}\right]
\end{align}
\end{enumerate}
\end{lemma}
\begin{IEEEproof}
Conditions (1)-(4) are dual feasibility (for the dual of $\mathcal{P}_{\textrm{dual}}^{\infty}$) and complementary slackness conditions; (5)-(6) are shown in Appendix~\ref{dualKKT}.
\end{IEEEproof}

Appendix \ref{algo:Pdual} presents an algorithm for solving $\mathcal{P}_{\textrm{dual}}^{\infty}$. 

\subsubsection{Downlink power allocation}\label{sec:power_allocation_lsa}
Once $\mathcal{P}_{\textrm{dual}}^{\infty}$ is solved, \eqref{eq:Wuk_finite} specifies the form of the optimal precoding vectors for mobiles in cells with strictly positive $\lambda$'s. 

Consider the following DL power allocation. For cell $k$, with $k \in \mathcal{K}^{\infty}_{sel}$, the base station uses power level $P_k$ as given in \eqref{eq:downlink_BSpower}, and allocates a fraction $\beta_k/N_t$ of this to each user in cell $k$. Thus, defining $p_k = P_k/\beta_k$, user $u$ is allocated the beamforming vector
\begin{align}
\mathbf{w}_{u,k} = \sqrt{\frac{p_k}{N_t}} \frac{\mathbf{v}_{u,k}}{\|\mathbf{v}_{u,k}\|} ,\label{eq:asymptoticW}
\end{align}
with
\begin{align}
&\mathbf{v}_{u,k} \nonumber \\
&
= \hspace{-1mm}\left(\mu_k^*\mathbf{I} + \hspace{-2mm}\sum_{j\in \mathcal{K}^{\infty}_{sel}} \frac{\lambda_j^*}{N_t} \hspace{-2mm}\sum_{u', (u',j) \neq (u,k)} \mathbf{h}_{u',j,k}^H\mathbf{h}_{u',j,k} \right)^{-1} \hspace{-4mm}\mathbf{h}_{u,k,k}^H.
\label{eq:asymptoticV}
\end{align}
Using this allocation,
\begin{align}
P_k = \frac{\sigma^2}{L} \sum_{k=1}^L \lambda_k^* \beta_k - x_k.
\end{align}
For $k \in \mathcal{K}^{\infty}_{sel}$, with $\mu_k^* > 0$, we have $x_k = 0$, so for such $k$ we have
\begin{align}
P_k = \frac{\sigma^2}{L} \sum_{k=1}^L \lambda_k^* \beta_k.
\end{align}
For $k \in \mathcal{K}^{\infty}_{sel}$, with $\mu_k^* = 0$, we have
\begin{align}
P_k \leq \frac{\sigma^2}{L} \sum_{k=1}^L \lambda_k^* \beta_k.
\end{align}
It follows that this power allocation achieves a primal value of $\sigma^2 \sum_{k=1}^L \lambda_k^* \beta_k$, which is the asymptotically optimal dual value. Thus, provided this power allocation is asymptotically feasible, it follows from the uniqueness of the solution to the highest layer primal optimization problem, that it must be the asymptotically optimal primal power allocation. Asymptotic feasibility is shown in the following lemmas.

\begin{lemma}
For $k \in \mathcal{K}^{\infty}_{sel}$,
\begin{align}
&p_k \epsilon_{k,k} m_k(-\mu_k^*, \boldsymbol{\lambda}^*) \left(\mu_k^*+\sum_{j \in \mathcal{K}^{\infty}_{sel}} \frac{\beta_j\lambda_j^* \epsilon_{j,k}}{(1+\lambda_j^* \epsilon_{j,k}\frac{\gamma_k}{\epsilon_{k,k} \lambda_k^*})^2}\right) = \nonumber \\
&\gamma_k \left(\sigma^2 + \sum_{j \in \mathcal{K}^{\infty}_{sel}} \frac{P_j \epsilon_{k,j}}{\left(1 + \lambda_k^* \epsilon_{k,j} \frac{\gamma_j}{\epsilon_{j,j} \lambda_j^*}\right)^2}\right).
\label{eq:asymptotic_equality}
\end{align}
\end{lemma}
\begin{IEEEproof}
Rearrange \eqref{eq:downlink_BSpower}, and substitute $m_k(-\mu_k^*, \boldsymbol{\lambda}^*)$ for $\left(\mu_k^*+\sum_{j \in \mathcal{K}^{\infty}_{sel}} \frac{\beta_j\lambda_j^* \epsilon_{j,k}}{1+\lambda_j^* \epsilon_{j,k}\frac{\gamma_k}{\epsilon_{k,k} \lambda_k^*}}\right)^{-1}$.
\end{IEEEproof}

\begin{lemma}
\label{lem:asymptotic_equality}
With $p_{uk} = p_k$ for all users $u$ in cell $k$, $k \in \mathcal{K}^{\infty}_{sel}$, we have that the left hand side of \eqref{eq:SINRConstraintPA} converges to the left hand side of \eqref{eq:asymptotic_equality}, and the right hand side of \eqref{eq:SINRConstraintPA} converges to the right hand side of \eqref{eq:asymptotic_equality}.
\end{lemma}
\begin{IEEEproof}
See Appendix~\ref{weak_convergence}.
\end{IEEEproof}

\begin{corollary}
With $p_{uk} = p_k$ for all users $u$ in cell $k$, $k \in \mathcal{K}^{\infty}_{sel}$, we have that the left hand side of \eqref{eq:SINRConstraintPA} converges to the same value that the right hand side of \eqref{eq:SINRConstraintPA} converges to, and hence that $\mbox{SINR}_{uk} \rightarrow \gamma_k$ for all users $u$ in cell $k$.
\end{corollary}
We conclude that the allocation of powers and beamforming vectors to the users served by the selfish BSs, as given in \eqref{eq:asymptoticW}, is asymptotically optimal for the highest level primal optimization problem.

If $\mathcal{K}^{\infty}_{alt}$ is nonempty, it remains to solve the lower level optimization problems. The first step is to compute the interference generated to users in $\mathcal{K}^{\infty}_{alt}$:

\begin{lemma}\label{lemma:noise}
Let $\mathcal{K}^{\infty}_{alt}$ denote the set of cells with optimal $\lambda_{u,k}$'s converging weakly to $0$\footnote{$\mathcal{K}^{\infty}_{alt}$ is obtained by solving $\mathcal{P}_{\textrm{dual}}^{\infty}$.}. Then
$\sigma^2_{u,k}$ (cf. \eqref{eq:Sigma_ukPA}) at users in cells in $\mathcal{K}^{\infty}_{alt}$, converge to constants $\sigma^2_k$ in the large system regime, given by
\begin{align}
\sigma^2_k = \sigma^2 + \sum_{j \in \mathcal{K}^{\infty}_{sel}} P_j \epsilon_{k,j},\label{eq:Sigma2K}
\end{align}
where $P_j$'s are as defined in \eqref{eq:downlink_BSpower}.
\end{lemma}
\begin{IEEEproof}
Refer to Appendix \ref{weak_convergence}.
\end{IEEEproof}

For cells with zero $\lambda_k^*$'s, the large system equivalent dual of \eqref{eq:L1Opt} will amount to solving a problem of the same form as $\mathcal{P}_{\textrm{dual}}^{\infty}$ with $L$ and $\beta_k$'s replaced by their values in the reduced space and the noise power $\sigma^2$ for cell $k$ replaced by $\sigma_k^2$, the asymptotic noise plus interference at any of its users, as specified by Lemma \ref{lemma:noise}. This will be illustrated in the next section for the two cell setup.

To summarize, unlike their finite system counterparts, which require full CSI, the asymptotically optimal $\lambda_k$'s, $\mu_k$'s and $p_k$'s require only statistical CSI knowledge to compute. Once these are determined, asymptotically optimal beamforming vectors only require \emph{local} CSI, in the form of channels from the BS in question to all users, to be implemented, as in \eqref{eq:asymptoticW}-\eqref{eq:asymptoticV}. As will be shown in Section \ref{sec:num_res}, they are useful even when the numbers of antennas and users per cell are quite small.

\section{Two cell case}\label{sec:two_cell}
For scenarios with only two cells, $\mathcal{P}_{\textrm{dual}}^{\infty}$ can be solved in a much simpler way (by examining three simple functions, $g_1$, $g_2$ and $h$, see below) and its feasibility can be characterized, as shown in Lemma \ref{lemm:2cellfeas} below. In this section, we further analyze the zero-forcing case and determine the downlink asymptotically optimal beamforming strategies and power allocations.

Let $c_k  = 1-\frac{\beta_k \gamma_k}{1+\gamma_k}$. The condition $c_k > 0$ corresponds to target SINR $\gamma_k$ being achievable under cell loading $\beta_k$ in the asymptotic regime, \emph{if cell $k$ were isolated}\footnote{If $c_k < 0$, then clearly the problem is infeasible.}. As shown in Appendix \ref{proof:2cellfeas}, in the two cell case, $\mathcal{P}_{\textrm{dual}}^{\infty}$ can be reduced to an optimization over a single parameter $\rho$ representing the ratio $\frac{\lambda_2}{\lambda_1}$. The following lemma characterizes its feasibility.

\begin{lemma}[Boundedness of $\mathcal{P}_{\textrm{dual}}^{\infty}$ in the two cell case]\label{lemm:2cellfeas}
Assume $c_k > 0, k = 1, 2$, then the optimum of $\mathcal{P}_{\textrm{dual}}^{\infty}$ will be bounded if one of the following holds:
\begin{enumerate}
\item $c_1 - \beta_2 \ge 0$ or $c_2 - \beta_1 \ge 0$,
\item $c_1 - \beta_2 < 0$, $c_2 - \beta_1 < 0$, and
\begin{align}
& \frac{\epsilon_{1,1} c_1\epsilon_{2,2} c_2}{\gamma_1\gamma_2} \ge \epsilon_{1,2} \epsilon_{2,1} \left[\beta_{1}-c_2\right]\left[\beta_{2}-c_1\right]. \label{eq:FeasibilityCBf}
\end{align}
\end{enumerate}
\end{lemma}
\begin{IEEEproof}
This is shown in Appendix \ref{proof:2cellfeas}, where we show that the problem in the two-cell case can be reduced to an optimization over a single parameter $\rho = \frac{\lambda_2}{\lambda_1}$ with simple upper and lower bound constraints.
\end{IEEEproof}

Note that the first condition for feasibility is independent of the average channel gains: It corresponds to the scenario that either cell is sufficiently underloaded as to be able to accommodate its own users {\it irrespective} of the target SINR level in the other cell. When this condition holds, the underloaded cell can zero-force its interference to the other cell, and the other cell's SINR target is then necessarily feasible, since it is effectively an isolated cell (recall that $c_k > 0$). If the first condition fails to hold (this will normally be the case), the further condition \eqref{eq:FeasibilityCBf} (which does depend on the $\epsilon_{k,j}$'s) needs to be satisfied.

\begin{figure*}[htb]
\center{
\subfigure[$\rho^* = \infty$ ($\beta_1 = .1, \beta_2 = .5, \gamma_1 = \gamma_2 = 5$)]{
\includegraphics[width=2.15in, height=2in, viewport = 30 100 560 645, clip]{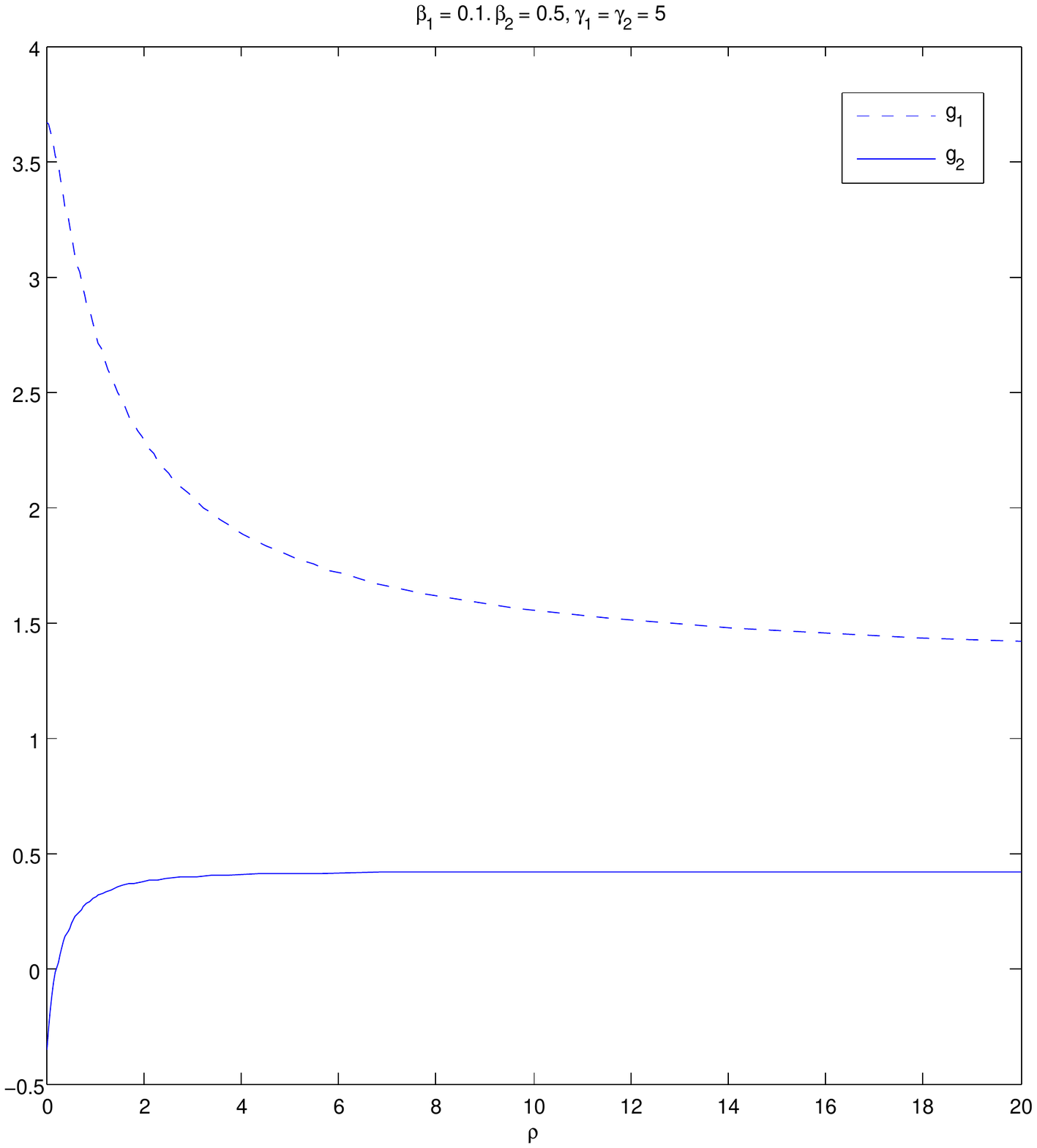}
}
\subfigure[$\rho^*$ at intersection ($\beta_1 = .55, \beta_2 = .5,  \gamma_1 = \gamma_2 = 5$)]{
\includegraphics[width=2.3in, height=2in, viewport = 30 100 560 655, clip]{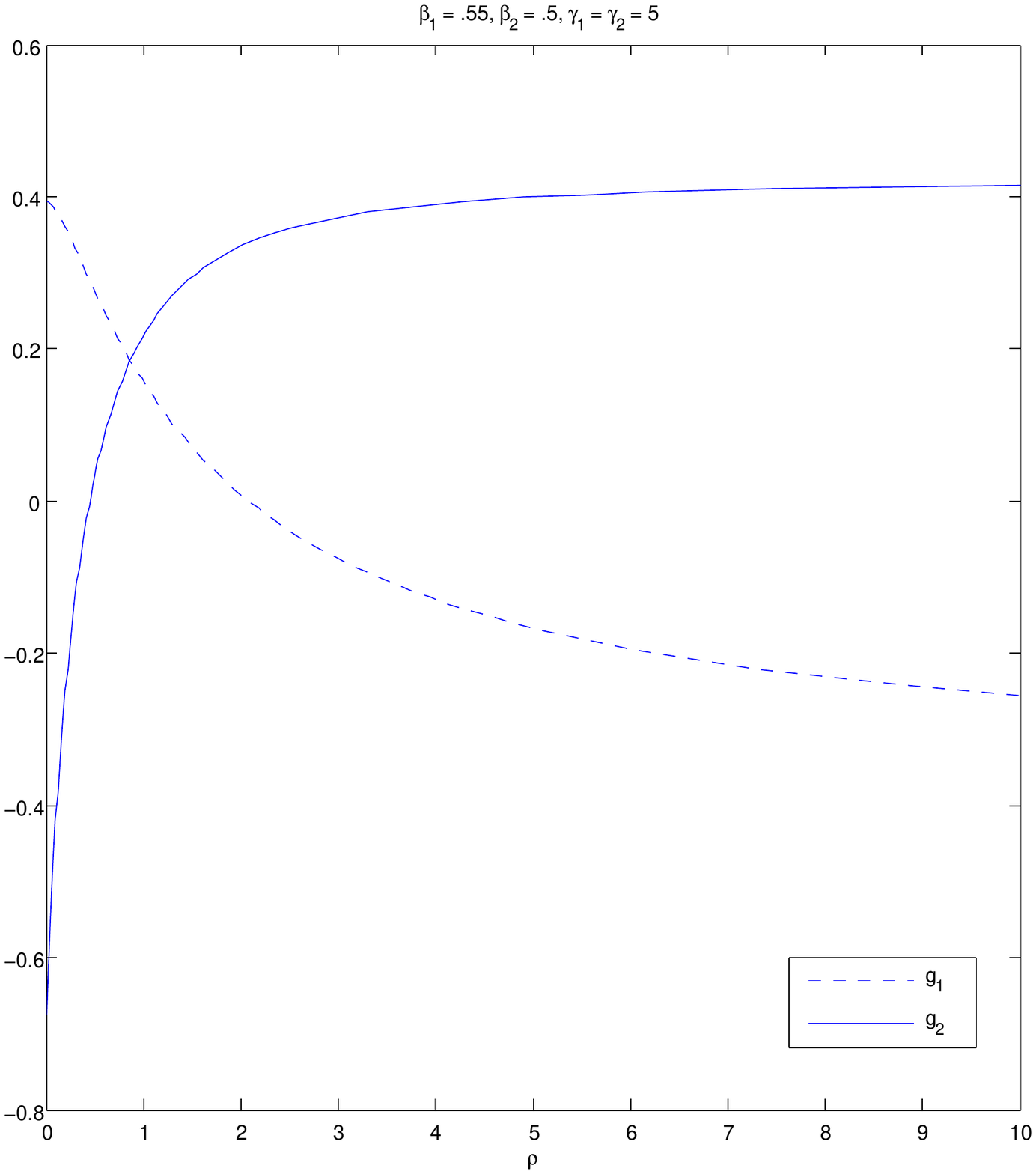}
}
\subfigure[$\rho^* = 0$ ($\beta_1 = .6, \beta_2 = .2,  \gamma_1 = 5, \gamma_2 = 2$)]{
\includegraphics[width=2.2in, height=2in, viewport = 30 100 555 640, clip]{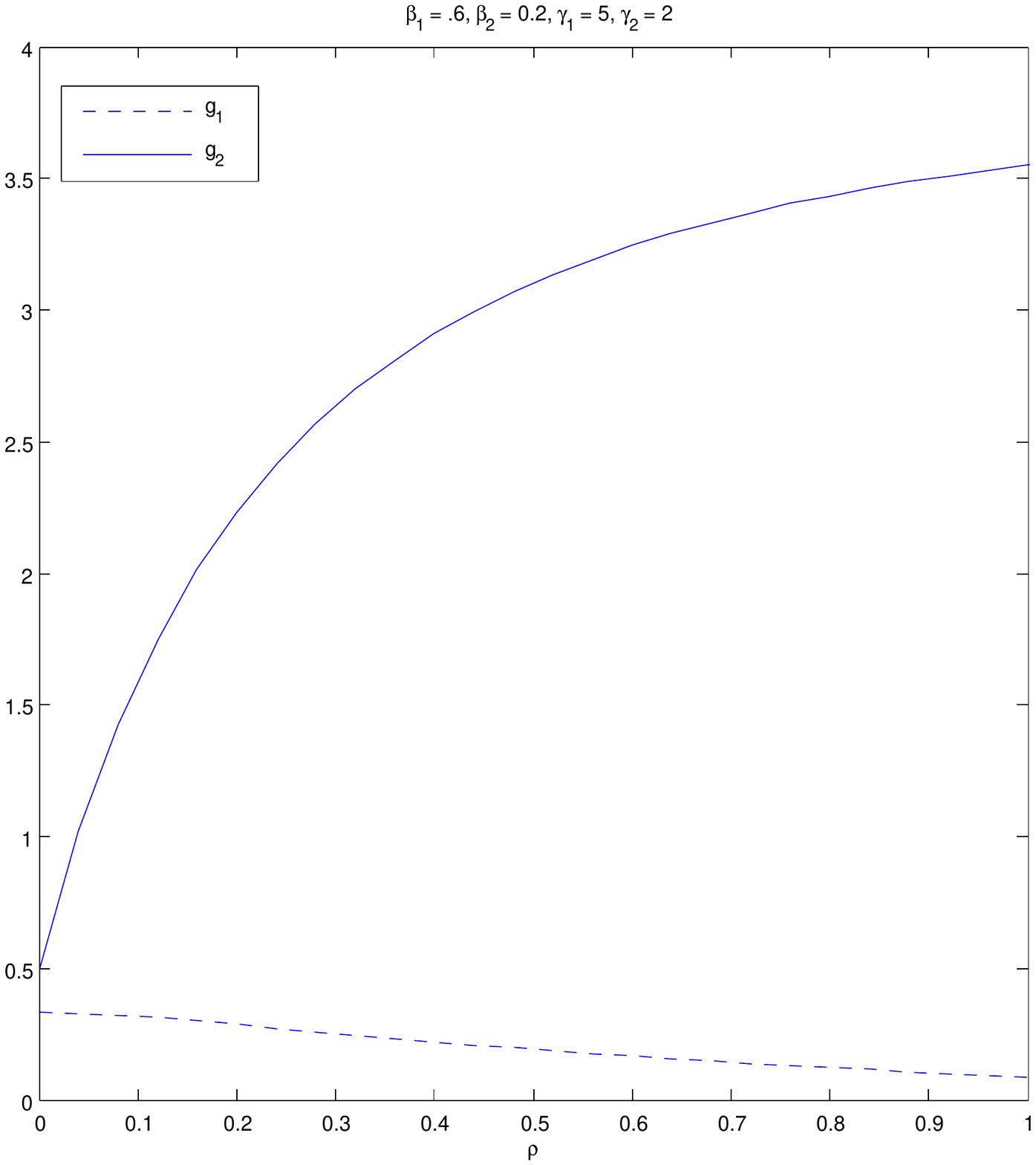}
}
\caption{Optimal $\rho$ for different cell loadings, target SINRs, $\epsilon_{1,1} = 2, \epsilon_{1,2} = .5, \epsilon_{2,1} = .7, \epsilon_{2,2} = 1.8$.} \label{fig:g1g2Optimal_rho}
}
\end{figure*}

\begin{theorem}[$\mathcal{P}_{\textrm{dual}}^{\infty}$ solution for two cells] \label{theo:2cell}
Let
\begin{align}
\rho_{lo} &= \left\{\begin{array}{ll} 0 & c_2 - \beta_1 \ge 0 \\ \frac{\epsilon_{1,2}\gamma_{2}}{\epsilon_{2,2} c_2}(\beta_1-c_2) & \textrm{otherwise} \end{array}\right. \label{eq:rho_min}\\
\rho_{hi} &= \left\{\begin{array}{ll}\infty & c_1- \beta_2 \ge 0 \\ \frac{\epsilon_{1,1} c_1}{\gamma_1 \epsilon_{2,1}(\beta_2-c_1)} & \textrm{otherwise} \end{array}\right. \label{eq:rho_max}
\end{align}
and define $h(.)$, $g_1(.)$ and $g_2(.)$ as follows:
\begin{align}
&h(\rho) \nonumber \\
&=\epsilon_{1,1} \left[\frac{c_1}{\gamma_1}- \frac{\beta_{2}\epsilon_{2,1}\rho}{\epsilon_{1,1} + \rho \epsilon_{2,1}  \gamma_1}\right]
+\rho \epsilon_{2,2} \left[\frac{c_2}{\gamma_2}- \frac{\beta_{1} \epsilon_{1,2}}{\epsilon_{2,2} \rho + \epsilon_{1,2}  \gamma_2}\right], \label{eq:Lambda1Cbf} \\
&g_1(\rho) \nonumber \\
&= \frac{\epsilon_{1,1}c_1}{\beta_1\gamma_1}
-  \frac{\frac{\beta_2}{\beta_1}\epsilon_{1,1} \epsilon_{2,1}^2 \gamma_1}{\left(\epsilon_{1,1}\frac{1}{\rho} +  \epsilon_{2,1}\gamma_1\right)^2} -
\frac{\epsilon_{2,2}^2 \epsilon_{1,2}}{\left(\epsilon_{2,2} + \frac{1}{\rho} \epsilon_{1,2}\gamma_2\right)^2}, \label{eq:g1} \\
&g_2(\rho) \nonumber \\
&=\frac{\epsilon_{2,2}c_2}{\beta_2\gamma_2}
-\frac{\frac{\beta_1}{\beta_2} \epsilon_{2,2} \epsilon_{1,2}^2 \gamma_2}{\left(\epsilon_{2,2}\rho + \epsilon_{1,2}\gamma_2\right)^2} -
\frac{\epsilon_{1,1}^2 \epsilon_{2,1}}{\left(\epsilon_{1,1} + \rho \epsilon_{2,1}\gamma_1\right)^2}. \label{eq:g2}
\end{align}
For feasible $\mathcal{P}_{\textrm{dual}}^{\infty}$, let $\rho^*$ be equal to
\begin{itemize}
\item $\rho_{lo}$, if $g_1(\rho_{lo})-g_2(\rho_{lo}) \le 0$, $g_1(\rho_{hi})-g_2(\rho_{hi}) < 0$,
\item $\rho_{hi}$, if $g_1(\rho_{lo})-g_2(\rho_{lo}) > 0$, $g_1(\rho_{hi})-g_2(\rho_{hi}) \ge 0$,
\item the value of $\rho$ at which $g_1(.)$ and $g_2(.)$ intersect, otherwise.
\end{itemize}
The optimal $(\lambda_1, \lambda_2)$ pair, $(\lambda_1^*, \lambda_2^*)$, is given by
\begin{itemize}
\item $\lambda_1^* = 0$, and $\lambda_2^* = \frac{2\gamma_2}{\epsilon_{2,2}c_2}$, if $\rho^* = \infty$ (cell 1 is zero-forcing cell 2);
\item $\lambda_1^* = \frac{2}{h(\rho^*)}$, $\lambda_2^* = \frac{2\rho^*}{h(\rho^*)}$, otherwise.
\end{itemize}

The optimal $(\mu_1, \mu_2)$ pair, $(\mu_1^*, \mu_2^*)$, is given by
\begin{align}
\mu_1^* &= \epsilon_{1,1} \lambda_1^* \left[\frac{c_1}{\gamma_1} -  \frac{\beta_{2}\epsilon_{2,1} \lambda_{2}^*}{\epsilon_{1,1}\lambda_1^* + \epsilon_{{2},1} \lambda_{2}^* \gamma_1}\right], \nonumber \\
\mu_2^* &= 2-\mu_1^* = \epsilon_{2,2} \lambda_2^* \left[\frac{c_2}{\gamma_2} -  \frac{\beta_{1}\epsilon_{1,2} \lambda_{1}^*}{\epsilon_{2,2}\lambda_2^* + \epsilon_{{1},2} \lambda_{1}^* \gamma_2}\right].
\end{align}
\end{theorem}
\begin{IEEEproof}
This is shown in Appendix \ref{proof:theo2cell}, by solving the equivalent problem in terms of $\rho = \frac{\lambda_2}{\lambda_1}$.
\end{IEEEproof}
Figure \ref{fig:g1g2Optimal_rho} illustrates the different cases that may arise. The monotonicity of both functions is key to obtaining the result. The values of $g_1$ and $g_2$ at $\rho_{lo}$ and $\rho_{hi}$ can easily be computed (by taking a limit if $\rho_{hi} = \infty$) to verify whether an intersection point exists. If it does, it can be found by a bisection method.

Optimal zero-forcing configurations are characterized in the following corollary.

\begin{corollary}[Zero-forcing optimality conditions for two cells]
It is asymptotically optimal for cell $k$ to zero-force if
\begin{align}
\frac{\epsilon_{k,k}c_k}{\beta_k\gamma_k}  - \frac{\epsilon_{\bar{k},\bar{k}}(c_{\bar{k}}-\beta_k)}{\beta_{\bar{k}}\gamma_{\bar{k}}} + \epsilon_{\bar{k},k} \le 0.
\end{align}
\end{corollary}
\begin{IEEEproof}
This follows from analyzing the cases in Theorem \ref{theo:2cell}, when $\rho_{lo} = 0$, alternatively, $\rho_{hi} = \infty$ and checking when these are optimal.
\end{IEEEproof}

\subsection{Downlink power allocation}
We now illustrate the different cases that may arise and the corresponding asymptotically optimal beamforming.
\subsubsection{$\rho^* \notin \{0, \infty \}$}
In this case, neither BS is zero-forcing the other's users, so that asymptotically optimal beamformers are of the form given in \eqref{eq:asymptoticW}-\eqref{eq:asymptoticV}, and
 the asymptotic power levels satisfy \eqref{eq:downlink_BSpower}.
Thus,
\begin{align}
&P_1 \frac{\epsilon_{1,1}}{\beta_1} \left[\frac{c_1}{\gamma_1} -  \frac{\beta_{2}\rho^*\epsilon_{2,1}
\left(\rho^* \epsilon_{2,1} \gamma_1\right)}{\left(\epsilon_{1,1}+\rho^* \epsilon_{2,1} \gamma_1\right)^2}\right]-\frac{\epsilon_{1,2} P_2}{\left(1+ \epsilon_{1,2}\frac{\gamma_2}{\epsilon_{2,2} \rho^*}\right)^2} \nonumber \\
&= \sigma^2 \nonumber \\
&- P_1\frac{\epsilon_{2,1}\epsilon_{1,1}^2}{\left(\epsilon_{1,1}+\rho^* \epsilon_{2,1}\gamma_1\right)^2}+P_2 \frac{\epsilon_{2,2}}{ \beta_2} \left[\frac{c_2}{\gamma_2} -  \frac{\beta_{1}\epsilon_{1,2}\left(\epsilon_{1,2} \gamma_2\right)}{\left(\epsilon_{2,2} \rho^*+\epsilon_{1,2} \gamma_2\right)^2}\right] \nonumber \\
&= \sigma^2.
\end{align}

If $\rho^*$ corresponds to the intersection of $g_1$ and $g_2$, the solution simplifies to
\begin{align}
P_1 = P_2 = \frac{\sigma^2}{g_1(\rho^*)} = \frac{\sigma^2}{g_2(\rho^*)} = \sigma^2 \frac{\beta_1 + \rho^* \beta_2}{h(\rho^*)}. 
\end{align}

\subsubsection{$\rho^* = \infty$}
Cell 1 will be zero-forcing to cell 2's users.
Equation \eqref{eq:downlink_BSpower} for cell 2 simplifies to
\begin{align}
P_2 =  \sigma^2 \beta_2 \frac{\gamma_2}{\epsilon_{2,2}c_2}.
\end{align}
The asymptotic noise plus interference due to cell 2's transmission term at users in cell 1, $\sigma_1^2$, will be given by (cf. \eqref{eq:Sigma2K})
\begin{align}
\sigma^2_1 = \sigma^2 + P_2 \epsilon_{1,2}.
\end{align}

We now focus on finding the asymptotically optimal beamforming vectors and power allocation for users in cell 1, i.e. solving problem \eqref{eq:L1Opt}, for $\mathcal{K}_{1, sel} = \{2\}$ and $\mathcal{K}_{1, alt} = \{1\}$. This becomes:
\begin{align}
\textrm{min. } & \phi P \nonumber \\
\textrm{s.t. } & \frac{1}{\gamma_1} |\mathbf{\bar{h}}_{u,1,1} \mathbf{\bar{w}}_{u,1}|^2 \ge \sigma_{u,1}^2 + \sum_{\bar{u} \neq u} |\mathbf{\bar{h}}_{u,1,1}\mathbf{\bar{w}}_{\bar{u},1}|^2  \nonumber \\
& \sum_{u=1}^{U_1} \|\mathbf{\bar{w}}_{u,1}\|^2\le \phi P, \nonumber
\end{align}
Since $BS$ 1 is zero-forcing to users in cell 2, $\mathbf{\bar{h}}_{u,1,1}$ are $N_t - U_2$ dimensional row vectors, and as noted previously, its entries are iid $\mathcal{CN}\left(0, \epsilon_{1,1}\right)$. Thus, in the definition of the large system dual, cell loading $\beta_1$ is reduced by a factor of $1-\beta_2$. Letting the new large system dual variables be denoted $\bar{\mu}_1$ and $\bar{\lambda}_1$, to distinguish them from $\mu_1$ and $\lambda_1$ in the optimization at the first recursion (both were zero), the large system dual becomes
\begin{align}
\textrm{max. } &  \sigma_1^2\frac{\beta_1}{1-\beta_2} \bar{\lambda}_{1} \nonumber  \\
 \textrm{s.t. } & \quad \bar{\mu}_1 \ge 0, \bar{\lambda}_{1} \ge 0, 
  1-\bar{\mu}_1 \ge 0  \nonumber \\
&  \bar{\lambda}_1 \le \frac{\gamma_1}{\epsilon_{1,1} \bar{m}_1\left(-\bar{\mu}_1, \bar{\lambda}_1\right)},
\end{align}
with
\begin{align}
\bar{m}_1(-\bar{\mu}_1, \bar{\lambda}_1)  = \frac{1}{\bar{\mu}_1 + \frac{\frac{\beta_1}{1-\beta_2}\bar{\lambda}_1 \epsilon_{1,1}}{1+\bar{\lambda}_1 \epsilon_{1,1} \bar{m}_1(-\bar{\mu}_1, \bar{\lambda}_1)}}.
\end{align}

It is trivial to verify that the optimal $\bar{\mu}_1$ will be equal to 1, and the optimal $\bar{\lambda}_1$ is equal to
\begin{align}
\bar{\lambda}_1  = \frac{1}{\epsilon_{1,1}\left[\frac{1}{\gamma_1}-\frac{\beta_1}{1-\beta_2}\frac{1}{1+\gamma_1}\right]}.
\end{align}

The total transmit power at BS 1 thus converges to (cf. \eqref{eq:downlink_BSpower})
\begin{align}
P_1 = \frac{\beta_1}{1-\beta_2} \bar{\lambda}_1 \sigma^2_1,
\end{align}
and the asymptotically optimal $\bar{\mathbf{w}}_{u,1}$ will be of the form
\begin{align}
\bar{\mathbf{w}}_{u,1} = \sqrt{\frac{p_{1}}{N_t-U_2}} \frac{\bar{\mathbf{v}}_{u,1}}{\|\bar{\mathbf{v}}_{u,1}\|}
\end{align}
where
\begin{align}
p_1 = \frac{1-\beta_2}{\beta_1} P_1 = \bar{\lambda}_1 \sigma^2_1,
\end{align}
and
\begin{align}
\bar{\mathbf{v}}_{u,1} = \left[\mathbf{I}  +
  \sum_{\bar{u} \neq u} \frac{\bar{\lambda}_1}{N_t-U_2}\mathbf{\bar{h}}_{\bar{u},1,1}^H\mathbf{\bar{h}}_{\bar{u},1,1}\right]^{-1}\mathbf{\bar{h}}_{u,1,1}^H.
\end{align}

\subsection{$\rho^* = 0$}
This is the case where BS 2 is zero-forcing to cell 1's users. Exactly the same derivations as in the previous subsection hold, with the roles of BS 1 and BS 2 interchanged.

\section{Numerical Results}\label{sec:num_res}

In this section, we illustrate the applicability of the above large system results to the finite system case. For the finite system, we will consider an overall rate maximization optimization problem, but power minimization can be considered as a subroutine to solve it, and in that way we can utilize the above large systems results. The rate maximization problem that we formulate in this section can be solved in a finite system, but it is computationally very intensive to do so. Our interest is in suboptimal solutions that can be obtained from the large systems analysis.

In the following, we use a rate profile, as introduced in \cite{mohseni_jsac06}, to characterize the rate region boundary of a multi-user channel, as an alternative to weighted sum rate maximization. Thus, let the rate profile $\boldsymbol{\alpha} = \left\{\alpha_1, \ldots, \alpha_L\right\}$, be such that $\sum_{k=1}^L \alpha_k = 1$: $\boldsymbol{\alpha}$ specifies how the sum rate is split across the cells. We impose the constraint that identical rates are maintained across users in the same cell, denote the rate in cell $k$ by $r_k$, and note that the sum rate in cell $k$, normalized by $N_t$, will be $\frac{U_k}{N_t} r_k$. For a given channel realization, and for a given $\boldsymbol{\alpha}$, we define the following optimization problem over the beamforming vectors $\mathbf{w}_{u,k,k}$, for $k = 1, \ldots, L$, $u = 1, \ldots, U_k$:
\begin{align}
\mathcal{P}_{\boldsymbol{\alpha}}: &~ \textrm{max. }  r \nonumber \\
& ~\textrm{s.t. }  \frac{U_k}{N_t} r_k = \alpha_k r, \quad k = 1, \ldots, L \nonumber \\
& ~\textrm{SINR}_{u,k} \ge 2^{r_k}-1, \nonumber \\
& ~\quad \quad  \quad \quad  k = 1, \ldots, L, u = 1, \ldots, U_k \nonumber \\
& ~\sum_{u=1}^{U_k} \|\mathbf{w}_{u,k}\|^2 \le P , \forall k= 1, \ldots, L.  \nonumber
\end{align}
This formulation takes into account fairness across mobiles in the network via the rate profile. However, subject to the fairness prescribed by $\boldsymbol{\alpha}$, all mobiles seek as much rate as possible. For a given channel realization, it can be solved by a bisection method over the maximal sum rate $r$. Feasibility of a fixed $r$ can be determined by solving $\mathcal{P}_{\textrm{primal}}$ with\footnote{From Shannon's capacity formula, $r_k = \log_2 (1+\gamma_k)$.} $\gamma_k = 2^{\alpha_k r \frac{N_t}{U_k}}-1$: if  $\mathcal{P}_{\textrm{primal}}$ is infeasible or if it is feasible but the corresponding optimal $\phi$ is strictly greater than 1, then $r$ cannot be achieved. Note that all mobiles get the same rate in each channel state, but the rate will vary across the channel states.

A large system equivalent to $\mathcal{P}_{\boldsymbol{\alpha}}$ can be solved along similar lines, where
in the considered asymptotic regime, feasibility of a given $r$ is ascertained by solving the corresponding $\mathcal{P}_{\textrm{dual}}^{\infty}$ with $\gamma_k = 2^{\frac{\alpha_k r}{\beta_k}}-1$.

For the numerical results, we will focus on a two cell example, for simplicity. Since the rates achieved depend on the channel state, we will measure average performance, averaged over the random channel parameters of the mobiles. In this way, we can construct an average rate region. We start by comparing the average rate region for a system with small number of antennas at the base stations ($N_t = 4$) and small number of users in each cell ($U_1 = 2$, $U_2 = 3$), to the rate region corresponding to the large system obtained by letting $N_t, U_1$ and $U_2$ grow large such that the ratio $U_k/N_t$ tend to their finite system values ($\frac{U_1}{N_t} \rightarrow \frac{2}{4}$ and $\frac{U_2}{N_t} \rightarrow \frac{3}{4}$). In both cases, channels of users in a given cell have identical  statistics, as specified by our model.

\begin{figure}[htb]
\center{
\includegraphics[width=3.3in, height=2.5in, viewport = 80 220 550 580, clip]{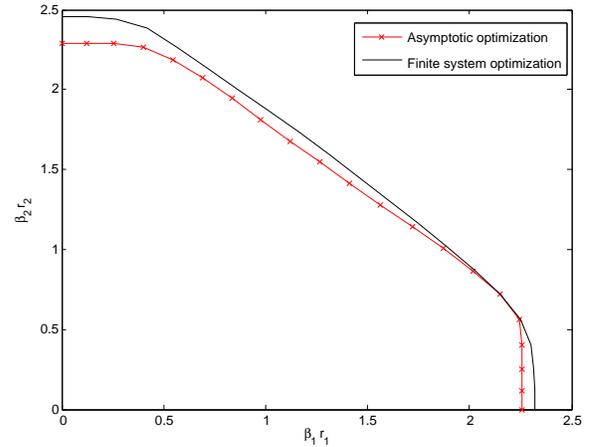}
\caption{Ergodic rate region comparison, finite system vs. asymptotic regime: $\frac{U_1}{N_t} = \beta_1 = 2/4$, $\frac{U_2}{N_t} = \beta_2 = 3/4$, $\epsilon_{1,1} =  2.1$, $\epsilon_{1,2}=.6$, $\epsilon_{2,1} = .8$, $\epsilon_{2,2} =1.6$, SNR = 10dB. In the finite system, $N_t = 4$.} \label{fig:erg2cell}
}
\end{figure}

For the finite system, the average rate region boundary is obtained by varying $\alpha_1$ from 0 to 1 ($\alpha_2 = 1-\alpha_1$), and for each value of $\alpha_1$, solving $\mathcal{P}_{\boldsymbol{\alpha}}$ for a large number of channel instances and averaging over the resulting instantaneous optimal rates.
The large system rate boundary is obtained by solving the large system equivalent of $\mathcal{P}_{\boldsymbol{\alpha}}$, as discussed above. As illustrated in Figure \ref{fig:erg2cell}, the much simpler to compute large system boundary provides a good approximation to the average rate region, even for quite a small system.

The large systems curve in Figure~\ref{fig:erg2cell} does not tell us how the large systems parameters would perform when used in a finite system, only that the large systems curve provides a good approximation to the average rate region of the finite system. We will now address the issue of how useful these parameters are in designing beamformers for the finite system.

Solving $\mathcal{P}_{\boldsymbol{\alpha}}$ in the large system case yields optimal rates, as well as asymptotically optimal variables to achieve them, which depend on the {\it statistics} of the user channels alone. What happens when we use the asymptotically optimal $\lambda_k$'s and $p_k$'s to obtain beamforming vectors in the finite system (cf. \eqref{eq:asymptoticW})?

\begin{figure}[htb]
\center{
\includegraphics[width=3.3in, height=2.5in, viewport = 0 120 650 660, clip]{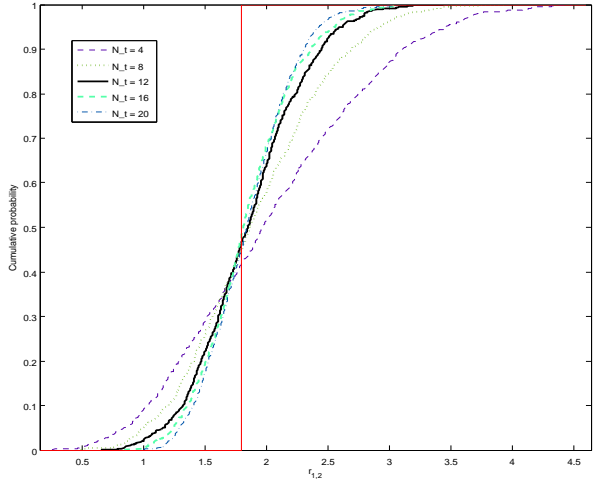}
\caption{CDF of the rate of a randomly selected user in cell 1: $\beta_2 = 2/4$, $\beta_3 = 3/4$, $\epsilon_{1,1} =  1$, $\epsilon_{1,2}=.2$, $\epsilon_{2,1} = .5$, $\epsilon_{2,2} =1.3$, for increasing $N_t$, SNR = 10dB.} \label{fig:outage2cell}
}
\end{figure}

In the large systems analysis, the optimal rates are deterministic, but if the corresponding beamforming structures are used in the finite case, the rates obtained are random variables, just as the optimal solution to $\mathcal{P}_{\boldsymbol{\alpha}}$ provides rates that vary with the channel state. Figure \ref{fig:outage2cell} illustrates the cumulative distribution function (cdf) of the rate supported by the proposed beamforming strategy \footnote{$\log_2 (1+\textrm{SINR}_{u,k})$, where $\textrm{SINR}_{u,k}$ is obtained from beamforming vectors constructed using the asymptotically optimal dual parameters and power levels} at the first user in cell 2, for $\alpha_1 = \alpha_ 2 = .5$, for increasing number of antennas (and users in each cell).
As the number of antennas tends to infinity, both instantaneous and average rates converge to the value predicted by the large systems analysis, but the convergence rate is quite slow.

It turns out that if we want a reasonably close approximation to the finite system average rate region but using beamformer structures obtained from the large systems analysis, we need to include some adaptive power control into the solution. Note that the finding the solution to $\mathcal{P}_{\boldsymbol{\alpha}}$ involves searching over both power levels and beamforming directions. The large systems beamformer provides both power levels and beamforming directions and is much simpler to compute. A compromise approach that provides a suboptimal solution to $\mathcal{P}_{\boldsymbol{\alpha}}$ may be obtained using only the asymptotically optimal $\lambda$'s to define the {\it directions} of the beamforming vectors.
Write beamforming vector $\mathbf{w}_{u,k,k}$ as
\begin{align}
\mathbf{w}_{u,k,k} =\sqrt{\frac{p_{u,k,k}}{N_t}} \mathbf{u}_{u,k,k},
\end{align}
where $\mathbf{u}_{u,k,k}$ specifies the direction of the beamforming vector ($\|\mathbf{u}_{u,k,k}\| = 1$).
Fixing these particular directions transforms $\mathcal{P}_{\boldsymbol{\alpha}}$ from an optimization over the $\mathbf{w}_{u,k,k}$'s to one over the power levels $p_{u,k,k}$ alone, thereby reducing complexity drastically.

\begin{figure}[htb]
\center{
\includegraphics[width=3.3in, height=2.5in, viewport = 20 150 600 650, clip]{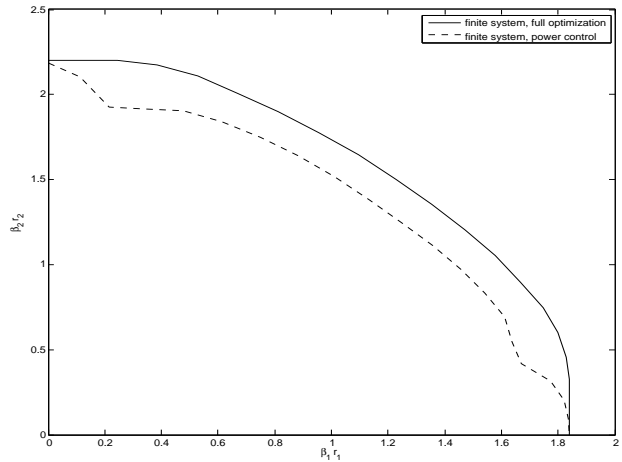}
\caption{Ergodic rate region comparison, full beamforming optimization vs. power control only: $N_t = 4$, $U_2 = 2$, $U_3 = 3$, $\epsilon_{1,1} =  1$, $\epsilon_{1,2}=.2$, $\epsilon_{2,1} = .5$, $\epsilon_{2,2} =1.3$, SNR = 10dB.} \label{fig:pc2cell}
}
\end{figure}

Figure \ref{fig:pc2cell} compares the resulting average rate region to the average rate region corresponding to the solution of $\mathcal{P}_{\boldsymbol{\alpha}}$.
The dips in the power control curve are due to the fact that at the corresponding value of $\boldsymbol{\alpha}$, the asymptotic analysis leads to strictly positive $\lambda$'s for both cells, i.e. both cells will always interfere with each other's transmission, whereas in the finite system, it is often optimal for one of the cells to zero-force the other's users. The curves show that while this approach is clearly suboptimal, the loss in capacity is not very significant, and the approach suggested here may be of practical interest.

\section{Conclusion}
In this paper, we have discussed a specific type of coordinated beamforming, namely min-max fairness in the power usage subject to target SINR constraints. The optimal beamforming strategies were characterized and shown to have an interesting nested zero-forcing structure.
In the asymptotic regime where the number of antennas at each BS and the number of users in each cell both grow large with their ratio tending to a finite constant, these problems simplify greatly and only statistical CSI is required to solve them.  The optimal parameters can be found by solving a convex optimization problem that only involves the statistical CSI. The optimal solution is characterized, and an algorithm is proposed that converges to the optimal transmit parameters, for feasible SINR targets. Following that, the individual base stations precode using the optimal parameters together with their own, instantaneous channel measurements.
The applicability of these asymptotic results to finite systems analysis is illustrated in a two cell example network, and a suboptimal approach to beamforming is presented that uses the large systems analysis to greatly simplify the beamformer design.

\appendices
\section{Lemmas for Asymptotic Problem}\label{app:asymp_lemmas}

Throughout what follows, let $\mathcal{S}(\boldsymbol{\mu}) = \{k | k \in \{1, \ldots, L\}, \mu_k = 0\}$ and let $\mathcal{S}^c(\boldsymbol{\mu})$ denote its complement in $\{1, \ldots, L\}$.

The feasibility constraint \eqref{eq:SINR_LAS_constraint} in $\mathcal{P}_{\textrm{dual}}^{\infty}$ becomes in vector notation
\begin{align}
\boldsymbol{\lambda} \le \mathbf{F}(\boldsymbol{\lambda}, \boldsymbol{\mu}, \boldsymbol{\gamma}), \label{eq:SINR_LAS_constraint_vec}
\end{align}
where $\mathbf{F}(\boldsymbol{\lambda}, \boldsymbol{\mu}, \boldsymbol{\gamma}) = \left\{F_k(\boldsymbol{\lambda}, \mu_k, \gamma_k) \right\}_{k=1}^L$, defined for $\boldsymbol{\lambda} \ge \boldsymbol{0}$, as given by \eqref{eq:defFk}, where $\boldsymbol{\mu}$ has at least one non-zero component.

In this appendix, we characterize the solution to the fixed-point equation $\boldsymbol{\lambda} = \mathbf{F}(\boldsymbol{\lambda}, \boldsymbol{\mu}, \boldsymbol{\gamma})$, which needs to be satisfied by the optimal solution of $\mathcal{P}_{\textrm{dual}}^{\infty}$.

We also characterize in Lemma \ref{lemm:Pinf_mu_G} below a property of the following subproblem of $\mathcal{P}_{\textrm{dual}}^{\infty}$, which will be used in the proof of Lemma \ref{lemma:nonsingular} (see Appendix \ref{proof:opt_nonsingular}):
\begin{align}
\mathcal{P}^{\infty}\left({\boldsymbol{\mu}}, \boldsymbol{\gamma}\right):~ & \begin{array}{cc} \textrm{max. } & \sigma^2 \sum_{k=1}^L \beta_k \lambda_{k} \\ \bol & \end{array} ~ \nonumber  \\
& \textrm{s.t. } \quad \mathbf{0} \le \boldsymbol{\lambda} \le \mathbf{F}(\boldsymbol{\lambda}, \boldsymbol{\mu}, \boldsymbol{\gamma}). \label{eq:Pinf_mu_G}
\end{align}

\noindent Before proceeding, we characterize $F_k(\boldsymbol{\lambda}, \mu_k, \gamma_k) $ as follows:
\begin{itemize}
\item If $\mu_k = 0$, and $\sum_{j=1}^L \beta_j I_{\lambda_j} \le 1$, then by definition, $F_k(\boldsymbol{\lambda}, \mu_k, \gamma_k) = 0$.
\item Otherwise, using the definition of $m_k$ as given by \eqref{eq:mkDef}, $y_k = F_k(\boldsymbol{\lambda}, \mu_k, \gamma_k)$ must satisfy
\begin{align}
g_k(y_k, \boldsymbol{\lambda}, \mu_k, \gamma_k) &\triangleq \frac{\gamma_k}{\epsilon_{k,k} y_k} \left[\mu_k + \sum_j \frac{\beta_j \epsilon_{j,k} \lambda_j}{1+  \frac{\gamma_k \epsilon_{j,k} \lambda_j}{\epsilon_{k,k} y_k}}\right]-1 \nonumber \\
&= 0,
 \label{eq:IF_FP}
\end{align}
The uniqueness of the root can be verified by noting that $g_k(y, \boldsymbol{\lambda}, \mu_k, \gamma_k)$ is strictly decreasing in $y$ since
\begin{align*}
\frac{\partial g_k}{\partial y} = -\frac{\gamma_k}{\epsilon_{k,k} y^2} \left[\mu_k + \sum_j \frac{\beta_j \epsilon_{j,k} \lambda_j}{\left(1+ \epsilon_{j,k} \lambda_j \frac{\gamma_k}{\epsilon_{k,k} y}\right)^2}\right] < 0.
\end{align*}
\end{itemize}

\begin{proposition}
\label{prop1}
Given $\boldsymbol{\mu} \ge \boldsymbol{0},$ such that $\mathcal{S}^c(\boldsymbol{\mu}) \neq \emptyset$, and $\boldsymbol{\gamma} > \boldsymbol{0}$, if $\boldsymbol{\lambda} \ge \boldsymbol{0}$ satisfies the fixed-point equation $\boldsymbol{\lambda} = \mathbf{F}(\boldsymbol{\lambda}, \boldsymbol{\mu}, \boldsymbol{\gamma})$, then the components in  $\mathcal{S}(\boldsymbol{\mu})$ are either all equal to zero or all strictly positive. The components in $\mathcal{S}^c(\boldsymbol{\mu})$ are all strictly positive.
\end{proposition}
\begin{IEEEproof}
Assume $\boldsymbol{\lambda}$ satisfies the fixed-point equation $\boldsymbol{\lambda} = \mathbf{F}(\boldsymbol{\lambda}, \boldsymbol{\mu}, \boldsymbol{\gamma})$.
If $\mu_k > 0$, then $F_k(\boldsymbol{\lambda}, \mu_k, \gamma_k) > 0$ implying that $\lambda_k > 0$.
If $\sum_{j=1}^L \beta_j I_{\lambda_j} > 1$, then for all $k \in  \mathcal{S}(\boldsymbol{\mu})$, $F_k(\boldsymbol{\lambda}, \mu_k, \gamma_k) > 0$ implying that $\lambda_k  > 0$.
Otherwise, for all $k \in \mathcal{S}(\boldsymbol{\mu})$, $F_k(\boldsymbol{\lambda}, \mu_k, \gamma_k) = 0$ implying that $\lambda_k   = 0$.
\end{IEEEproof}

\begin{proposition}\label{prop2} Given $\boldsymbol{\mu} \ge \boldsymbol{0},$ such that $\mathcal{S}^c(\boldsymbol{\mu}) \neq \emptyset$, and $\boldsymbol{\gamma} > \boldsymbol{0}$,  $\mathbf{F}(\boldsymbol{\lambda}, \boldsymbol{\mu}, \boldsymbol{\gamma})$ has the following properties for $\boldsymbol{\lambda} \ge 0$:
\begin{itemize}
\item Non-negativity: $\mathbf{F}(\boldsymbol{\lambda}, \boldsymbol{\mu}, \boldsymbol{\gamma}) \ge 0$; the inequality is strict for $k \in \mathcal{S}^c(\boldsymbol{\mu})$ and, whenever $\sum_{j=1}^L \beta_j I_{\lambda_j} > 1$, for $k \in \mathcal{S}(\boldsymbol{\mu})$  as well.
\item Monotonicity: if $\boldsymbol{\lambda} \ge \boldsymbol{\lambda}'$, $\boldsymbol{\lambda} \neq \boldsymbol{\lambda}'$, $\mathbf{F}(\boldsymbol{\lambda}, \boldsymbol{\mu}) \ge \mathbf{F}(\boldsymbol{\lambda}', \boldsymbol{\mu})$. The inequality is strict for all components $k \in \mathcal{S}^c(\boldsymbol{\mu})$; it is also strict for components with $k \in \mathcal{S}(\boldsymbol{\mu})$ if $\sum_{j=1}^L \beta_j I_{\lambda_j} > 1$.
\item Scalability: For all $\alpha > 1$, $\alpha \mathbf{F}(\boldsymbol{\lambda}, \boldsymbol{\mu}, \boldsymbol{\gamma})  \ge \mathbf{F}(\alpha \boldsymbol{\lambda}, \boldsymbol{\mu}, \boldsymbol{\gamma})$; The inequality is strict for components in $\mathcal{S}^c(\boldsymbol{\mu})$ and tight otherwise.
\end{itemize}
\end{proposition}
\begin{IEEEproof}
The non-negativity follows from the definition of $\mathbf{F}(\boldsymbol{\lambda}, \boldsymbol{\mu}, \boldsymbol{\gamma})$.

To prove monotonicity, note that if $k \in  \mathcal{S}(\boldsymbol{\mu})$ and $\sum_{j=1}^L \beta_j I_{\lambda'_j} \le 1$ but $\sum_{j=1}^L \beta_j I_{\lambda_j} > 1$, then $F_k(\boldsymbol{\lambda}, \mu_k, \gamma_k) > 0 = F_k(\boldsymbol{\lambda}', \mu_k, \gamma_k)$.
Otherwise, if $k \in  \mathcal{S}^c(\boldsymbol{\mu})$ or $k \in  \mathcal{S}(\boldsymbol{\mu})$ and $\sum_{j=1}^L \beta_j I_{\lambda'_j} > 1$, then both $F_k(\boldsymbol{\lambda}, \mu_k, \gamma_k)$ and $F_k(\boldsymbol{\lambda}', \mu_k, \gamma_k)$ are strictly positive and obtained by solving \eqref{eq:IF_FP}.

We drop dependence on $\mu_k$ and $\gamma_k$ from $g_k(.)$ and $F_k(.)$ to simplify notation and show that
\begin{align*}
&g_k(y, \boldsymbol{\lambda}) - g_k(y, \boldsymbol{\lambda}') \\
&= \frac{\gamma_k}{\epsilon_{k,k} y} \sum_j \beta_j \epsilon_{j,k}  \left[\frac{\lambda_j}{1+ \epsilon_{j,k} \lambda_j \frac{\gamma_k}{\epsilon_{k,k} y}}
-\frac{\lambda'_j}{1+ \epsilon_{j,k} \lambda'_j \frac{\gamma_k}{\epsilon_{k,k} y}}
\right] \\
&= \frac{\gamma_k}{\epsilon_{k,k} y} \sum_j \beta_j \epsilon_{j,k}  \left[\frac{\left(\lambda_j
- \lambda'_j\right)
}{\left(1+ \epsilon_{j,k} \lambda_j \frac{\gamma_k}{\epsilon_{k,k} y}\right)\left(1+ \epsilon_{j,k} \lambda'_j \frac{\gamma_k}{\epsilon_{k,k} y}\right)}
\right] \\
 &> 0,
\end{align*}
since all the $\lambda_j - \lambda'_j \ge 0$ and at least one of the inequalities is strict.
As a result,
\begin{align}
&g_k(F_k(\boldsymbol{\lambda}'), \boldsymbol{\lambda}) - g_k(F_k(\boldsymbol{\lambda}'), \boldsymbol{\lambda}') \nonumber \\
&= g_k(F_k(\boldsymbol{\lambda}'), \boldsymbol{\lambda}) > 0 = g_k(F_k(\boldsymbol{\lambda}), \boldsymbol{\lambda}).
\end{align}
Thus, $F_k(\boldsymbol{\lambda}') < F_k(\boldsymbol{\lambda})$ as $g_k(y, \boldsymbol{\lambda})$ is strictly decreasing in $y$.

To prove the scalability property, we start by noting that $\forall \alpha > 1$, $\sum_{j=1}^L \beta_j I_{\lambda_j} = \sum_{j=1}^L \beta_j I_{\alpha \lambda_j}$.
\begin{itemize}
\item If $k \in  \mathcal{S}(\boldsymbol{\mu})$:
\begin{itemize}
\item $F_k(\alpha \boldsymbol{\lambda}, \mu_k, \gamma_k) = \alpha  F_k(\boldsymbol{\lambda}, \mu_k, \gamma_k) = 0$ if $\sum_{j=1}^L \beta_j I_{\lambda_j} \le 1$;
\item  if $\sum_{j=1}^L \beta_j I_{\lambda_j}> 1$, $F_k(\alpha \boldsymbol{\lambda}, \mu_k, \gamma_k)$ is also equal to $\alpha F_k(\boldsymbol{\lambda}, \mu_k, \gamma_k)$. This follows from \eqref{eq:IF_FP} with $\mu_k = 0$.
\end{itemize}
\item If $k \in  \mathcal{S}^c(\boldsymbol{\mu})$, $F_k(\alpha \boldsymbol{\lambda}, \mu_k, \gamma_k) < \alpha F_k(\boldsymbol{\lambda}, \mu_k, \gamma_k)$ since
\begin{align*}
&g_k\left(F_k(\alpha  \boldsymbol{\lambda}, \mu_k, \gamma_k), \alpha \boldsymbol{\lambda}, \mu_k, \gamma_k\right) \\
& = 0 \\
&= g_k\left(F_k(\boldsymbol{\lambda}, \mu_k, \gamma_k), \boldsymbol{\lambda}, \mu_k, \gamma_k\right) \\
&= g_k\left(\alpha F_k(\boldsymbol{\lambda}, \mu_k, \gamma_k), \alpha \boldsymbol{\lambda}, \alpha \mu_k, \gamma_k\right) \\
&< g_k\left(\alpha F_k(\boldsymbol{\lambda}, \mu_k, \gamma_k), \alpha \boldsymbol{\lambda}, \mu_k, \gamma_k\right).
\end{align*}
\end{itemize}
\end{IEEEproof}

\begin{lemma}\label{lemm:FP}
Given $\boldsymbol{\mu} \ge \boldsymbol{0},$ such that $\mathcal{S}^c(\boldsymbol{\mu}) \neq \emptyset$, and $\boldsymbol{\gamma} > \boldsymbol{0}$,
\begin{itemize}
\item If $\mathcal{S}(\boldsymbol{\mu}) = \emptyset$ or  $\mathcal{S}(\boldsymbol{\mu}) \neq \emptyset$ with $\sum_{j=1}^L \beta_j I_{\mu_j} > 1$,
there is at most one $\boldsymbol{\lambda} \ge \boldsymbol{0}$ satisfying the
fixed-point equation $\boldsymbol{\lambda} = \mathbf{F}(\boldsymbol{\lambda}, \boldsymbol{\mu}, \boldsymbol{\gamma})$. Moreover, this $\boldsymbol{\lambda} > \boldsymbol{0}$.
\item If $\mathcal{S}(\boldsymbol{\mu}) \neq \emptyset$ with $\sum_{j=1}^L \beta_j I_{\mu_j} \le 1$, there are at most two $\boldsymbol{\lambda} \ge \boldsymbol{0}$ satisfying the
fixed-point equation $\boldsymbol{\lambda} = \mathbf{F}(\boldsymbol{\lambda}, \boldsymbol{\mu}, \boldsymbol{\gamma})$:
\begin{itemize}
\item If there is only one fixed point $\boldsymbol{\lambda}$, then $\lambda_k = 0$ if $k \in \mathcal{S}(\boldsymbol{\mu})$.
\item If there are two fixed points, $\boldsymbol{\lambda}^{(2)}$ and $\boldsymbol{\lambda}^{(1)}$, then $\boldsymbol{\lambda}^{(2)} > \boldsymbol{\lambda}^{(1)}$ and $\lambda_k^{(1)} = 0$ if $k \in \mathcal{S}(\boldsymbol{\mu})$. This can only occur if $\sum_{j=1}^L \beta_j > 1$.
\end{itemize}
\end{itemize}
\end{lemma}
\begin{IEEEproof}
Note the difference to Theorem 1 in \cite{yates95}, where the fixed-point solution, if it exists, is unique.

We focus on the proof of the second part, where $\mathcal{S}(\boldsymbol{\mu}) \neq \emptyset$ and  with $\sum_{j=1}^L \beta_j I_{\mu_j} \le 1$: the proof of the first part follows similar lines.

Assume there are two vectors $\boldsymbol{\lambda}^{(1)}$ and $\boldsymbol{\lambda}^{(2)}$ that satisfy the fixed-point equation. Without loss of generality, based on Proposition \ref{prop1}, one of the following cases occurs:
\begin{enumerate}
\item for every $k \in \mathcal{S}(\boldsymbol{\mu})$: $\lambda_k^{(1)} = \lambda_k^{(2)} = 0$. In this case, assume without loss of generality that there exists $j$ in $\mathcal{S}^c(\boldsymbol{\mu})$ such that $\lambda^{(1)}_j < \lambda^{(2)}_j$. Hence, there exists $\alpha > 1$ such that
$\alpha \boldsymbol{\lambda}^{(1)} \ge \boldsymbol{\lambda}^{(2)}$ with $\alpha \lambda^{(1)}_j = \lambda^{(2)}_j$. The monotonicity and scalability properties in Proposition \ref{prop2} imply that
\begin{align*}
\alpha \lambda_j^{(1)} &= \alpha  F_j(\boldsymbol{\lambda}^{(1)}, \mu_j, \gamma_j) \\
&> F_j(\alpha \boldsymbol{\lambda}^{(1)}, \mu_j, \gamma_j) > F_j(\boldsymbol{\lambda}^{(2)}, \mu_j, \gamma_j) = \lambda_j^{(2)},
\end{align*}
thus leading to a contradiction. Thus, there can be at most one $\boldsymbol{\lambda} \ge \boldsymbol{0}$, having components in $\mathcal{S}(\boldsymbol{\mu})$ equal to zero, satisfying the fixed-point equation.
\item for every $k \in \mathcal{S}(\boldsymbol{\mu})$: $\lambda_k^{(1)} > 0$ and $\lambda_k^{(2)} > 0$. The same  argument to that in case 1) can be used to obtain a contradiction. Thus, there can be at most one $\boldsymbol{\lambda} > \boldsymbol{0}$, satisfying the fixed-point equation.
\item for every $k \in \mathcal{S}(\boldsymbol{\mu})$: $\lambda_k^{(1)} = 0$ and $\lambda_k^{(2)} > 0$. Assume we can find an index in $\mathcal{S}^c(\boldsymbol{\mu})$ such that $\lambda^{(2)}_j < \lambda^{(1)}_j$. Then, exactly as in the previous two cases, a contradiction may be reached. If there is no such $j$, then it must be that $\boldsymbol{\lambda}^{(2)} > \boldsymbol{\lambda}^{(1)}$, since we can rule out the inequality being tight for any of the components from the monotonicity property.
\end{enumerate}

Thus, if there are two distinct vectors $\boldsymbol{\lambda}^{(2)}$ and $\boldsymbol{\lambda}^{(1)}$ satisfying the fixed-point equation, then $\boldsymbol{\lambda}^{(2)} > \boldsymbol{\lambda}^{(1)}$, and the components of $\boldsymbol{\lambda}^{(1)}$ in $\mathcal{S}(\boldsymbol{\mu})$ are equal to zero.

The proof is completed by showing that if there exists a vector $\boldsymbol{\lambda}^{(2)} > \boldsymbol{0}$ satisfying the fixed-point equation, then there must be a $\boldsymbol{\lambda}^{(1)}$ having components in $\mathcal{S}(\boldsymbol{\mu})$ equal to zero, which also satisfies the fixed-point equation.

Assume such a $\boldsymbol{\lambda}^{(2)}$ exists, and define $\tilde{\boldsymbol{\lambda}}^{(2)}$ as follows:
\begin{align*}
\tilde{\lambda}^{(2)}_k = \left\{\begin{array}{ll}
0 & k \in \mathcal{S}(\boldsymbol{\mu}) \\
\lambda_k^{(2)} & k \notin \mathcal{S}(\boldsymbol{\mu}) \end{array} \right..
\end{align*}
Thus, $\tilde{\boldsymbol{\lambda}}^{(2)} \le \boldsymbol{\lambda}^{(2)}$ and the inequality is strict for all $k \in \mathcal{S}(\boldsymbol{\mu})$. By the monotonicity property in Proposition \ref{prop2},
\begin{align*}
\mathbf{F}\left(\tilde{\boldsymbol{\lambda}}^{(2)}, \boldsymbol{\mu}, \boldsymbol{\gamma}\right) < \mathbf{F}\left(\boldsymbol{\lambda}^{(2)}, \boldsymbol{\mu}, \boldsymbol{\gamma}\right) = \boldsymbol{\lambda}^{(2)},
\end{align*}
and since $\sum_{j=1}^L \beta_j I_{\mu_j} \le 1$, the components in the left-hand side corresponding to $\mu_k = 0$ are equal to zero. As a result,
\begin{align*}
\mathbf{F}\left(\tilde{\boldsymbol{\lambda}}^{(2)}, \boldsymbol{\mu}, \boldsymbol{\gamma}\right) \le \tilde{\boldsymbol{\lambda}}^{(2)},
\end{align*}
where the inequalities corresponding to $k \in \mathcal{S}(\boldsymbol{\mu})$ are tight. Using this result, and the monotonicity property, the following  sequence
\begin{align*}
\boldsymbol{\lambda}(0) &= \tilde{\boldsymbol{\lambda}}^{(2)}, \\
\boldsymbol{\lambda}(n) &= \mathbf{F}\left(\boldsymbol{\lambda}(n-1), \boldsymbol{\mu}, \boldsymbol{\gamma}\right), n = 1, 2, \ldots
\end{align*}
will be non-increasing, and each $\boldsymbol{\lambda}(n)$ will have components corresponding to $k \in \mathcal{S}(\boldsymbol{\mu})$ equal to zero. Since the sequence $\boldsymbol{\lambda}(n)$ is also bounded below by zero, it must converge to a fixed point (which must be unique as a result of part 1)), which will also have components in $\mathcal{S}(\boldsymbol{\mu})$ equal to zero.
\end{IEEEproof}

\begin{lemma}\label{lemm:Pinf_mu_G}
Let $\mathcal{P}^{\infty}\left({\boldsymbol{\mu}}, \boldsymbol{\gamma}\right)$ be as defined in \eqref{eq:Pinf_mu_G}.
Let $\boldsymbol{\mu} \ge 0, \boldsymbol{\gamma} > 0$, be such that $\mathcal{S}^c(\boldsymbol{\mu})$ is non-empty, $\sum_{j=1}^L \beta_j I_{\mu_j} \le 1$ and $\sum_{j=1}^L \beta_j > 1$. Also let $\boldsymbol{\mu}' \ge \boldsymbol{\mu}$, $\boldsymbol{\gamma}' \ge \boldsymbol{\gamma}$, be such that both inequalities are tight for components $k \in \mathcal{S}(\boldsymbol{\mu})$ , and strict otherwise, (which implies $\mathcal{S}(\boldsymbol{\mu}')=\mathcal{S}(\boldsymbol{\mu})$).
Then if $\mathcal{P}^{\infty}\left({\boldsymbol{\mu}}, \boldsymbol{\gamma}\right)$ is bounded and its unique optimum $\boldsymbol{\lambda}^{opt}$  is such that $\lambda_k^{opt} = 0$ for all $k \in \mathcal{S}(\boldsymbol{\mu})$, then the solution, $\boldsymbol{\lambda}$, to $\mathcal{P}^{\infty}\left(\boldsymbol{\mu}', \boldsymbol{\gamma}'\right)$ will also have $\lambda_k  = 0$ for all $k \in \mathcal{S}(\boldsymbol{\mu}')$, {\it i.e.} in this case, to solve $\mathcal{P}^{\infty}\left(\boldsymbol{\mu}', \boldsymbol{\gamma}'\right)$, its enough to take $\lambda_k  = 0$ for all $k \in \mathcal{S}(\boldsymbol{\mu}')$ and optimize over the remaining components.
\end{lemma}
\begin{IEEEproof}
$\mathcal{P}^{\infty}\left({\boldsymbol{\mu}}, \boldsymbol{\gamma}\right)$ is a convex optimization problem\footnote{The proof is similar to the proof of the convexity of $\mathcal{P}_{\textrm{dual}}^{\infty}$ in Appendix \ref{proof:SINR_LAS_constraint}.} and is always feasible. From its KKT conditions we can show that if the problem is bounded, its optimum must satisfy the fixed-point equation $\boldsymbol{\lambda}^{opt} = \mathbf{F}(\boldsymbol{\lambda}^{opt}, \boldsymbol{\mu}, \boldsymbol{\gamma})$\footnote{This is proved similarly to the proof of Lemma~\ref{lem:KKT_for_dual} in Appendix~\ref{dualKKT}.}.

Given the conditions on $\boldsymbol{\mu}$ and $\{\beta_k\}_{k=1}^L$ in the statement of the lemma, Lemma \ref{lemm:FP} establishes that
 fixed-point equations $\boldsymbol{\lambda} = \mathbf{F}(\boldsymbol{\lambda}, \boldsymbol{\mu}, \boldsymbol{\gamma})$ and $\boldsymbol{\lambda} = \mathbf{F}(\boldsymbol{\lambda}, \boldsymbol{\mu}', \boldsymbol{\gamma}')$ each could have up to two solutions; moreover, if two solutions exist, then the first has components in $ \mathcal{S}(\boldsymbol{\mu})$ equal to zero and is strictly dominated by the second.
To prove the lemma, we show that the conditions on $\mathcal{P}^{\infty}\left({\boldsymbol{\mu}}, \boldsymbol{\gamma}\right)$ in Lemma~\ref{lemm:Pinf_mu_G} imply
that there is no strictly positive $\boldsymbol{\lambda}$ that is feasible for $\mathcal{P}^{\infty}\left(\boldsymbol{\mu}', \boldsymbol{\gamma}'\right)$. For this will imply that only the unique solution to $\boldsymbol{\lambda} = \mathbf{F}(\boldsymbol{\lambda}, \boldsymbol{\mu}', \boldsymbol{\gamma}')$ with $\lambda_k=0$ for $k \in \mathcal{S}(\boldsymbol{\mu}')$ can be feasible for $\mathcal{P}^{\infty}\left(\boldsymbol{\mu}', \boldsymbol{\gamma}'\right)$.

Since $\mathcal{P}^{\infty}\left({\boldsymbol{\mu}}, \boldsymbol{\gamma}\right)$ is bounded and its optimum solution $\boldsymbol{\lambda}^{opt}$ has components  in $\mathcal{S}(\boldsymbol{\mu})$  equal to zero, then there can be no $\boldsymbol{\tilde{\lambda}}$, such that $\boldsymbol{0} < \boldsymbol{\tilde{\lambda}} \le  \mathbf{F}(\boldsymbol{\tilde{\lambda}}, \boldsymbol{\mu}, \boldsymbol{\gamma})$. This can be seen because if there were such a vector, then either the problem will be unbounded or we can construct a strictly positive solution to the equation $\boldsymbol{\lambda} = \mathbf{F}(\boldsymbol{\lambda}, \boldsymbol{\mu}, \boldsymbol{\gamma})$ by starting with $\boldsymbol{\lambda}(0) = \boldsymbol{\tilde{\lambda}}$, letting $\boldsymbol{\lambda}(n) = F(\boldsymbol{\lambda}(n-1), \boldsymbol{\mu}, \boldsymbol{\gamma})$, and taking the limit of the increasing\footnote{This follows from the monotonicity property in Proposition \ref{prop2}.} sequence generated by this iteration. By Lemma~\ref{lemm:FP}, this $\boldsymbol{\lambda}$ dominates any other solution to the equation $\boldsymbol{\lambda} = \mathbf{F}(\boldsymbol{\lambda}, \boldsymbol{\mu}, \boldsymbol{\gamma})$, contradicting the optimality of $\boldsymbol{\lambda}^{opt}$.

Recalling the definition of $F_k(\boldsymbol{\lambda}, \mu_k, \gamma_k)$, the above condition, i.e. that there is no $\boldsymbol{\tilde{\lambda}}$, such that $\boldsymbol{0} < \boldsymbol{\tilde{\lambda}} \le  \mathbf{F}(\boldsymbol{\tilde{\lambda}},\boldsymbol{\mu}, \boldsymbol{\gamma})$, is equivalent to the following region being empty
\begin{align}
& 0 < \lambda_k \le F_k(\boldsymbol{\lambda}, \mu_k, \gamma_k), \quad \forall k,
\end{align}
implying that \footnote{Recall that $g_k(y, \boldsymbol{\lambda}, \mu_k, \gamma_k)$ is strictly decreasing in $y$ and $y_k = F_k(\boldsymbol{\lambda}, \mu_k, \gamma_k)$ in this case is the unique positive solution of $g_k(y_k, \boldsymbol{\lambda}, \mu_k, \gamma_k) = 0$.}
\begin{align}
& 0 < \lambda_k , \quad \forall k, \\
& 0 \le g_k(\lambda_k, \boldsymbol{\lambda}, \mu_k, \gamma_k), \forall k.
\end{align}
In other words, the region ${\cal R}(\boldsymbol{\mu})$, defined by the constraints
\begin{align}
& 0 < \lambda_k , \quad \forall k, \\
& 0 \le \frac{\gamma_k}{\epsilon_{k,k} \lambda_k} \left[\mu_k + \sum_j \frac{\beta_j \epsilon_{j,k} \lambda_j}{1+  \frac{\gamma_k \epsilon_{j,k} \lambda_j}{\epsilon_{k,k} \lambda_k}}\right]-1, \forall k,
\end{align}
is empty. Distinguishing between $k \in \mathcal{S}(\boldsymbol{\mu})$ and $k \in \mathcal{S}^c(\boldsymbol{\mu})$, the constraints defining ${\cal R}(\boldsymbol{\mu})$ can be written:
\begin{align}
 &\sum_j \frac{\beta_j}{\frac{\epsilon_{k,k} \lambda_k}{\gamma_k \epsilon_{j,k} \lambda_j}+ 1} \ge 1  & \forall k \in \mathcal{S}(\boldsymbol{\mu}) \label{eq:con1}\\
&\frac{\gamma_k \mu_k }{\epsilon_{k,k} \lambda_k} + \sum_j \frac{\beta_j}{1+ \frac{\epsilon_{k,k} \lambda_k}{\gamma_k \epsilon_{j,k} \lambda_j }} \ge 1  & \forall k \in \mathcal{S}^c(\boldsymbol{\mu}) \label{eq:con2}\\
& \lambda_k > 0 & \forall k. \label{eq:con3}
\end{align}
Define the region ${\cal R}'$ (which doesn't depend on $\boldsymbol{\mu}$) by the constraints:
\begin{align}
 &\sum_j \frac{\beta_j}{\frac{\epsilon_{k,k} \lambda_k}{\gamma_k \epsilon_{j,k} \lambda_j}+ 1} \ge 1  & \forall k \in \mathcal{S}(\boldsymbol{\mu}) \\
& \lambda_k > 0 & \forall k. \label{eq:con3}
\end{align}
By definition ${\cal R}(\boldsymbol{\mu}) \subseteq {\cal R}'$, but if $\boldsymbol{\lambda}$ is an element of ${\cal R}'$, then we can construct an element in ${\cal R}(\boldsymbol{\mu})$ by scaling $\boldsymbol{\lambda}$ by a sufficiently small positive scalar. Thus, ${\cal R}(\boldsymbol{\mu})$ is empty if and only if ${\cal R}'$ is empty. The latter statement is true for any other value of $\boldsymbol{\mu}$, including $\boldsymbol{\mu}'$. Since we have  ${\cal R}(\boldsymbol{\mu})$ is empty it therefore follows that  ${\cal R}(\boldsymbol{\mu}')$ is also empty.
Thus, there is no strictly positive $\boldsymbol{\lambda}$ that is feasible for $\mathcal{P}^{\infty}\left(\boldsymbol{\mu}', \boldsymbol{\gamma}'\right)$, which completes the proof.
\end{IEEEproof}

\section{KKT conditions}\label{app:KKT}
The KKT conditions of problem $\mathcal{P}_{\textrm{primal}}$ are the following.
\begin{itemize}
\item Stationarity constraints:
\begin{align}
&\left[\boldsymbol{\Sigma}_{u,k}-\frac{\lambda_{u,k}}{\gamma_{k}N_t}\mathbf{h}_{u,k,k}^H\mathbf{h}_{u,k,k}\right]\mathbf{w}_{u,k}  = \boldsymbol{0}, \\
&\sum_k \left(1-\mu_k\right) = 0.
\end{align}
\item Feasibility constraints:
\begin{align}
&\sum_{u=1}^{U_k} \|\mathbf{w}_{u,k}\|^2 \le \phi P, \label{eq:feas1} \\
&\frac{1}{\gamma_k} |\mathbf{h}_{u,k,k} \mathbf{w}_{u,k}|^2 \ge \sigma^2 + \sum_{(\bar{u}, j) \neq (u,k)} |\mathbf{h}_{u,k,j} \mathbf{w}_{\bar{u},j}|^2. \label{eq:feas2}
\end{align}
\item Dual feasibility constraints:
\begin{align}
& \mu_k \ge 0, \frac{\lambda_{u,k}}{N_t} \ge 0. \label{eq:dual_feas}
\end{align}
\item Complementary slackness constraints:
\begin{align}
&\mu_k \left[\sum_{u=1}^{U_k} \|\mathbf{w}_{u,k}\|^2 - \phi P\right] = 0, \label{eq:comp1} \\
& \frac{\lambda_{u,k}}{N_t}\left[\frac{ |\mathbf{h}_{u,k,k} \mathbf{w}_{u,k}|^2}{\gamma_k} - \sigma^2 + \sum_{(\bar{u}, j) \neq (u,k)} |\mathbf{h}_{u,k,j} \mathbf{w}_{\bar{u},j}|^2\right] \nonumber \\
&= 0. \label{eq:comp2}
\end{align}
\end{itemize}

\section{Proof of Lemma \ref{lemma:nonsingular}}
\label{proof:opt_nonsingular}
The stationarity constraint \eqref{eq:KKT_delL} can be rewritten as (we follow derivations in \cite{wiesel_sp06, yu_sp07, dahrouj_tw10})
\begin{align}
\boldsymbol{\Sigma}_{u,k} \mathbf{w}_{u,k} = \frac{\lambda_{u,k}}{\gamma_{k}N_t}\mathbf{h}_{u,k,k}^H\mathbf{h}_{u,k,k}\mathbf{w}_{u,k}
\end{align}
If $\boldsymbol{\Sigma}_{u,k}$ is nonsingular, we can invert it so that
\begin{align}
\mathbf{w}_{u,k} = \left(\frac{\lambda_{u,k}}{\gamma_{k}N_t}\mathbf{h}_{u,k,k}\mathbf{w}_{u,k} \right)\boldsymbol{\Sigma}_{u,k}^{-1} \mathbf{h}_{u,k,k}^H.
\end{align}

\noindent Multiplying \eqref{eq:KKT_delL} by $\mathbf{w}_{u,k}^H$,
\begin{align}
&\mathbf{w}_{u,k}^H \left[\boldsymbol{\Sigma}_{u,k}-\frac{\lambda_{u,k}}{\gamma_{k}N_t}\mathbf{h}_{u,k,k}^H\mathbf{h}_{u,k,k}\right]\mathbf{w}_{u,k}  = 0
\end{align}
Plugging in the above value of $\mathbf{w}_{u,k}$ completes the proof.

\section{Proof of Lemma \ref{lemma:singular}}
\label{proof:opt_singular}
Let $r \le N_t$ denote the rank of rank-deficient optimal $\boldsymbol{\Sigma}_{u,k}$, and let $\mathbf{U}_{u,k}\mathbf{D}_{u,k} \mathbf{U}_{u,k}^H$ be its eigenvalue decomposition, such that $\mathbf{U}_{u,k}$ is an $N_t \times N_t$ unitary matrix, $\mathbf{D}_{u,k}$ is a diagonal matrix such that its first $r$ diagonal elements of $\mathbf{D}_{u,k}$ are strictly positive and the remaining $N_t-r$ diagonal elements are zero. The stationarity constraint on $\mathbf{w}_{u,k}$ is equivalent to
\begin{align}
\mathbf{D}_{u,k} \mathbf{U}_{u,k}^H \mathbf{w}_{u,k} = \frac{\lambda_{u,k}}{\gamma_k N_t} \mathbf{U}_{u,k}^H \mathbf{h}_{u,k,k}^H\mathbf{h}_{u,k,k}\mathbf{w}_{u,k}.
\end{align}
By definition of $\mathbf{D}_{u,k}$, the last $N_t-r$ elements of the left-hand side are equal to zero. Since with probability 1, none of the entries of $\mathbf{U}_{u,k}^H \mathbf{h}_{u,k,k}^H$ will be zero due to the randomness and independence of the channel realizations, and recalling that $\mathbf{h}_{u,k,k} \mathbf{w}_{u,k}$ must be strictly positive, the corresponding elements in the right-hand side of the equation will only be zero if $\lambda_{u,k} = 0$.\footnote{This imposes the constraint that the first $r$ elements of $\mathbf{D}_{u,k} \mathbf{U}_{u,k}^H \mathbf{w}_{u,k}$ are also zero, the zero-forcing condition!}

We now show that the optimal $\lambda_{u',k}$'s will also be zero for all other users in the cell. Consider any user $u' \neq u$ in the same cell. Its beamforming vector must satisfy the KKT condition
\begin{align}
\boldsymbol{\Sigma}_{u',k} \mathbf{w}_{u',k} = \frac{\lambda_{u',k}}{\gamma_k N_t}\mathbf{h}_{u',k,k}^H \mathbf{h}_{u',k,k} \mathbf{w}_{u',k}.
\end{align}

Since $\lambda_{u,k} = 0$, $\boldsymbol{\Sigma}_{u',k} = \boldsymbol{\Sigma}_{u,k}- \frac{\lambda_{u', k}}{N_t}
\mathbf{h}_{u',k,k}^H\mathbf{h}_{u',k,k}
$, so that the above becomes
\begin{align}
\boldsymbol{\Sigma}_{u,k} \mathbf{w}_{u',k} = \left(1+\frac{1}{\gamma_k}\right)\frac{\lambda_{u',k}}{N_t}\mathbf{h}_{u',k,k}^H \mathbf{h}_{u',k,k} \mathbf{w}_{u',k}.
\end{align}
But $\mathbf{h}_{u',k,k} \mathbf{w}_{u',k} > 0$, and $\boldsymbol{\Sigma}_{u,k}$ is rank-deficient, so that, using the same argument as in the proof of $\lambda_{u,k} = 0$, we can show that $\mathbf{w}_{u',k}$ must lie in the null space of $\boldsymbol{\Sigma}_{u,k}$ and $\lambda_{u',k}$ must be zero.
As a result, $\boldsymbol{\Sigma}_{u',k} = \boldsymbol{\Sigma}_{u,k}$. This completes the proofs of 2)-4).

\section{Proof of Theorem \ref{theo:LAS_dual}}\label{proof:LAS_dual}
\noindent Define $\mathcal{P}\left(\boldsymbol{\mu}, \boldsymbol{\gamma}\right)$ as
\begin{align}
\mathcal{P}\left(\boldsymbol{\mu}, \boldsymbol{\gamma}\right): & \begin{array}{cc} \textrm{max. } & \sigma^2 \sum_{u,k} \frac{\lambda_{u,k}}{N_t} \\ \bol & \end{array} \nonumber  \\
&  \lambda_{u,k} \ge 0, ~ \min_{\mathbf{v}_{u,k}} \frac{\mathbf{v}_{u,k}^H\boldsymbol{\Sigma}_{u,k}\mathbf{v}_{u,k}}{\frac{1}{\gamma_{k}N_t}|\mathbf{v}_{u,k}^H\mathbf{h}_{u,k,k}^H|^2} \ge \lambda_{u,k} , ~\forall u, k
\end{align}
where $\boldsymbol{\Sigma}_{u,k} = \mu_k \mathbf{I} + \sum_{(\bar{u}, j) \neq (u,k)} \frac{\lambda_{\bar{u}, j}}{N_t}
\mathbf{h}_{\bar{u},j,k}^H\mathbf{h}_{\bar{u},j,k},$ and note that\footnote{We find it useful to index $\mathcal{P}_{\textrm{dual}}$ by its objective target SINR vector.} $\mathcal{P}_{\textrm{dual}}\left(\boldsymbol{\gamma}\right) = \max._{\boldsymbol{\mu}\ge \boldsymbol{0}, \sum_{k} \left(1-\mu_k\right) \ge 0} \mathcal{P}\left(\boldsymbol{\mu}, \boldsymbol{\gamma}\right)$. 
Denote the unique solution to $\mathcal{P}\left(\boldsymbol{\mu}, \boldsymbol{\gamma}\right)$ by $\boldsymbol{\lambda}\left(\boldsymbol{\mu}, \boldsymbol{\gamma}\right).$ The following result is easily proven using Yates' monotonicity framework:
\begin{lemma}\label{lbl:fsm} Finite system monotonicity:
\begin{enumerate}
\item Suppose $\boldsymbol{\mu}^{(1)} \le \boldsymbol{\mu}^{(2)}$ with $\boldsymbol{\mu}^{(2)}$ strictly positive, and $\boldsymbol{\gamma}$ fixed.
Then $\boldsymbol{\lambda}\left(\boldsymbol{\mu}^{(1)}, \boldsymbol{\gamma}\right) \le \boldsymbol{\lambda}\left(\boldsymbol{\mu}^{(2)}, \boldsymbol{\gamma}\right)$.
\item Suppose $ \boldsymbol{\gamma}^{(1)} \le  \boldsymbol{\gamma}^{(2)}$, and $\boldsymbol{\mu}$ is strictly positive,
then $\boldsymbol{\lambda}\left(\boldsymbol{\mu}, \boldsymbol{\gamma}^{(1)}\right) \le \boldsymbol{\lambda}\left(\boldsymbol{\mu}, \boldsymbol{\gamma}^{(2)}\right)$.
\end{enumerate}
\end{lemma}

The following argument follows along similar lines to Theorem~2 in \cite{ZH10_archive}, but note that \cite{ZH10_archive} only considers a two cell system.

Consider the {\it optimal} dual variables $\boldsymbol{\mu}, \boldsymbol{\lambda}$, 
where optimality refers to the dual CBf optimization problem $\mathcal{P}_{\textrm{dual}}$ (see \eqref{eq:SINR_CBf_UL}). Due to the dual
feasibility constraints, the sequence of $\boldsymbol{\mu}$ \footnote{The superscript $N_t$ is implicit here: {\it i.e.} we write $\boldsymbol{\mu}$ in place of the more accurate $\boldsymbol{\mu}^{(N_t)}$.}
is contained in a compact set and so its probability distribution function, $F^{(N_t)}$, forms a tight sequence \cite{billingsley99}. Let $F$ denote a limit point, so that $F^{(N_t)} \Rightarrow F$ along a convergent subsequence.

For the purpose of obtaining a contradiction, let $\delta > 0$ be small enough so that if $\bar{\mu}_i > 0$, then $\bar{\mu}_i-\delta > 0$, and let $\bar{\boldsymbol{\mu}}$ be such that 
$F\left(\prod_{i=1}^L (\bar{\mu}_i - \delta, \bar{\mu}_i + \delta)\right) > 0$. Define $B(\delta)$ to be the event that $\mu_i \in (\bar{\mu}_i-\delta, \bar{\mu}_i + \delta)~\forall i$, then by the second Borel-Cantelli lemma, event $B(\delta)$ will occur infinitely often.\footnote{We write $B(\delta)$ in place of the more accurate $B^{(N_t)}(\delta)$.} 
Corresponding to $\bar{\boldsymbol{\mu}}$, let $\bar{\boldsymbol{\lambda}}$ be the unique solution to the convex optimization problem $\mathcal{P}^{\infty}\left(\bobmu, \boldsymbol{\gamma}\right)$, as defined in \eqref{eq:Pinf_mu_G}. Recall that $\mathcal{P}^{\infty}_{\textrm{dual}} \equiv \mathcal{P}^{\infty}_{\textrm{dual}}(\bog) = \max_{\boldsymbol{\mu} \ge 0, \sum_{k=1}^L \mu_k \le L} \mathcal{P}^{\infty}\left(\boldsymbol{\mu}, \boldsymbol{\gamma}\right)$.

For $\delta > 0$ let $\bar{\boldsymbol{\mu}}(-\delta)$ and $\bar{\boldsymbol{\mu}}(\delta)$ be defined by
\begin{align}
\bar{\mu}_k(-\delta) = \left\{\begin{array}{ll}\bar{\mu}_k - \delta & \bar{\mu}_k > 0 \\ 0 & \bar{\mu}_k = 0 \end{array} \right.,
\end{align}
and
\begin{align}
\bar{\mu}_k(\delta) = \bar{\mu}_k+\delta.
\end{align}
Similarly, let $\boldsymbol{\gamma}(-\delta)$ and  $\boldsymbol{\gamma}(\delta)$ be defined as follows:
\begin{align}
\gamma_k(-\delta) = \gamma_k - \delta \\
\gamma_k(\delta) = \gamma_k + \delta
\end{align}

Let $\bar{\boldsymbol{\lambda}}(-\delta)$ and $\bar{\boldsymbol{\lambda}}(\delta)$ denote the solutions of
$\mathcal{P}^{\infty}\left(\bobmumd, \boldsymbol{\gamma}(-\delta)\right)$ and $\mathcal{P}^{\infty}\left(\bobmud, \boldsymbol{\gamma}(\delta)\right)$, respectively. Since $\bobmud > 0$, it follows that $\bobld > 0$. Moreover, using random matrix theory, we can show that with these constant strategies, asymptotically, for users in cell $k$,
\begin{align}
&\min_{\mathbf{v}_{u,k}} \frac{\mathbf{v}_{u,k}^H\boldsymbol{\Sigma}_{u,k}(-\delta)\mathbf{v}_{u,k}}{\frac{1}{N_t}|\mathbf{v}_{u,k}^H\mathbf{h}_{u,k,k}^H|^2} \ge \frac{\bar{\lambda}_{k}(-\delta)}{\gamma_{k}(-\delta)}, \\
&\min_{\mathbf{v}_{u,k}} \frac{\mathbf{v}_{u,k}^H\boldsymbol{\Sigma}_{u,k}(\delta)\mathbf{v}_{u,k}}{\frac{1}{N_t}|\mathbf{v}_{u,k}^H\mathbf{h}_{u,k,k}^H|^2} \ge \frac{\bar{\lambda}_{k}(\delta)}{\gamma_{k}(\delta)},
\end{align}
where
\begin{align}
&\boldsymbol{\Sigma}_{u,k}(-\delta) = \bar{\mu}_k(-\delta) \mathbf{I} + \sum_{(\bar{u}, j) \neq (u,k)} \frac{\bar{\lambda}_{j}(-\delta)}{N_t}
\mathbf{h}_{\bar{u},j,k}^H\mathbf{h}_{\bar{u},j,k} \nonumber \\
&\boldsymbol{\Sigma}_{u,k}(\delta) = \bar{\mu}_k(\delta) \mathbf{I} + \sum_{(\bar{u}, j) \neq (u,k)} \frac{\bar{\lambda}_{j}(\delta)}{N_t}
\mathbf{h}_{\bar{u},j,k}^H\mathbf{h}_{\bar{u},j,k}, \nonumber
\end{align}
hold with equality.
 In fact, for target SINR vector $\boldsymbol{\gamma}$, fixing $\boldsymbol{\mu}$ at
$\bar{\boldsymbol{\mu}}$ and fixing $\boldsymbol{\lambda}$  at the $\bar{\boldsymbol{\lambda}}$ obtained by solving $\mathcal{P}^{\infty}\left(\bar{\boldsymbol{\mu}}, \boldsymbol{\gamma}\right)$, two cases arise\footnote{with $\delta$ positive or negative.}:
\begin{itemize}
\item If the constant strategy $\bar{\lambda}_k(\delta) = 0$ (this occurs iff $\bar{\mu}_k(\delta) = 0$ and $\sum_{j} \beta_j I_{\bar{\lambda}_{j}(\delta)}< 1$), then both sides of the above inequality will be equal to 0, for $N_t$ sufficiently large, since the corresponding $\boldsymbol{\Sigma}_{u,k}(\delta)$ will be rank-deficient ($\sum_{j=1} U_j I_{\bar{\lambda}_{j}(\delta)} < N_t$ almost surely as $\frac{U_j}{N_t} \rightarrow \beta_j$, $j = 1, \ldots, L$), so that zero-forcing minimizes the left-hand side of the inequality.
\item Otherwise, the left-hand side of the inequalities are equal to
$\frac{1}{\frac{1}{N_t} \mathbf{h}_{u,k,k}\boldsymbol{\Sigma}_{u,k}^{-1}(\delta)\mathbf{h}_{u,k,k}^H}$. This will converge almost surely to $\frac{\bar{\lambda}_k(\delta)}{\gamma_k(\delta)}$; see Appendix \ref{app:RMT} for details.
\end{itemize}

Since $\gamma_{k}(-\delta) \le  \gamma_{k}$ implies that $\frac{1}{\gamma_{k}(-\delta)} \ge \frac{1}{\gamma_{k}}$, one can easily verify that $\bar{\boldsymbol{\lambda}}(-\delta)$ yields, for large enough $N_t$ a feasible solution to $\mathcal{P}\left(\boldsymbol{\bar{\mu}}(-\delta), \boldsymbol{\gamma}\right)$.
Since $\boldsymbol{\bar{\mu}}(-\delta)$ satisfies the feasibility constraints on $\boldsymbol{\mu}$ of $\mathcal{P}_{\textrm{dual}}\left(\boldsymbol{\gamma}\right)$,
 $\sigma^2 \sum_{u,k} \frac{\bar{\lambda}_k(-\delta)}{N_t} \rightarrow \sigma^2 \sum_k \beta_k \bar{\lambda}_k(-\delta)$ lower bounds the optimum of  $\mathcal{P}_{\textrm{dual}}\left(\boldsymbol{\gamma}\right)$ along the subsequence and on the events $B(\delta)$.

Since $\bomu \leq \bobmu + \delta \bone$ along the subsequence, we have by Lemma~\ref{lbl:fsm} (1) that
\begin{equation}
\bol(\bomu,\bog) \leq \bol(\bobmu + \delta \bone, \bog), \label{eq:bound1}
\end{equation}
 and by Lemma~\ref{lbl:fsm} (2) we have that
\begin{equation}
 \bol(\bobmu + \delta \bone, \bog) \leq \bol(\bobmu + \delta \bone, \bog + \delta \bone). \label{eq:bound2}
 \end{equation}
  But for $N_t$ sufficiently large, all $\gamma_{u,k}$ in cell $k$ will be arbitrarily close to $\gamma_k + \delta$, under the policy $(\bobld, \bobmud)$, so it follows from \eqref{eq:bound1},\eqref{eq:bound2} that the $\lambda_{u,k}$ under the optimal policy, $\bol(\bomu,\bog)$, will be upper bounded by $\bal_k(\delta)$, along the subsequence and on the event $B(\delta)$.

Finally, we note that the function $\bar{\boldsymbol{\lambda}}(\cdot): {\mathbb R} \rightarrow {\mathbb R}^L$ defined above is continuous in $\delta$ (for $|\delta|$ sufficiently small). This follows from the implicit function theorem, and is proven in Appendix~\ref{app:implict_function}. Thus, $\bobld \rightarrow \bobl$ as $\delta \rightarrow 0$, which implies that the dual objective value is approaching $\sigma^2 \sum_k \beta_k \bar{\lambda}_k$ along the subsequence and on the events $B(\delta)$. If this is not the maximum value of $\mathcal{P}_{\textrm{dual}}^{\infty}$ then we get a contradiction of the optimality along the subsequence, on the events $B(\delta)$, since a deterministic approach using the optimal solution to $\mathcal{P}_{\textrm{dual}}^{\infty}$ would be better with nonzero probability. By Lemma~\ref{lem:dual_uniqueness} that solution is unique, and hence the optimal solution must converge to it.

\section{Convexity proof of constraint \eqref{eq:SINR_LAS_constraint}}\label{proof:SINR_LAS_constraint}
As mentioned in Appendix \ref{app:asymp_lemmas}, constraints \eqref{eq:SINR_LAS_constraint} can be rewritten as
\begin{align}
\boldsymbol{\lambda} \le \mathbf{F}(\boldsymbol{\lambda}, \boldsymbol{\mu}, \boldsymbol{\gamma}).
\end{align}
 Thus, $F_k(\boldsymbol{\lambda}, \mu_k, \gamma_k) = 0$ only corresponds to a feasible set of variables  \emph{if and only if} $\lambda_k$ is zero.

Let $\left(\boldsymbol{\lambda}^{(i)}, \boldsymbol{\mu}^{(i)}\right) = \left(\{\lambda^{(i)}_k\}_{k=1}^L, \mu_k^{(i)}\right)$, $i = 0, 1$ correspond to feasible sets of parameters with respect to the constraints.

Taking a convex combination $\left(\boldsymbol{\lambda}, \boldsymbol{\mu}\right)$ such that
\begin{align}
\lambda_j &= \theta \lambda_j^{(0)} + \bar{\theta} \lambda_j^{(1)}, \quad j = 1, \ldots, L \\
\mu_j &= \theta \mu_j^{(0)} + \bar{\theta} \mu_j^{(1)}, \quad j = 1, \ldots, L,
\end{align}
where $0 < \theta < 1$, $\bar{\theta} = 1 - \theta$,
and focussing on the $k$th constraint, the following cases arise:
\begin{itemize}
\item From Proposition \ref{prop2}, $F_k(\boldsymbol{\lambda}, \mu_k, \gamma_k) = 0$ iff
\begin{align}
&\mu_k = \theta \mu_k^{(0)} + \bar{\theta} \mu_k^{(1)} = 0 \nonumber \\
& \sum_{j}  \beta_j I_{\theta \lambda_j^{(0)} + \bar{\theta} \lambda_j^{(1)}} \le 1 \nonumber
\end{align}
This can only hold if $\mu_k^{(i)} = 0$, and $\sum_{j} \beta_j I_{\lambda_j^{(i)}} \le 1$ for $i = 0, 1$, i.e. if
$F_k(\boldsymbol{\lambda}^{(i)}, \mu_k^{(i)}) = 0$ for $i = 0, 1$. As noted above, this can only correspond to a feasible set if $\lambda_{k}^{(i)} = 0$ for $i = 0, 1$.
The convex combination in this case clearly also corresponds to a feasible set of parameters.
\item $F_k(\boldsymbol{\lambda}^{(i)}, \mu_k^{(i)}, \gamma_k) = 0$ for $i=0,1$ but $ \sum_{j}  \beta_j I_{\theta \lambda_j^{(0)} + \bar{\theta} \lambda_j^{(1)}} > 1$. In this case,
$\lambda_j = 0$ and $F_k(\boldsymbol{\lambda}, \mu_k) > 0$, and the constraint clearly holds.
\item Only one of the $F_k(\boldsymbol{\lambda}^{(i)}, \mu_k^{(i)}, \gamma_k) = 0$. Without loss of generality assume this $i$ is $1$, then $\mu_k^{(1)}$ and $\lambda_k^{(1)}$ must both be zero. Moreover,
$F_k(\boldsymbol{\lambda}, \mu_k, \gamma_k) > 0$ and is the unique strictly positive zero of $g_k\left(y, \boldsymbol{\lambda}, \theta \mu_k^{(0)}, \gamma_k \right)$, defined in Appendix \ref{app:asymp_lemmas}.

Let $y_0 \triangleq F_k(\boldsymbol{\lambda}^{(0)}, \mu_k^{(0)}, \gamma_k)$. Then, by definition (refer to Appendix \ref{app:asymp_lemmas}), $g_k\left(y_0, \boldsymbol{\lambda}^{(0)}, \mu_k^{(0)}, \gamma_k \right) = 0$. Thus,
\begin{align*}
&g_k\left(\theta y_0, \boldsymbol{\lambda}, \mu_k, \gamma_k\right) \nonumber \\
&=
g_k\left(\theta y_0, \boldsymbol{\lambda}, \theta \mu_k^{(0)}, \gamma_k \right)  - g_k\left(y_0, \boldsymbol{\lambda}^{(0)}, \mu_k^{(0)}, \gamma_k \right)
\\
&=  \sum_j \frac{\frac{\gamma_k}{\epsilon_{k,k}y_0} \beta_j \epsilon_{j,k}\bar{\theta} \lambda_j^{(1)}
}{\left(\theta + \epsilon_{j,k} \lambda_j \frac{\gamma_k}{\epsilon_{k,k}y_0}\right) \left(1+ \epsilon_{j,k} \lambda_j^{(0)} \frac{\gamma_k}{\epsilon_{k,k} y_0}\right)} \\
&> 0, ~{\textrm{ provided at least one of the $\lambda_j^{(1)}$'s is not zero.}}
\end{align*}
Since $g_k$ is strictly decreasing in $y$, this implies that
$g_k\left(y, \boldsymbol{\lambda}, \mu_k, \gamma_k \right)$'s zero is strictly greater than $\theta y_0$. In other words,
\begin{align}
F_k\left(\boldsymbol{\lambda}, \theta \mu_k^{(0)}, \gamma_k\right) > \theta F_k(\boldsymbol{\lambda}^{(0)}, \mu_k^{(0)}, \gamma_k).
\end{align}
Since $\lambda_k^{(0)} \le F_k\left(\boldsymbol{\lambda}^{(0)}, \mu_k^{(0)}, \gamma_k\right)$, this implies that
\begin{align}
\lambda_k &= \theta \lambda_k^{(0)} \nonumber \\
&\le \theta F_k(\boldsymbol{\lambda}^{(0)}, \mu_k^{(0)}, \gamma_k) <
F_k\left(\boldsymbol{\lambda}, \mu_k, \gamma_k\right), \label{eq:OneZeroOnly}
\end{align}
i.e., the new set of parameters is also feasible.
\item $y_i \triangleq F_k\left(\boldsymbol{\lambda}^{(i)}, \mu_k^{(i)}, \gamma_k\right) > 0$, $i = 0, 1$.
\begin{align}
&g_k\left(\theta y_0 + \bar{\theta} y_1, \boldsymbol{\lambda}, \mu_k, \gamma_k \right) \nonumber \\
&=
\frac{\gamma_k}{\epsilon_{k,k} \left(\theta y_0 + \bar{\theta} y_1\right)} \nonumber \\
&~~~~\left[\theta \mu_k^{(0)} + \bar{\theta} \mu_k^{(1)} + \sum_j \frac{\beta_j \epsilon_{j,k} \lambda_j}{1+ \frac{\epsilon_{j,k} \lambda_j  \gamma_k}{\epsilon_{k,k} \left(\theta y_0 + \bar{\theta} y_1\right)}}\right]-1 \nonumber
\\
&\stackrel{(a)}{=}
\frac{\gamma_k}{\epsilon_{k,k} \left(\theta y_0 + \bar{\theta} y_1\right)} \nonumber \\
&~~~~\left[ -\sum_j \frac{\beta_j \epsilon_{j,k} \theta \lambda_j^{(0)}}{1+ \epsilon_{j,k} \lambda_j^{(0)} \frac{\gamma_k}{\epsilon_{k,k} y_0}}
-\sum_j \frac{\beta_j \epsilon_{j,k} \bar{\theta} \lambda_j^{(1)}}{1+ \epsilon_{j,k} \lambda_j^{(1)} \frac{\gamma_k}{\epsilon_{k,k} y_1}} \right. \nonumber \\
& \left.
 + \sum_j \frac{\beta_j \epsilon_{j,k} \left(\theta \lambda_j^{(0)} + \bar{\theta} \lambda_j^{(1)}\right)}{1+ \frac{\epsilon_{j,k} \lambda_j  \gamma_k}{\epsilon_{k,k} \left(\theta y_0 + \bar{\theta} y_1\right)}}\right] \nonumber
\\
&=
\frac{\gamma_k^2 \theta \bar{\theta}}{\epsilon_{k,k}^2 \left(\theta y_0 + \bar{\theta} y_1\right)^2 y_0 y_1}\nonumber \\
&~~~\sum_j  \frac{\beta_j \epsilon_{j,k}^2 \left(\lambda_j^{(0)} y_1 -  \lambda_j^{(1)} y_{0}\right)^2}{\left(1+ \frac{\epsilon_{j,k} \lambda_j  \gamma_k}{\epsilon_{k,k} \left(\theta y_0 + \bar{\theta} y_1\right)}\right) \prod_{i=0}^1 \left(1+ \epsilon_{j,k} \lambda_j^{(i)} \frac{\gamma_k}{\epsilon_{k,k} y_{i}}\right)}, \label{eq:pos}
\end{align}
where in $(a)$ we eliminated the $\mu_k^{(i)}$'s by using $g_k\left(y_i, \boldsymbol{\lambda}^{(i)}, \mu_k^{(i)}, \gamma_k \right) = 0$.
Clearly, \eqref{eq:pos} is nonnegative. We can verify that it is \emph{strictly positive unless $\left(\boldsymbol{\lambda}^{(i)}, \mu_k^{(i)}\right)$, $i = 0, 1$  are colinear}.
As a result, since $g_k(y, \boldsymbol{\lambda}, \mu_k, \gamma_k)$ is strictly decreasing in $y$, $F_k(\boldsymbol{\lambda}, \mu_k, \gamma_k)$ must be greater than $\theta y_0 + \bar{\theta} y_1$, leading to
\begin{align}
\lambda_k &= \theta \lambda_k^{(0)} + \bar{\theta}  \lambda_k^{(1)} \nonumber \\
&\le \theta y_0 + \bar{\theta}  y_1 \le F_k(\boldsymbol{\lambda}, \mu_k, \gamma_k),\label{eq:TwoNonZero}
\end{align}
where the inequality is strict unless $\left(\boldsymbol{\lambda}^{(i)}, \mu_k^{(i)}\right)$, $i = 0, 1$  are colinear.
Thus the convex combination is also a feasible set of parameters.
\end{itemize}
This completes the proof of the convexity of the constraint.

\section{Proof of uniqueness of solution to $\mathcal{P}_{\textrm{dual}}^{\infty}$ \label{proof:dual_uniqueness}}
Suppose there are two distinct solutions, $(\bomu^{(b)}, \bol^{(b)})$, $b = 0,1$, to problem $\mathcal{P}_{\textrm{dual}}^{\infty}$.
Consider the following convex combination, with $\theta \in (0, 1)$, $\bar{\theta} = 1 - \theta$,
\begin{align}
\bomu(\theta) = \theta \bomu^{(0)} + \bar{\theta} \bomu^{(1)} \\
\bol(\theta) = \theta \bol^{(0)} + \bar{\theta} \bol^{(1)} \\
f_k(\theta) = \lambda_k(\theta) - F_k(\theta),
\end{align}
where, with an abuse of notation, we used $F_k(\theta)$ to denote $F_k\left(\bol(\theta), \mu_k(\theta), \gamma_k\right)$ (cf. \eqref{eq:defFk}).

By the convexity of the dual feasible region, $(\bomu(\theta),\bol(\theta))$ is dual feasible for all $\theta \in (0,1)$. By optimality of $(\bomu^{(b)}, \bol^{(b)})$, $b = 0,1$, we have $f_k(0) = f_k(1) = 0$ for all $k$. By convexity of the dual feasible region, we have $f_k(\theta) \leq 0$ for all $\theta \in (0,1)$.

Define $V(\bol) = \sum_{k=1}^L \beta_k \lambda_k$. Since we are assuming that both $(\bomu^{(b)}, \bol^{(b)})$, $b = 0,1$ are optimal solutions,
$V(\bol^{(0)}) = V(\bol^{(1)})$. Let $V_{opt}$ denote their common value. It is clear that $V(\bol(\theta)) = V_{opt}$ is constant for $\theta \in (0, 1)$.

Going back to the analysis in Appendix \ref{proof:SINR_LAS_constraint}, three different cases arise, for a given component of $\bol(\theta)$:
\begin{itemize}
\item if both $\lambda_k^{(0)}$ and  $\lambda_k^{(1)}$ are zero, then so is $\lambda_k(\theta)$; moreover, this is only possible if $\mu_k^{(0)} = \mu_k^{(1)} = 0$, so that $\mu_k(\theta) = 0$. If in addition,
$\sum_{j}  \beta_j I_{\theta \lambda_j^{(0)} + \bar{\theta} \lambda_j^{(1)}} \le 1$, then the corresponding $F_k(\theta)$ will be $0$, leading to $f_k(\theta) = 0$.
Otherwise, $f_k(\theta) < 0$.  Obviously, not all the components of $\bol(\theta)$ will be zero, so even if $f_k(\theta)=0$ for all $k$'s corresponding to zero components,
 one of the remaining two cases will apply to at least one other component.
\item if only $\lambda_k^{(b)}$ is $> 0$ ($b=0$ or $b=1$), then by \eqref{eq:OneZeroOnly} and the optimality of $(\bomu^{(b)}, \bol^{(b)})$, $b = 0, 1$,
\begin{align}
\lambda_k(\theta) = \theta \lambda_k^{(b)} = \theta F_k(b) < F_k(\theta),
\end{align}
implying that $\left(\bomu(\theta), \bol(\theta)\right)$ is an interior point of the feasibility set that achieves $V_{opt}$, a contradiction.
\item if both $\lambda_k^{(b)} > 0, ~b = 0, 1$, then by \eqref{eq:TwoNonZero} and the optimality of $(\bomu^{(b)}, \bol^{(b)})$, $b = 0, 1$,
\begin{align}
\lambda_k(\theta) &= \theta \lambda_k^{(0)} + \bar{\theta}  \lambda_k^{(1)} = \theta F_k(0) + \bar{\theta}  F_k(1) \le F_k(\theta),
\end{align}
where the inequality is strict unless $(\mu_k^{(b)}, \bol^{(b)})$, $b = 0, 1$ are colinear (cf. the proof of \eqref{eq:TwoNonZero}).
The inequality being strict leads to a contradiction, as in the previous case, since it would imply that $\left(\bomu(\theta), \bol(\theta)\right)$ is an interior point of the feasibility region.
On the other hand, the inequality is tight only if $(\mu_k^{(b)}, \bol^{(b)})$, $b = 0, 1$ are colinear: in this case, unless they are identical, they clearly cannot both be optimal, i.e. another contradiction.
\end{itemize}

\section{Proof of Lemma~\ref{lem:KKT_for_dual} \label{dualKKT}}
We differentiate \eqref{eq:Lag} with respect to $\lambda_k$, $k = 1, \ldots, L$, and $\mu_k$, $k = 1, \ldots, L$ to obtain the corresponding stationarity constraints:
\begin{align}
\frac{\partial L}{\partial \lambda_k}&= \sigma^2 \beta_k + \sum_l z_l \frac{\partial F_l\left(\boldsymbol{\lambda}, \mu_l, \gamma_l\right)}{\partial \lambda_k} - z_k \le 0,\label{eq:dLagdLambda} \\
\frac{\partial L}{\partial \mu_k} &= x_k - z + z_k\frac{\partial F_k\left(\boldsymbol{\lambda}, \mu_k, \gamma_k\right)}{\partial \mu_k}  = 0,\label{eq:dLagdMu}
\end{align}
where the inequality in \eqref{eq:dLagdLambda} is tight if the optimal $\lambda_k> 0$.
The remaining KKT conditions correspond to the complementary slackness conditions and are given by:
\begin{align}
&\mu_k x_k = 0, ~k = 1, \ldots, L\label{eq:Slack1} \\
&z\sum_k(1-\mu_k) = 0 \label{eq:Slack3}\\
&z_k \left(F_k\left(\boldsymbol{\lambda}, \mu_k, \gamma_k\right)-\lambda_k\right) = 0,  ~k = 1, \ldots, L \label{eq:Slack4}
\end{align}

To get more insight into constraints \eqref{eq:dLagdLambda} and \eqref{eq:dLagdMu}, we need to obtain expressions for $\frac{\partial F_l\left(\boldsymbol{\lambda}, \mu_k, \gamma_k\right)}{\partial \lambda_k}$ and
$\frac{\partial F_k\left(\boldsymbol{\lambda}, \mu_k, \gamma_k\right)}{\partial \mu_k}$.
We start by considering the case where $F_k\left(\boldsymbol{\lambda}, \mu_k, \gamma_k\right)$ is the unique strictly positive solution to fixed-point equation \eqref{eq:IF_FP}. Thus,
\begin{align}
\mu_k + \sum_j \frac{\beta_j \epsilon_{j,k} \lambda_j}{1+ \frac{\epsilon_{j,k} \lambda_j \gamma_k}{\epsilon_{k,k} F_k\left(\boldsymbol{\lambda}, \mu_k, \gamma_k\right)}} = \frac{\epsilon_{k,k} F_k\left(\boldsymbol{\lambda}, \mu_k, \gamma_k\right)}{\gamma_k}.
\end{align}
Differentiating both sides with respect to $\lambda_l$, we get
\begin{align}
& \frac{\partial F_k\left(\boldsymbol{\lambda}, \mu_k, \gamma_k\right)}{\partial \lambda_l}= \frac{\frac{\gamma_k}{\epsilon_{k,k}} \frac{\beta_l \epsilon_{l,k}}{\left(1+ \frac{\epsilon_{l,k} \lambda_l \gamma_k}{\epsilon_{k,k} F_k\left(\boldsymbol{\lambda}, \mu_k, \gamma_k\right)}\right)^2}}{1-
\sum_j \frac{\beta_j \epsilon_{j,k}^2 \lambda_j^2}{\left(
\frac{\epsilon_{k,k} F_k\left(\boldsymbol{\lambda}, \mu_k, \gamma_k\right)}{\gamma_k}
+ \epsilon_{j,k} \lambda_j \right)^2}}. \label{diff:FkLambdaL}
\end{align}

Similarly,
\begin{align}
&  \frac{\partial F_k\left(\boldsymbol{\lambda}, \mu_k, \gamma_k\right)}{\partial \mu_k} \nonumber \\
&= \frac{\frac{\gamma_k}{\epsilon_{k,k}}}{\left[1 -
\sum_j \frac{\beta_j \epsilon_{j,k}^2 \lambda_j^2}{\left(\frac{\epsilon_{k,k} F_k\left(\boldsymbol{\lambda}, \mu_k, \gamma_k\right)}{\gamma_k}+ \epsilon_{j,k} \lambda_j \right)^2}
 \right]} \\
&= \frac{F_k\left(\boldsymbol{\lambda}, \mu_k, \gamma_k\right)}{\left[
\mu_k + \sum_j \frac{\beta_j \epsilon_{j,k} \lambda_j}{\left(1+ \epsilon_{j,k} \lambda_j \frac{\gamma_k}{\epsilon_{k,k} F_k\left(\boldsymbol{\lambda}, \mu_k, \gamma_k\right)}\right)^2}
 \right]}.\label{diff:FkMuk}
\end{align}

Thus, whenever $F_k\left(\boldsymbol{\lambda}, \mu_k, \gamma_k\right)$ is the unique strictly positive solution to fixed-point equation \eqref{eq:IF_FP},
\begin{align}
\frac{\partial F_k\left(\boldsymbol{\lambda}, \mu_k, \gamma_k\right)}{\partial \lambda_l}=
 \frac{\partial F_k\left(\boldsymbol{\lambda}, \mu_k, \gamma_k\right)}{\partial \mu_k} \frac{\beta_l \epsilon_{l,k}}{\left(1+ \frac{\epsilon_{l,k} \lambda_l \gamma_k}{\epsilon_{k,k} F_k\left(\boldsymbol{\lambda}, \mu_k, \gamma_k\right)}\right)^2}.
\end{align}

Moreover, we can verify that
\begin{align}
 \frac{\partial F_k\left(\boldsymbol{\lambda}, \mu_k, \gamma_k\right)}{\partial \mu_k} &= \frac{\frac{\gamma_k}{\epsilon_{k,k}}}{\left[1 - \sum_{j}\beta_j I_{\lambda_j}\right]},
\end{align}
when  $\mu_k = 0$, $1 - \sum_{j}\beta_j I_{\lambda_j} \ge 0$. We may combine the two cases by writing:
\begin{align}
 \frac{\partial F_k\left(\boldsymbol{\lambda}, \mu_k, \gamma_k\right)}{\partial \mu_k}
&= \frac{\frac{\gamma_k}{\epsilon_{k,k}}}{\left[1 - \sum_{j, \lambda_j > 0} \frac{\beta_j \left(\lambda_j \epsilon_{j,k}\right)^2}{\left(\frac{1}{\bar{m}_k}+\lambda_j \epsilon_{j,k} \right)^2}  \right]} \ge 0.
\end{align}
For $\frac{\partial F_k\left(\boldsymbol{\lambda}, \mu_k, \gamma_k\right)}{\partial  \lambda_l}$, we need to distinguish between different cases:
\begin{align}
&\frac{\partial F_k\left(\boldsymbol{\lambda}, \mu_k, \gamma_k\right)}{\partial  \lambda_l} = \nonumber \\
&\left\{\begin{array}{ll}
\frac{\beta_l \epsilon_{l,k} \frac{\partial F_k\left(\boldsymbol{\lambda}, \mu_k, \gamma_k\right)}{\partial \mu_k}}{\left(1+ \frac{\epsilon_{l,k} \lambda_l \gamma_k}{\epsilon_{k,k} F_k\left(\boldsymbol{\lambda}, \mu_k, \gamma_k\right)}\right)^2}
 & \mu_k > 0,  \textrm{ or } \\
& \left(\mu_k = 0, \sum_{j}  \beta_j I_{\lambda_j} > 1 \right) \\
0 & \mu_k = 0, \sum_{j \neq l}  \beta_j I_{\lambda_j} + \beta_l \le 1 \\
\lim_{h \rightarrow 0^+} \frac{F_k(\boldsymbol{\lambda}(h), \mu_k, \gamma_k)}{h}
  & \\
 = \frac{\epsilon_{l,k} \gamma_k}{\epsilon_{k,k}} \frac{\beta_l+
\sum_{j \neq l} \beta_j I_{\lambda_j}-1}{1 -
\sum_{j \neq l} \beta_j I_{\lambda_j}} & \mu_k = 0, \lambda_l = 0, \sum_{j}  \beta_j I_{\lambda_j} \le 1  \\
& \textrm{ and } \beta_l + \sum_{j}  \beta_j I_{\lambda_j} > 1
\end{array}\right. \label{dInvMLambda}
\end{align}
where we used $\boldsymbol{\lambda}(h)$ to correspond to $\boldsymbol{\lambda}$ with the $l$th entry replaced with $h > 0$.
$F_k(\boldsymbol{\lambda}(h), \mu_k, \gamma_k)$ is the unique positive solution of $g_k\left(y, \boldsymbol{\lambda}(h), \mu_k, \gamma_k\right) = 0$.
\begin{IEEEproof}
We derive the limit in \eqref{dInvMLambda} as follows. \eqref{eq:IF_FP} for $\boldsymbol{\lambda}(h)$ and $\mu_k = 0$ is equivalent to
\begin{align}
\sum_{j \neq l} \frac{\beta_j \epsilon_{j,k} \lambda_j}{
\frac{\epsilon_{k,k} F_k(\boldsymbol{\lambda}(h), \mu_k, \gamma_k)}{\gamma_k} + \epsilon_{j,k} \lambda_j }
+ \frac{\beta_l \epsilon_{l,k} h}{\frac{\epsilon_{k,k} F_k(\boldsymbol{\lambda}(h), \mu_k, \gamma_k)}{\gamma_k}+ \epsilon_{l,k} h}
 = 1.
\end{align}
Equivalently,
\begin{align}
h
&=\frac{\frac{\epsilon_{k,k} F_k(\boldsymbol{\lambda}(h), \mu_k, \gamma_k)}{\gamma_k}
\left(1- \sum_{j \neq l} \frac{\beta_j \epsilon_{j,k} \lambda_j}{
\frac{\epsilon_{k,k} F_k(\boldsymbol{\lambda}(h), \mu_k, \gamma_k)}{\gamma_k} + \epsilon_{j,k} \lambda_j }\right)}{\epsilon_{l,k}  \left(\beta_l  - 1
+ \sum_{j \neq l} \frac{\beta_j \epsilon_{j,k} \lambda_j}{
\frac{\epsilon_{k,k} F_k(\boldsymbol{\lambda}(h), \mu_k, \gamma_k)}{\gamma_k} + \epsilon_{j,k} \lambda_j }
\right)}.
\end{align}

Now define $h(x) = \frac{1}{\epsilon_{l,k}} \frac{x \left[1 -
\sum_{j \neq l} \frac{\beta_j\lambda_j \epsilon_{j,k}}{(x+\lambda_j \epsilon_{j,k})}\right]}{\left[\beta_l + \sum_{j \neq l} \frac{\beta_j\lambda_j \epsilon_{j,k}}{(x+\lambda_j \epsilon_{j,k})} - 1\right]}$.
This is an increasing function as long $\beta_l + \sum_{j \neq l} \frac{\beta_j\lambda_j \epsilon_{j,k}}{(x+\lambda_j \epsilon_{j,k})} - 1 \ge 0$, which by assumption holds strictly at zero.

\begin{align}
&h'(x) \nonumber \\
&=
\frac{1}{\epsilon_{l,k}} \frac{\beta_l\left[1 -
\sum_{j \neq l} \frac{\beta_j(\lambda_j \epsilon_{j,k})^2}{(x+\lambda_j \epsilon_{j,k})^2}\right]-\left[1 -
\sum_{j \neq l} \frac{\beta_j\lambda_j \epsilon_{j,k}}{(x+\lambda_j \epsilon_{j,k})}\right]^2
}{\left[\beta_l + \sum_{j \neq l} \frac{\beta_j\lambda_j \epsilon_{j,k}}{(x+\lambda_j \epsilon_{j,k})} - 1\right]^2}
\end{align}

Thus, for $x$ small, we can approximate $h(x)$ as follows:
\begin{align}
h(x) &\approx h(0) + h'(0) x + \mathcal{O}(x^2) \nonumber \\
&=
\frac{1}{\epsilon_{l,k}} \frac{\left[1 -
\sum_{j \neq l} \beta_j I_{\lambda_j}\right]
}{\left[\beta_l + \sum_{j \neq l} \beta_jI_{\lambda_j} - 1\right]}
 x + \mathcal{O}(x^2)
\end{align}
The desired limit is equivalent to taking the limit as $x \rightarrow 0$ of $\frac{\gamma_k}{\epsilon_{k,k}}\frac{x}{h(x)}$.
\end{IEEEproof}

Thus, both $\frac{\partial F_l\left(\boldsymbol{\lambda}, \mu_l, \gamma_l\right)}{\partial \lambda_k}$ and $\frac{\partial F_k\left(\boldsymbol{\lambda}, \mu_k, \gamma_k\right)}{\partial \mu_k}$ are nonnegative. As a result, $z_k$ will be strictly positive ($z_k \ge \sigma^2 \beta_k$) at the optimum, as will $z$.
This means that, at the optimum,
\begin{align}
&\boldsymbol{\lambda} = \mathbf{F}\left(\boldsymbol{\lambda}, \boldsymbol{\mu}, \boldsymbol{\gamma}\right) \\
&\sum_k \left(1-\mu_k\right) = 0.
\end{align}

\noindent Now, distinguish between two cases:
\subsection{At the optimum, $1-\sum_{j \in \mathcal{K}^{\infty}_{sel}} \beta_j \ge 0$}
In this case, if $\mu_k = 0$, then $F_k\left(\boldsymbol{\lambda}, \mu_k, \gamma_k\right) = 0$, and therefore the optimal $\lambda_k$ must be equal to 0 (cf. Proposition \ref{prop1} in Appendix \ref{app:asymp_lemmas}).
I.e. $\mathcal{K}^{\infty}_{sel}$ is equal to $\mathcal{S}^c\left(\boldsymbol{\mu}\right)$ as defined in Appendix \ref{app:asymp_lemmas}.
If $\lambda_k > 0$, $\frac{\partial L}{\partial \lambda_k}= 0$ and $\frac{\partial L}{\partial \mu_k}= 0$, so that  
\begin{align}
&\sigma^2 \beta_k + \sum_{l \in \mathcal{K}^{\infty}_{sel}} z_l \frac{\partial F_l\left(\boldsymbol{\lambda}, \mu_l, \gamma_l\right)}{\partial \mu_l}
\frac{\beta_k \epsilon_{k,l}}{\left(1+ \frac{\epsilon_{k,l} \lambda_k \gamma_l}{\epsilon_{l,l} F_l\left(\boldsymbol{\lambda}, \mu_l, \gamma_l\right)}\right)^2}
 = z_k \label{eq1} \\
&z-x_k = z_k \frac{\partial F_k\left(\boldsymbol{\lambda}, \mu_k, \gamma_k\right)}{\partial \mu_k} \label{eq2}
\end{align}

Making use of \eqref{eq1}, \eqref{eq2} becomes
\begin{align}
&\sigma^2 \beta_k + \sum_{l \in \mathcal{K}^{\infty}_{sel}} \frac{\beta_k \epsilon_{k,l} \left(z-x_l\right)}{\left(1+ \frac{\epsilon_{k,l} \lambda_k \gamma_l}{\epsilon_{l,l} F_l\left(\boldsymbol{\lambda}, \mu_l, \gamma_l\right)}\right)^2}
 = \frac{z-x_k}{\frac{\partial F_k\left(\boldsymbol{\lambda}, \mu_k, \gamma_k\right)}{\partial \mu_k}}.
\end{align}

Replacing $\frac{\partial F_k\left(\boldsymbol{\lambda}, \mu_k, \gamma_k\right)}{\partial \mu_k}$ by its value (cf. \eqref{diff:FkMuk}), and using the fact that at the optimum $\lambda_k = F_k\left(\boldsymbol{\lambda}, \mu_k, \gamma_k\right)$, this simplifies to
\begin{align}
&\sigma^2 \beta_k + \sum_{l \in \mathcal{K}^{\infty}_{sel}} \frac{\beta_k \epsilon_{k,l} \left(z-x_l\right)}{\left(1+ \frac{\epsilon_{k,l} \lambda_k \gamma_l}{\epsilon_{l,l} \lambda_l}\right)^2} \nonumber \\
&= \left(z-x_k\right) \frac{\left[
\mu_k + \sum_j \frac{\beta_j \epsilon_{j,k} \lambda_j}{\left(1+ \epsilon_{j,k} \lambda_j \frac{\gamma_k}{\epsilon_{k,k} \lambda_k}\right)^2}
 \right]}{\lambda_k}.
\end{align}

This can be rearranged as
\begin{align}
&\sigma^2 \beta_k \lambda_k + \sum_{l \in \mathcal{K}^{\infty}_{sel}} \frac{\beta_k \epsilon_{k,l}  \lambda_k\left(z-x_l\right)}{\left(1+ \frac{\epsilon_{k,l} \lambda_k \gamma_l}{\epsilon_{l,l} \lambda_l}\right)^2}
-  \sum_j \frac{\beta_j \epsilon_{j,k} \lambda_j \left(z-x_k\right)}{\left(1+ \epsilon_{j,k} \lambda_j \frac{\gamma_k}{\epsilon_{k,k} \lambda_k}\right)^2}
 \nonumber \\
&= \left(z-x_k\right)  \mu_k. \label{eq:LPos2}
\end{align}
Since, $x_k \mu_k = 0$ at the optimum, this further simplifies to
\begin{align}
&\sigma^2 \beta_k \lambda_k + \sum_{l \in \mathcal{K}^{\infty}_{sel}} \frac{\beta_k \epsilon_{k,l} \lambda_k \left(z-x_l\right)}{\left(1+ \frac{\epsilon_{k,l} \lambda_k \gamma_l}{\epsilon_{l,l} \lambda_l}\right)^2}
-  \sum_j \frac{\beta_j \epsilon_{j,k} \lambda_j \left(z-x_k\right)}{\left(1+ \epsilon_{j,k} \lambda_j \frac{\gamma_k}{\epsilon_{k,k} \lambda_k}\right)^2}
 \nonumber \\
&= z \mu_k. \label{eq3}
\end{align}

\noindent Summing \eqref{eq3} over $k \in \mathcal{K}^{\infty}_{sel}$,
\begin{align}
&\sum_{k  \in \mathcal{K}^{\infty}_{sel}} \sigma^2 \beta_k \lambda_k = z\sum_{k \in \mathcal{K}^{\infty}_{sel}}  \mu_k
\end{align}
Since at the optimum $\sum_k (1-\mu_k) = 0$, this implies
\begin{align}
z = \frac{1}{L} \sum_{k} \sigma^2 \beta_k \lambda_k \label{eq:Iter2}
\end{align}

Reordering the cells so that the first $|\mathcal{K}^{\infty}_{sel}|$ indices correspond to the cells with strictly positive $\lambda_k$'s,
we can rewrite \eqref{eq:LPos2} in matrix notation as $\mathbf{A} \mathbf{p} = \mathbf{b}$, where $\mathbf{A} \in \mathbb{R}^{|\mathcal{K}^{\infty}_{sel}|\times|\mathcal{K}^{\infty}_{sel}|}$, such that
\begin{align}
\left[\mathbf{A}\right]_{k,k} &= \left[\mu_k+\sum_{j\neq k, j \in \mathcal{K}^{\infty}_{sel}} \frac{\beta_j\lambda_j \epsilon_{j,k}}{\left(1+\lambda_j \epsilon_{j,k}\frac{\gamma_k}{\epsilon_{k,k} \lambda_k}\right)^2}\right], \nonumber \\
& \quad\quad k = 1, \ldots, \mathcal{K}^{\infty}_{sel} \\
\left[\mathbf{A}\right]_{k,l} &= -\frac{\epsilon_{k,l} \lambda_k \beta_k \epsilon_{l,l}^2 \lambda_l^2}{\left(\epsilon_{l,l} \lambda_l+\lambda_k \epsilon_{k,l} \gamma_l\right)^2}, ~l,k = 1, \ldots, \mathcal{K}^{\infty}_{sel}, j \neq k \\
\left[\mathbf{p}\right]_k &= z - x_k = P_k, ~ k = 1, \ldots, \mathcal{K}^{\infty}_{sel} \\
\left[\mathbf{b}\right]_k &= \lambda_k \beta_k \sigma^2, ~ k = 1, \ldots, \mathcal{K}^{\infty}_{sel}
\end{align}
Since at least one of the $\mu_k$'s is strictly positive,  $\mathbf{A}$ is irreducibly diagonally dominant and thus non-singular, so that $\mathbf{p} = \mathbf{A}^{-1} \mathbf{b}$ will be unique. This allows us (\eqref{eq:Iter2} gives $z$) to obtain the corresponding $x_k$'s. 

Finally, $\lambda_k = F_k\left(\boldsymbol{\lambda}, \mu_k, \lambda_k\right)$ can be rewritten as
\begin{align}
\lambda_k = \frac{\gamma_k}{\epsilon_{k,k}}\left[\mu_k+\sum_{j \in \mathcal{K}^{\infty}_{sel}} \frac{\beta_j\lambda_j \epsilon_{j,k}}{1+\lambda_j \epsilon_{j,k}\frac{\gamma_k}{\epsilon_{k,k} \lambda_k}}\right], \label{eq:Iter3}
\end{align}
from which the lemma follows in this case.

\subsection{At the optimum, $1-\sum_{j \in \mathcal{K}^{\infty}_{sel}} \beta_j < 0$}
In this case, even if $\mu_k = 0$, $F_k\left(\boldsymbol{\lambda}, \mu_k, \gamma_k\right) > 0$, and therefore the optimal $\lambda_k > 0$ for all the cells (cf. Proposition \ref{prop1} in Appendix \ref{app:asymp_lemmas}): this implies that all the cells will be ``selfish", i.e.
$\mathcal{K}^{\infty}_{sel} = \{1, \ldots, L\}$. \eqref{eq:dLagdLambda}  and \eqref{eq:dLagdMu} become:
\begin{align}
&\sigma^2 \beta_k + \sum_{l=1}^L z_l \frac{\partial F_l\left(\boldsymbol{\lambda}, \mu_l, \gamma_l\right)}{\partial \mu_l}
\frac{\beta_k \epsilon_{k,l}}{\left(1+ \frac{\epsilon_{k,l} \lambda_k \gamma_l}{\epsilon_{l,l} F_l\left(\boldsymbol{\lambda}, \mu_l, \gamma_l\right)}\right)^2}
 = z_k  \\
&x_k -z + z_k \frac{\partial F_k\left(\boldsymbol{\lambda}, \mu_k, \gamma_k\right)}{\partial \mu_k} = 0,
\end{align}
where $\frac{\partial F_k\left(\boldsymbol{\lambda}, \mu_k, \gamma_k\right)}{\partial \mu_k}$ is given by \eqref{diff:FkMuk}.

Exactly the same analysis of the KKT conditions as for $\mathcal{K}^{\infty}_{sel}$  when $1-\sum_{j \in \mathcal{K}^{\infty}_{sel}} \beta_j \ge 0$ holds here, which proves the lemma.

\section{Solving $\mathcal{P}_{\textrm{dual}}^{\infty}$}\label{algo:Pdual}

Solving $\mathcal{P}_{\textrm{dual}}^{\infty}$ is equivalent to solving
\begin{align}
\textrm{max.}_{\mu_k \ge 0, \sum_k \left(1-\mu_k\right) \ge 0} g\left(\boldsymbol{\mu}\right)
\end{align}
where
\begin{align}
g\left(\boldsymbol{\mu}\right) = \max_{\boldsymbol{\lambda}, \lambda_k \ge 0, F_k\left(\boldsymbol{\lambda}, \mu_k, \gamma_k\right) \ge \lambda_k} \sigma^2 \sum_{k} \lambda_k \beta_k. \label{eq:subproblem}
\end{align}

One can verify that $g(.)$ is a concave function of $\boldsymbol{\mu}$, by noting that due to the convexity of the constraints, for two feasible $\boldsymbol{\mu}$ and $\boldsymbol{\mu}'$, letting $\boldsymbol{\lambda}$ and $\boldsymbol{\lambda}'$ denote the $\{\lambda_k\}$ vectors achieving their optima, respectively, then $\theta \boldsymbol{\lambda} + \left(1-\theta\right) \boldsymbol{\lambda}'$ corresponds to a feasible point for $\theta \boldsymbol{\mu} + (1-\theta) \boldsymbol{\mu}'$, for any $\theta \in (0, 1)$. 

One way to solve the problem is as follows:
\begin{itemize}
\item 
Initialize $\bomu$ at a point on the boundary of the feasibility region (i.e. satisfying $\mu_j \ge 0, \sum_{j=1}^L \mu_j = L$.
\item Repeat until no more improvement in the objective function: \\
For $i = 1, \ldots, L$, $j = 1, \ldots, L$, $j \neq i$,
\begin{itemize}
\item Let $c_{ij} = \mu_i^{(0)} + \mu_j^{(0)}$, where $\mu_k^{(0)}$ denotes the $k$th entry of $\bomu^{(0)}$, the value of $\bomu$ at the beginning of this step.
\item Solve
\begin{align}
\max_{
\begin{matrix}
\bomu_{-i,j} = \bomu^{(0)}_{-i,j}, \\
\mu_i + \mu_j = c_{ij}, \\
\mu_i \ge 0, \mu_j \ge 0
\end{matrix}} \max_{\begin{matrix} \bol, \lambda_k \ge 0, \\
 F_k\left(\boldsymbol{\lambda}, \mu_k, \gamma_k\right) \ge \lambda_k \end{matrix}} \sigma^2 \sum_{k} \lambda_k \beta_k, \nonumber
\end{align}
where $\bomu_{-i,j}$ denotes the components of $\bomu$ other than $i$ and $j$.
Solving the above problem requires a line search. For each value of $\bomu$, the inner optimization corresponds to solving the fixed-point equation:
\begin{align}
\boldsymbol{\lambda} = \mathbf{F}\left(\boldsymbol{\lambda}, \boldsymbol{\mu}, \boldsymbol{\gamma}\right),
\end{align}
which can be done iteratively.

\item Update $\bomu$ with the optimal values $\mu_i$ and $\mu_j$.
\end{itemize}
\end{itemize}
This generates a sequence of $\bomu$ vectors that converges to the optimal $\bomu^*$.

\section{Asymptotic optimality of downlink power allocation and weak convergence of interference}\label{weak_convergence}
One can show that (see the derivations in \cite{ZH10_archive} for example), as $N_t \rightarrow \infty$, $\frac{U_k}{N_t} \rightarrow \beta_k$, $k = 1, \ldots, L$, if $\lambda_k > 0$,
\begin{align}
&\frac{1}{N_t} \mathbf{h}_{u,k,k}\mathbf{v}_{u,k} \rightarrow \epsilon_{k,k} m_k(-\mu_k, \boldsymbol{\lambda}), ~~a.s. ~\forall u, k.
\end{align}
and\footnote{This uses the fact that \\$\frac{1}{N_t} \textrm{Tr}\left[ \left(\mu_k\mathbf{I} + \sum_{j=1, \lambda_j > 0}^L \frac{\lambda_j}{N_t} \sum_{u'} \mathbf{h}_{u',j,k}^H\mathbf{h}_{u',j,k} \right)^{-2}\right]$ converges weakly to
$\mathbb{E}_l \left[\frac{1}{(\mu_k + l)^2}\right] = -\frac{d}{d\mu_k} \mathbb{E}_l \left[\frac{1}{\mu_k + l}\right] = -\frac{dm_k(-\mu_k, \boldsymbol{\lambda})}{d\mu_k}$, where $l$ denotes an eigenvalue of  $\sum_{j=1, \lambda_j > 0}^L \frac{\lambda_j}{N_t} \sum_{u'} \mathbf{h}_{u',j,k}^H\mathbf{h}_{u',j,k}$.}
\begin{align}
\frac{1}{N_t} \|\mathbf{v}_{u,k}\|^2 \rightarrow - \epsilon_{k,k} \frac{dm_k(-\mu_k, \boldsymbol{\lambda})}{d\mu_k},
\end{align}
and
\begin{align}
&\frac{1}{N_t} \sum_{u'=1, (u',j) \neq (u,j)}^{U_j} \frac{1}{N_t}  \left|\mathbf{h}_{u',j,k} \mathbf{v}_{u,k}\right|^2 \nonumber \\
&\rightarrow \left\{\begin{array}{ll}
-\epsilon_{j,k}\epsilon_{k,k} \frac{\frac{dm_k(-\mu_k, \boldsymbol{\lambda})}{d\mu_k}}{\left(1+\lambda_j \epsilon_{j,k} m_k(-\mu_k, \boldsymbol{\lambda})\right)^2} & \lambda_j > 0
\\ -\epsilon_{j,k}\epsilon_{k,k} \frac{dm_k(-\mu_k, \boldsymbol{\lambda})}{d\mu_k} & \lambda_j = 0
\end{array}\right.\label{eq:Interference}
\end{align}
From \eqref{eq:mkDef},
\begin{align}
\frac{dm_k(-\mu_k, \boldsymbol{\lambda})}{d\mu_k} = -\frac{m_k(-\mu_k, \boldsymbol{\lambda})}{\mu_k + \sum_{j} \frac{ \beta_j\lambda_j \epsilon_{j,k}}{\left(1+\lambda_j \epsilon_{j,k} m_k(-\mu_k, \boldsymbol{\lambda})\right)^2}}.
\end{align}

Plugging these values in the left hand side of \eqref{eq:SINRConstraintPA}, using the proposed per user allocation of $p_k/N_t$, yields the stated result in Lemma~\ref{lem:asymptotic_equality}. Finally, it is easy to verify, using \eqref{eq:Interference}, that \eqref{eq:Sigma_ukPA} asymptotically converges to
\begin{align}
\sigma^2_k = \sigma^2 + \sum_{j, \lambda_j > 0} P_j \epsilon_{k,j}.
\end{align}
which gives Lemma~\ref{lemma:noise}.

\section{Proof of Lemma \ref{lemm:2cellfeas}}\label{proof:2cellfeas}
Clearly, $\mathcal{P}_{\textrm{dual}}^{\infty}$ is always feasible. However, we are only interested in the case where its optimal value is bounded.
Since constraint \eqref{eq:SINR_LAS_constraint} must be met with equality at the optimum, instead of solving $\mathcal{P}_{\textrm{dual}}^{\infty}$, we can restrict ourselves to solving the problem with constraint $\sum_{k=1}^2 \mu_k = 2$\footnote{From the KKT conditions, the associated Lagrange coefficients $z$ is always strictly positive.}.

If neither $\lambda_k$'s is zero at the optimum,
\begin{align}
&\lambda_k = \frac{\gamma_k}{\epsilon_{k,k}  m_k\left(-\mu_k, \boldsymbol{\lambda}\right)}, ~k = 1, 2 \\
&m_k(-\mu_k, \boldsymbol{\lambda}) = \frac{1}{\mu_k + \sum_{j=1}^2 \frac{ \beta_j\lambda_j \epsilon_{j,k}}{1+\lambda_j \epsilon_{j,k} m_k(-\mu_k, \boldsymbol{\lambda})}}, ~k = 1, 2. \label{eq:LambdaMu}
\end{align}
Combining these two equations, we write
\begin{align}
&\mu_k  = \frac{\epsilon_{k,k} \lambda_k}{\gamma_k}\left[1 - \frac{\beta_k\gamma_k}{1+\gamma_k} - \frac{ \beta_{\bar{k}}\lambda_{\bar{k}} \epsilon_{{\bar{k}},k}}{\frac{\epsilon_{k,k} \lambda_k}{\gamma_k}+\lambda_{\bar{k}} \epsilon_{{\bar{k}},k}}\right]. \label{eq:mu_kLambdaFunc}
\end{align}

 Using \eqref{eq:mu_kLambdaFunc}, the constraint $\sum_{k} (1-\mu_k) \le 1$ becomes
\begin{align}
\sum_{k=1}^2 \frac{\epsilon_{k,k} \lambda_k}{\gamma_k}\left[1 - \frac{\beta_k\gamma_k}{1+\gamma_k} - \frac{ \beta_{\bar{k}}\lambda_{\bar{k}} \epsilon_{{\bar{k}},k}}{\frac{\epsilon_{k,k} \lambda_k}{\gamma_k}+\lambda_{\bar{k}} \epsilon_{{\bar{k}},k}}\right] \le 2.
\end{align}
Similarly, the positivity constraint on $\mu_k$ becomes ($\frac{\epsilon_{k,k} \lambda_k}{\gamma_k}+\lambda_{\bar{k}} \epsilon_{{\bar{k}},k} > 0$, since at least one of the $\lambda_k$'s is strictly positive; also $c_k = 1- \frac{\beta_k \gamma_k}{1+\gamma_k}$):
\begin{align}
&
\frac{\epsilon_{k,k}}{\gamma_k}c_k \lambda_k + \left[c_k -  \beta_{\bar{k}}\right]\lambda_{\bar{k}} \epsilon_{{\bar{k}},k} \ge 0. \label{eq:posMu}
\end{align}

The problem thus becomes
\begin{align}
\textrm{max. } & \sigma^2 \sum_{k=1}^2 \beta_k \lambda_{k}  \nonumber \\
\textrm{s.t. }& \lambda_{k} \ge 0, ~~k = 1, 2 \nonumber \\
& \frac{\epsilon_{k,k} c_k}{\gamma_k} \lambda_k  + \lambda_{\bar{k}} \epsilon_{{\bar{k}},k} \left[c_k- \beta_{\bar{k}}\right] \ge 0, k = 1, 2 \label{eq:2cellCon1} \\
& \sum_{k=1}^2 \epsilon_{k,k} \lambda_k \frac{\frac{\epsilon_{k,k} c_k}{\gamma_k} \lambda_k  + \lambda_{\bar{k}} \epsilon_{{\bar{k}},k} \left[c_k- \beta_{\bar{k}}\right]}{\epsilon_{k,k} \lambda_k+\epsilon_{\bar{k},k} \lambda_{\bar{k}}\gamma_k} \le 2.  \label{eq:L2}
\end{align}
Note that constraints \eqref{eq:2cellCon1} are crucial for feasibility, since if one has a $\lambda_1$ and $\lambda_2$ that satisfy them, one can always scale them appropriately to ensure \eqref{eq:L2} with \eqref{eq:2cellCon1} still holding.
The conditions in the Lemma are necessary and sufficient to ensure that the set defined by constraints \eqref{eq:2cellCon1} is non-empty.
Assuming this is the case, \eqref{eq:L2} ensures none of the $\lambda_k$'s grows unbounded.

The above derivation ignored the fact that one of the $\lambda_k$'s may be zero at the optimum. We thus need to show that this solution is still contained in this new problem formulation.
 Assume without loss of generality, that the zero $\lambda_k$ corresponds to $k = 1$. For a bounded problem, this corresponds to $\bar{m}_1(-\mu_1, \boldsymbol{\lambda}) = \infty$, which requires $\mu_1 = 0$ and $\beta_{2} \le 1$, so that
\begin{align*}
&\mu_{2} = 2,
&\lambda_{2} = \frac{\gamma_{2}}{\epsilon_{2, 2}m_{2}(-2, \boldsymbol{\lambda})} = \frac{\gamma_{2}}{\epsilon_{2, 2}}\frac{2}{c_2}
\end{align*}
Clearly, both $\mu_k$'s in this case satisfy \eqref{eq:mu_kLambdaFunc}.

\eqref{eq:2cellCon1} only allows for the optimal $\lambda_1$ to be zero if $c_1 - \beta_2 \ge 0$: this is more restrictive than $\beta_2 \le 1$.
The proof is completed by noting that in the case where $\beta_2 \le 1$ but $c_1 - \beta_2 < 0$, either the problem is unbounded or has a strictly better solution.
In fact, $c_1 - \beta_2 < 0$ implies $1 < \beta_1 + \beta_2$, which makes it possible to have $\mu_1 = 0$  but both $\lambda_k$'s strictly positive: if the set corresponding to  \eqref{eq:2cellCon1} turns out to be empty, the problem will be unbounded; otherwise, we can solve \eqref{eq:LambdaMu} with $\mu_1 = 0, \mu_2 = 2$, for strictly positive $\lambda_1$ and $\lambda_2$ and verify that indeed the corresponding objective function is higher.

\section{Proof of Theorem \ref{theo:2cell}}\label{proof:theo2cell}
In terms of $\rho = \frac{\lambda_2}{\lambda_1}$ and $\lambda_1$, constraint \eqref{eq:L2} is equal to
\begin{align}
 h(\rho) \le \frac{2}{\lambda_1},
\end{align}
where $h(.)$ is as specified in \eqref{eq:Lambda1Cbf}.

We already know that this must be met with equality at the optimum (from the analysis in Appendix \ref{dualKKT}, $\sum_k \mu_k = L$ at the optimum).
 This provides us with a way to rewrite the problem in terms of $\rho$ alone.
At the optimum $h(\rho) = \frac{2}{\lambda_1}$, and the objective function, in terms of $\rho$, will be equal to
\begin{align}
\sigma^2 \frac{2\left(\beta_1 + \rho \beta_2\right)}{h(\rho)}
\end{align}
We define the value at $\infty$ in terms of a limit
\begin{align}
&\lim_{\rho\rightarrow \infty}\sigma^2 \frac{2\left(\beta_1 + \rho \beta_2\right)}{h(\rho)} \nonumber \\
&=\lim_{\rho\rightarrow \infty}\sigma^2 \frac{2\left(\beta_1 + \rho \beta_2\right)}{\frac{\epsilon_{1,1} c_1}{\gamma_1}- \frac{\beta_{2}  \epsilon_{1,1}\epsilon_{2,1}\rho}{\epsilon_{1,1} + \rho \epsilon_{2,1}  \gamma_1}+\rho\frac{\epsilon_{2,2} c_2}{\gamma_2}- \frac{\beta_{1} \epsilon_{2,2} \epsilon_{1,2}\rho}{\epsilon_{2,2} \rho + \epsilon_{1,2}  \gamma_2}} \nonumber \\
&= \sigma^2 \frac{2\beta_2\gamma_2}{\epsilon_{2,2} c_2}.
\end{align}

Thus an equivalent problem in terms of $\rho$ alone is
\begin{align}
\textrm{max. } \frac{\beta_1+\rho \beta_2}{h(\rho)} \quad \textrm{s.t. } \rho_{lo} \le \rho \le \rho_{hi},
\end{align}
where
$\rho_{lo}$ and $\rho_{hi}$ are as defined in \eqref{eq:rho_min} and \eqref{eq:rho_max}, respectively.

Taking the derivative of the objective function wrt $\rho$, we get
\begin{align}
\frac{\beta_2 h(\rho)-\left(\beta_1+\rho \beta_2\right)h'(\rho)}{h^2(\rho)}.
\end{align}
Thus the sign of the derivative depends on the numerator only.

Using $
h'(\rho) = \frac{\epsilon_{2,2} c_2}{\gamma_2}
 - \frac{\beta_{2}  \epsilon_{1,1}^2\epsilon_{2,1}}{(\epsilon_{1,1} + \rho \epsilon_{2,1}  \gamma_1)^2}- \frac{\beta_{1} \epsilon_{2,2} \epsilon_{1,2}^2  \gamma_2}{(\epsilon_{2,2} \rho + \epsilon_{1,2}  \gamma_2)^2}$, we get that
\begin{align}
& \beta_2 h(\rho)-\left(\beta_1+\rho \beta_2\right)h'(\rho) = \beta_2 g_1(\rho) - \beta_1 g_2(\rho),
\end{align}
where $g_1(.)$ and $g_2(.)$ are as defined in \eqref{eq:g1} and \eqref{eq:g2}, respectively.

\begin{align}
&\frac{\beta_2 h(\rho)-\left(\beta_1+\rho \beta_2\right)h'(\rho)}{h^2(\rho)} = \beta_1 \beta_2 \left(g_1(\rho) - g_2(\rho)\right)
\end{align}
One can verify that $g_1$ is strictly decreasing in $\rho$, whereas $g_2$ is strictly increasing.

3 different cases arise:
\begin{itemize}
\item $g_1(\rho_{lo}) - g_2(\rho_{lo}) > 0, g_1(\rho_{hi}) - g_2(\rho_{hi}) \ge 0$, thus $g_1(\rho) - g_2(\rho) > 0$ over the entire feasible range, and the optimum will be at $\rho_{hi}$.
\item $g_1(\rho_{lo}) - g_2(\rho_{lo}) \le 0, g_1(\rho_{hi}) - g_2(\rho_{hi}) < 0$, thus $g_1(\rho) - g_2(\rho) < 0$ over the entire feasible range, and the optimum will be at $\rho_{lo}$.
\item $g_1(\rho_{lo}) - g_2(\rho_{lo}) > 0, g_1(\rho_{hi}) - g_2(\rho_{hi}) < 0$, thus there must be some interior point at which $g_1(\rho) = g_2(\rho)$ and which maximizes the objective function.
\end{itemize}
Once the optimal $\rho$ is obtained, it is easy to get the optimal $\lambda_1$ and $\lambda_2$.
This concludes the proof.

\section{RMT Background}\label{app:RMT}
We are interested in
\begin{align}
\frac{1}{N_t} \mathbf{h}_{u,k,k} \left(\mu_k \mathbf{I} + \sum_{(u', j) \neq (u,k)} \mathbf{h}_{u',j,k}^H \mathbf{h}_{u',j,k}\right)^{-1} \mathbf{h}_{u,k,k}^H, \label{eq:SINR_UL}
\end{align}
for two cases:
\begin{itemize}
\item $\mu_k > 0$, \eqref{eq:SINR_UL} is equal to
\begin{align}
\frac{1}{N_t}  \mathbf{h}_{u,k,k} \left(\mu_k \mathbf{I} + \sum_{(u', j)  \neq (u,k)}\frac{\lambda_j}{N_t} \mathbf{h}_{u',j,k}^H \mathbf{h}_{u',j,k}\right)^{-1} \mathbf{h}_{u,k,k}^H.
\end{align}
Standard random matrix theory results (see Theorem 4.1 in \cite{tse_it1999} or Theorem II.1 in \cite{dai_sp03} for example, see also \cite{ZH10_archive}) show that in this case, we get that this quantity converges almost surely to $\epsilon_{k,k} m_k(-\mu_k, \boldsymbol{\lambda})$ as given in \eqref{eq:mkDef}.
\item $\mu_k = 0$, $\lambda_k > 0$ with $\sum_{j=1}^L \beta_j I_{\lambda_j} > 1$. Thus, \eqref{eq:SINR_UL} is equal to
\begin{align}
\frac{1}{\lambda_k} \frac{\lambda_k}{N_t} \mathbf{h}_{u,k,k} \left(\sum_{(u', j) \neq (u,k)} \frac{\lambda_j}{N_t} \mathbf{h}_{u',j,k}^H \mathbf{h}_{u',j,k}\right)^{-1} \mathbf{h}_{u,k,k}^H
\end{align}
For large $N_t$, there exists $c > 0$, such that the minimum eigenvalue of $\left(\sum_{(u', j) \neq (u,k)} \frac{\lambda_j}{N_t} \mathbf{h}_{u',j,k}^H \mathbf{h}_{u',j,k}\right)$ is bounded away from  $c$ with probability 1; See the discussion of Assumption 1 in \cite{wagner_arxiv10}.
This allows us to apply Lemma 5.1 in \cite{liang_it07} to show \eqref{eq:conv_muzero}.
\begin{figure*}
\begin{align}
\max_{u \le U_k} \left|\frac{1}{N_t}  \mathbf{h}_{u,k,k} \left(\sum_{(u', j)  \neq (u,k)}\frac{\lambda_j}{N_t} \mathbf{h}_{u',j,k}^H \mathbf{h}_{u',j,k}\right)^{-1} \mathbf{h}_{u,k,k}^H - \frac{\epsilon_{k,k}}{N_t} \textrm{tr} \left(\sum_{(u', j)  \neq (u,k)}\frac{\lambda_j}{N_t} \mathbf{h}_{u',j,k}^H \mathbf{h}_{u',j,k}\right)^{-1} \right| \stackrel{a.s.}{\rightarrow} 0 \label{eq:conv_muzero}
\end{align}
\end{figure*}
Also, since the minimum eigenvalue of $\left(\sum_{(u', j) \neq (u,k)} \frac{\lambda_j}{N_t} \mathbf{h}_{u',j,k}^H \mathbf{h}_{u',j,k}\right)$ is bounded away from  $c$ with probability 1, we can write
\begin{align}
&\frac{1}{N_t} \textrm{tr} \left(\sum_{(u', j) \neq (u,k)} \frac{\lambda_j}{N_t} \mathbf{h}_{u',j,k}^H \mathbf{h}_{u',j,k}\right)^{-1} \nonumber \\
&= \lim_{\mu\rightarrow 0} \frac{1}{N_t} \textrm{tr} \left(\mu\mathbf{I} + \sum_{(u', j) \neq (u,k)} \frac{\lambda_j}{N_t} \mathbf{h}_{u',j,k}^H \mathbf{h}_{u',j,k}\right)^{-1},
\end{align}
which converges almost surely, in the considered asymptotic regime to $\lim_{\mu \rightarrow 0} m_k(-\mu, \boldsymbol{\lambda}) = m_k(0, \boldsymbol{\lambda}) < \infty$. 
\end{itemize}
Thus, in both cases, \eqref{eq:SINR_UL} converges a.s. as $N_t \rightarrow \infty$, with $\frac{U_k}{N_t} \rightarrow \beta_k < \infty$ to $\epsilon_{k,k} m_k(-\mu_k, \boldsymbol{\lambda})$. Since the $\lambda_k$'s obtained by solving $\mathcal{P}^{\infty}\left(\boldsymbol{\mu}, \boldsymbol{\gamma}\right)$ are such that $\epsilon_{k,k} m_k(-\mu_k, \boldsymbol{\lambda}) = \frac{\gamma_k}{\lambda_k}$, this completes the proof.

\section{Continuity of $\bobld$}
\label{app:implict_function}

In this section, we show that the function  $\bar{\boldsymbol{\lambda}}(\cdot)$ defined in Appendix~\ref{proof:LAS_dual} is continuous in $\delta$ for sufficiently small $|\delta|$. 

For $\delta > 0$ we have that $\boldsymbol{\bar{\mu}}(\delta)$ and $\boldsymbol{\bar{\lambda}}(\delta)$ are both strictly positive. By Lemma~\ref{lemm:FP} in Appendix~\ref{app:asymp_lemmas}, $\boldsymbol{\bar{\lambda}}(\delta)$ is the unique solution to
$$\boldsymbol{\bar{\lambda}}(\delta) = \mathbf{F}(\boldsymbol{\bar{\lambda}}(\delta), \boldsymbol{\bar{\mu}}(\delta), \boldsymbol{\gamma}(\delta)),$$
where $\mathbf{F}$ is defined at the beginning of Appendix \ref{app:asymp_lemmas}. 
Thus, if we define the functions
\begin{align}
&\phi_k(\boldsymbol{\lambda}, \delta) \nonumber \\
&= \lambda_k  - \gamma_k(\delta)\frac{\bar{\mu}_k(\delta)}{\epsilon_{k,k}} - \lambda_k \sum_{j, \lambda_j > 0} \frac{\beta_j \lambda_j \epsilon_{j,k}\gamma_k(\delta)}{\epsilon_{k,k} \lambda_k +\lambda_j \epsilon_{j,k} \gamma_k(\delta)} \nonumber \\
&= \lambda_k\left[1 - \sum_{j, \lambda_j > 0} \frac{\beta_j \lambda_j \epsilon_{j,k}\gamma_k(\delta)}{\epsilon_{k,k} \lambda_k +\lambda_j \epsilon_{j,k} \gamma_k(\delta)} \right]  - \gamma_k(\delta)\frac{\bar{\mu}_k(\delta)}{\epsilon_{k,k}},  \label{defphik}
\end{align}
then
\begin{equation}
\phi(\bar{\boldsymbol{\lambda}}(\delta), \delta) = {\bf 0}.
\label{eq:ifeqn}
\end{equation}
The function $\phi$ is continuous, and we will show next that for $\delta > 0$, $\phi(\cdot, \delta)$ is locally one-to-one. The implicit function theorem \cite{kumagai80} then tells us that $\bobld$ is the unique, continuous function satisfying \eqref{eq:ifeqn}.

Let $\nabla \boldsymbol{\Phi}$ denote the matrix whose $(k,j)$th entry is the partial derivative $\frac{\partial \phi_k}{\partial \lambda_j}$.
To show that the function $\phi(\cdot, \delta)$ is locally one-to-one, we need to prove that $\nabla \boldsymbol{\Phi}$ is full-rank for $\boldsymbol{\lambda}$ in an open neighborhood of $\bobld$.

For $j \neq k$:
\begin{align}
\frac{\partial \phi_k}{\partial \lambda_j} =   - \frac{\beta_j \epsilon_{j,k} \epsilon_{k,k} \lambda_k^2 \gamma_k(\delta)}{\left(\epsilon_{k,k} \lambda_k +\lambda_j \epsilon_{j,k} \gamma_k(\delta)\right)^2}.
\end{align}
Moreover, for $j = k$:
\begin{align}
\frac{\partial \phi_k}{\partial \lambda_k} &= 1  -
\frac{\beta_k \gamma_k(\delta)}{1+\gamma_k(\delta)}
-  \sum_{j\neq k} \beta_j \left(\frac{\lambda_j \epsilon_{j,k} \gamma_k(\delta) }{\epsilon_{k,k} \lambda_k +\lambda_j \epsilon_{j,k} \gamma_k(\delta)}\right)^2.
\end{align}

Since multiplying a row or a column of a matrix by a non-zero constant does not change its rank, we verify that $\nabla \boldsymbol{\Phi}$ is full-rank by showing that $\textrm{diag}(\boldsymbol{\lambda}) \nabla \boldsymbol{\Phi}$ is. Since all its off-diagonal entries are negative, $\textrm{diag}(\boldsymbol{\lambda}) \nabla \boldsymbol{\Phi}$ will be non-singular if it is irreducibly diagonally dominant. This will be the case if
\begin{align}
&\lambda_k \frac{\partial \phi_k}{\partial \lambda_k} - \sum_{j\neq k} \lambda_j \left|\frac{\partial \phi_k}{\partial \lambda_j}\right| \nonumber \\
&=\lambda_k \left[1
-  \sum_{j} \beta_j \frac{\lambda_j \epsilon_{j,k} \gamma_k(\delta)
}{\epsilon_{k,k} \lambda_k +\lambda_j \epsilon_{j,k} \gamma_k(\delta)}\right] > 0.
\label{eq:diag_dom}
\end{align}

Since $\phi(\bobld,\delta)=0$ and $\bobmud$ is strictly positive when $\delta > 0$, it follows from \eqref{defphik} that \eqref{eq:diag_dom} holds for $\boldsymbol{\lambda} = \bobld$ when $\delta > 0$, and there must be some open neighborhood of $\bobld$ in which \eqref{eq:diag_dom} will also hold. As a result, we may apply the implicit function theorem. This argument establishes that $\bobld$ is continuous in $\delta$ for $\delta > 0$, and it is clear that $\bobld \downarrow \bobl$ as $\delta \downarrow 0$.

From Lemma~\ref{lemm:FP}, there are two cases to consider: Either $\bobl$ is strictly positive, or $\mathcal{S}(\bobmu) \neq \emptyset$ and $\bal_k = 0$ for all $k \in \mathcal{S}(\bobmu)$. In the case that $\bobl$ is strictly positive, the above argument can be extended to show that $\bobld$ is continuous for $|\delta|$ sufficiently small. In this case, the condition \eqref{eq:diag_dom} is true for $\boldsymbol{\lambda} = \bobl$ and $\delta = 0$.

Let us now consider the other case in which $\mathcal{S}(\bobmu) \neq \emptyset$ and $\bal_k = 0$ for all $k \in \mathcal{S}(\bobmu)$. In this case, we cannot show that $\nabla \boldsymbol{\Phi}$ defined above is full rank. However, we can take a similar approach if we focus on just the components in the set $\mathcal{S}(\bobmu)^c$, since $\bal_k > 0$ for $k \in \mathcal{S}(\bobmu)^c$. To this end, define, for $\delta > 0$,
\begin{align}
\tilde{\mu}_k(+\delta) = \left\{\begin{array}{ll} \bamu_k + \delta & k \in \mathcal{S}(\bobmu)^c \\ 0 & k \in \mathcal{S}(\bobmu) \end{array} \right.. \\
\tilde{\mu}_k(-\delta) = \left\{\begin{array}{ll} \bamu_k - \delta & k \in \mathcal{S}(\bobmu)^c \\ 0 & k \in \mathcal{S}(\bobmu) \end{array} \right..
\end{align}
and,
\begin{align}
\tilde{\gamma}_k(+\delta) &= \left\{\begin{array}{ll} \gamma_k + \delta & k \in \mathcal{S}(\bobmu)^c \\ \gamma_k & k \in \mathcal{S}(\bobmu) \end{array}\right. \\
\tilde{\gamma}_k(-\delta) &= \gamma_k - \delta.
\end{align}
Define $\botld$ to be the optimal solution to $\mathcal{P}^{\infty}\left(\botmud, \botgd\right)$. By Lemma~\ref{lemm:Pinf_mu_G} in Appendix~\ref{app:asymp_lemmas}, when $\delta > 0$, we have $\tl_k(\delta) = 0$ for $k \in \mathcal{S}(\bobmu)$ and $\tl_k(\delta) > 0$ for $k \in \mathcal{S}(\bobmu)^c$. We also have
$$\tl_k(\delta) = F_k(\botld, \botmud, \botgd),$$
where $F_k$ is defined in \eqref{eq:defFk}. Note that $F_k = 0$ for $k \in \mathcal{S}(\bobmu)$. Notice also that $\botld = \bobld$ for $\delta < 0$. This is not true for $\delta > 0$, but by monotonicity (see Proposition~\ref{prop2} in Appendix~\ref{app:asymp_lemmas}) we have that $\botld \leq \bobld$ for $\delta > 0$.

From now on we restrict attention to the components in $\mathcal{S}(\bobmu)^c$. So let $\botld$ and $\botmud$ now refer to the strictly positive vectors indexed by just these components\footnote{This is an abuse of notation: Later, we recover the original $L$-length vectors by appending zeros to the components in $\mathcal{S}(\bobmu)$.}. The same argument as above can then be used to show that $\botld$ is continuous in $\delta$ for $|\delta|$ sufficiently small. To see this, we need to show that the function $\tilde{\phi}(\bol,\delta)$, defined by
\begin{align}
&\tilde{\phi}_k(\boldsymbol{\lambda}, \delta) \nonumber \\
&= \lambda_k\left[1 - \sum_j \frac{\beta_j \lambda_j \epsilon_{j,k}\tilde{\gamma}_k(\delta)}{\epsilon_{k,k} \lambda_k +\lambda_j \epsilon_{j,k} \tilde{\gamma}_k(\delta)} \right]  - \tilde{\gamma}_k(\delta)\frac{\tilde{\mu}_k(\delta)}{\epsilon_{k,k}},
\end{align}
where $k \in \mathcal{S}(\bobmu)^c$ and $\bol$ is a vector of dimension $|\mathcal{S}(\bobmu)^c|$, has the property that $\tilde{\phi}(\cdot, \delta)$ is locally one-to-one. The same argument as above shows that $\textrm{diag}(\boldsymbol{\lambda}) \nabla \boldsymbol{\tilde{\Phi}}$ is nonsingular, since
\begin{align}
&\lambda_k \frac{\partial \tilde{\phi}_k}{\partial \lambda_k} - \sum_{j\neq k} \lambda_j \left|\frac{\partial \tilde{\phi}_k}{\partial \lambda_j}\right| \nonumber \\
&=\lambda_k \left[1
-  \sum_j \beta_j \frac{\lambda_j \epsilon_{j,k} \tilde{\gamma}_k(\delta)
}{\epsilon_{k,k} \lambda_k +\lambda_j \epsilon_{j,k} \tilde{\gamma}_k(\delta)}\right] > 0.
\label{eq:diag_dom2}
\end{align}
for $\bol$ in an open neighbourhood of $\botld$ for small $|\delta|$. Thus, $\botld$ is continuous in $\delta$ for $|\delta|$ sufficiently small, by the implicit function theorem. This statement remains true for the full $L$-dimensional vector, $\botld$, since the components of the full vector, in $\mathcal{S}(\bobmu)$, are fixed at zero.

Recall that $\botld = \bobld$ for $\delta < 0$ and $\botld \leq \bobld$ for $\delta > 0$. Since $\bobld \downarrow \bobl$ as $\delta \downarrow 0$, and $\botld \rightarrow \bobl$ as $\delta \downarrow 0$, it follows that $\bobld$ is continuous in $\delta$ for $|\delta|$ sufficiently small.

\bibliographystyle{IEEEbib}
\bibliography{IEEEabrv,zakhour_biblio3}
\end{document}

%% file: L3N3U2.pdf_t
\begin{picture}(0,0)%
\includegraphics{L3N3U2.pdf}%
\end{picture}%
\setlength{\unitlength}{3947sp}%
\begingroup\makeatletter\ifx\SetFigFont\undefined%
\gdef\SetFigFont#1#2#3#4#5{%
  \reset@font\fontsize{#1}{#2pt}%
  \fontfamily{#3}\fontseries{#4}\fontshape{#5}%
  \selectfont}%
\fi\endgroup%
\begin{picture}(7271,7261)(5189,-6560)
\put(6751,-2236){\makebox(0,0)[lb]{\smash{{\SetFigFont{12}{14.4}{\rmdefault}{\mddefault}{\updefault}{\color[rgb]{0,0,0}$\mathbf{h}_{1,1,2}$}%
}}}}
\put(7576,-61){\makebox(0,0)[lb]{\smash{{\SetFigFont{12}{14.4}{\rmdefault}{\mddefault}{\updefault}{\color[rgb]{0,0,0}$\mathbf{h}_{1,1,1}$}%
}}}}
\put(7801,-736){\makebox(0,0)[lb]{\smash{{\SetFigFont{12}{14.4}{\rmdefault}{\mddefault}{\updefault}{\color[rgb]{0,0,0}$\mathbf{h}_{2,1,1}$}%
}}}}
\end{picture}%